\documentclass[aps,prb,twocolumn,amsmath,amssymb,superscriptaddress,scrartcl,eqsecnum,longbibliography,nofootinbib]{revtex4-2}

\usepackage[T1]{fontenc}
\usepackage[latin9]{inputenc}
\usepackage{color}
\usepackage{verbatim}
\usepackage{amsmath}
\usepackage{amssymb}
\usepackage{graphicx}
\usepackage{mathtools}

\usepackage{tikz}
\usepackage{diagbox}
\usepackage[normalem]{ulem}

\makeatletter
\PassOptionsToPackage{caption=false}{subfig} 
\usepackage{hyperref}
\hypersetup{
breaklinks=true,
colorlinks=true,
citecolor=blue,
linkcolor=black,
filecolor=black,
urlcolor=black,
}
\IfFileExists{lmodern.sty}{\usepackage{lmodern}}{}

\renewcommand{\bar}{\overline}
\renewcommand{\tilde}{\widetilde}
\renewcommand{\hat}{\widehat}

\renewcommand{\Re}{\operatorname{Re}}
\renewcommand{\Im}{\operatorname{Im}}

\newcommand{\Tr}{\operatorname{Tr}}

\definecolor{darkred}{rgb}{0.5,0.,0.}

\makeatletter

\newcommand*{\wideboxed}[1]{\setlength{\fboxsep}{1ex}%
  \fbox{\m@th$\displaystyle#1$}}
\makeatother

\def\be{\begin{equation}}
\def\ee{\end{equation}}

\makeatother

\begin{document}

\title{Universal time evolution of string order parameter in quantum critical systems\\
with boundary invertible or non-invertible symmetry breaking}

\author{Ruhanshi Barad}
\thanks{These two authors contributed equally.}

\affiliation{School of Physics, Georgia Institute of Technology, Atlanta, GA 30332, USA}

\author{Qicheng Tang}
\thanks{These two authors contributed equally.}

\affiliation{School of Physics, Georgia Institute of Technology, Atlanta, GA 30332, USA}
\affiliation{Department of Physics, School of Science, Westlake University, Hangzhou 310030, China}
\affiliation{Institute of Natural Sciences, Westlake Institute for Advanced Study, Hangzhou 310024, China}

\author{Wei Zhu}

\affiliation{Department of Physics, School of Science, Westlake University, Hangzhou 310030, China}
\affiliation{Institute of Natural Sciences, Westlake Institute for Advanced Study, Hangzhou 310024, China}

\author{Xueda Wen}
\email{xueda.wen@physics.gatech.edu}

\affiliation{School of Physics, Georgia Institute of Technology, Atlanta, GA 30332, USA}

\begin{abstract}

The global symmetry, either invertible or non-invertible, has been extensively studied in two dimensional conformal field theories in recent years. When the theory is defined on a manifold with open boundaries, however, many interesting conformal boundary conditions will fully or partially break such global symmetry. In this work, we study the effect of symmetry-breaking boundaries or interfaces when the system is out of equilibrium. We show that the boundary or interface symmetry-breaking can be detected by the time evolution of string order parameters, which are constructed from the symmetry operators that implement the symmetry transformations. While the string order parameters are independent of time if the symmetry is preserved over the whole system, they evolve in time in a universal way if the boundary or interface breaks the symmetry. More explicitly, in the presence of boundary or interface symmetry-breaking, the string order parameters decay exponentially in time after a global quantum quench, and decay as a power-law in time after a local quantum quench. 
We also generalize our study to the case when the string order parameters are defined in a subsystem, which are related to the full counting statistics. It is found there are also universal features in the time evolution of string order parameters in this case. We verify our field theory results by studying the time evolution of these two different types of string order parameters in lattice models.
\end{abstract}
\maketitle

\tableofcontents

\section{Introduction}

Quantum critical systems with conformal boundaries or defects are ubiquitous and play an important role both in condensed matter physics and in string theory. For example, they have interesting applications in different areas from impurity problems and boundary modes in condensed matter physics to D-branes in string theory\cite{Cardy1984,Cardy1986,Cardy1989,Affleck_Ludwig_1991_A,Affleck_Ludwig_1991,1995_Affleck_review,polchinski2005string,1996_D_brane,Karch2000}.

When the system has a global symmetry, one can find that
once the system is put on a manifold with boundaries, 
many interesting boundary conditions could break the global symmetry while preserving the conformal symmetry\cite{Fuchs_1999,2000Fuchs_2,birke1999symmetry,
2001Moore,2001MooreB,2001_Gaberdiel,2002Quella,Quella_2007,2021Tong,2022_higherD_CFT,2024Herzog}. A well known and perhaps the simplest example is the quantum transverse-field Ising model at the critical point, the low energy physics of which is described by an Ising conformal field theory (CFT). 
This model has a global $\mathbb Z_2$ symmetry when there are no boundaries. If we introduce boundaries,
however, this $\mathbb Z_2$ global symmetry may be broken 
-- there are three distinct conformally invariant boundary conditions in Ising CFT, and two of them break the global $\mathbb Z_2$ symmetry, as we will review shortly.
Such symmetry breaking boundaries as well as interfaces have been extensively studied in (1+1) dimensional CFTs \cite{Fuchs_1999,2000Fuchs_2,birke1999symmetry,
2001Moore,2001MooreB,2001_Gaberdiel,2002Quella,Quella_2007,2021Tong}, and most recently in higher dimensional CFTs \cite{2022_higherD_CFT,2024Herzog}.

Then a natural questions arises: How to detect such symmetry-breaking boundaries or more generally symmetry-breaking interfaces? If there is no boundary or interface symmetry breaking, it is known that the symmetry transformation is implemented by a non-local operator acting on the whole system.
Such non-local operators correspond to codimension-1 topological defects \cite{2015_Kapustin}. Now, by introducing the symmetry breaking boundary or interface, such topological defects can no longer move freely if they touch the boundary/interface. In other words, the symmetry-breaking boundary/interface add constraints to the dynamics of topological defects. It is expected that such topological defects may provide us information of the symmetry-breaking boundary/interface. Or in the opposite way, the presence of the symmetry-breaking boundary/interface may give rise to universal features in the dynamics of topological defects.

The main focus of this work is to study the interplay between such non-local topological defects and the symmetry-breaking boundary/interface in the context of \textit{non-equilibrium physics}. Our work is inspired by the recent study in \cite{2023BSB}, where it was found that when the system is in the ground state with a boundary invertible symmetry breaking, the (log of) string order parameter, which is constructed from the symmetry operator of an invertible global symmetry, depends on the total length $L$ of the system as $-\log L$. If the symmetry is preserved over the whole system, then the (log of) string order parameter becomes zero. That is, one can detect the boundary symmetry breaking via the string order parameter when the system is in the ground state.

In this work, we hope to understand if one can detect the boundary/interface symmetry breaking when the system is in a more general state, e.g., in an out-of-equilibrium state. 
In recent years, we have seen many interesting applications of using time-dependent \textit{non-local} physical quantities, such as the entanglement entropy for an extensive subsystem, to detect out-of-equilibrium phases, e.g., in the measurement induced phase transition\cite{2019_Skinner,2018_Fisher, Li_2019_MIPT}, many-body localizations\cite{MBL_2012,MBL_2013,MBL_2008,2017_Fan}, the time-dependent driven conformal field theories\cite{2018_WenWu,2020Fan_peak}. The time evolution of the entanglement entropy have different scaling behaviors in different out-of-equilibrium phases, and therefore serve as an order parameter to distinguish different phases.
Now, for the out-of-equilibrium quantum critical systems with symmetry-breaking boundaries or interfaces, we hope to detect such symmetry breaking based on the time evolution of non-local order parameters that can be systematically constructed. 
As a remark, if one simply uses the entanglement entropy in this problem, one can find its scaling behavior cannot detect the boundary symmetry breaking, since the boundary information is hidden only in the sub-leading term. The order parameter we hope to study in this work should be able to detect the boundary symmetry breaking in its leading term of time evolution.

In short, in this work we will answer the following two questions, which are closely related to each other:
\begin{enumerate}
    \item 
When there is a symmetry breaking at the boundary or interface of the Hamiltonian, can we detect such symmetry breaking with certain order parameters when the system is out of equilibrium?
    
    \item 
For the order parameter that can detect the boundary or interface symmetry breaking, is its time evolution universal (i.e., independent of the concrete models)?

\end{enumerate}

Before we introduce the concrete setup to study these two questions, 
in the rest of this introduction, let us first introduce how to construct the non-local order parameter based on the symmetry operator that implements a symmetry transformation. 
Then we introduce several examples on the symmetry-breaking conformal boundary conditions in $(1+1)$ dimensional CFTs. Based on these two ingredients, we introduce our setup and summarize the main results of this work.

\subsection{Symmetry operator in $(1+1)d$ quantum critical systems}
\label{Sec:Intro_String}

Depending on the motivations, there may be different ways of constructing string operators in quantum many-body systems. 
For example, in the context of symmetry protected topological phases (SPTs) \cite{2009_Gu_Wen,SPT_2013}, different phases have the same symmetry.  To distinguish these different phases in one dimension, one may design the string operators that are able to detect the projective representations of the symmetries of a given state\cite{1989StringOp_z2, 2012Pollmann}.  
We recommend the recent talk by Oshikawa on the review of applications of nonlocal string order parameters in  SPT phases as well as their very recent progress \cite{Oshikawa_2024}.
These include the applications of string order parameters in both gapped SPT phases  \cite{1989StringOp_z2, 2012Pollmann,kennedy1992hidden,KT_1992,Oshikawa_1992,2013_Else,2013_Quella,2023_Oshikawa} and gapless SPT phases \cite{2021_Ruben,2021_Ryan,2021_gaplessSPT,2023_Li,2023_Wen_Potter,2023Nov_Wen_Potter}.

\medskip
In this work, to detect the boundary or interface symmetry breaking for a general out-of-equilibrium state, the string operators we use are simply from the \textit{symmetry operators} in a CFT with global symmetries, including both invertible and non-invertible ones. In (1+1)d CFT, both invertible and non-invertible (0-form) global symmetries are related to topological defect lines. We give a brief introduction to such concepts by following the nice review\cite{2023_Shao}, without going to the vast number of references on this topic. We refer the readers to the excellent reviews \cite{2022_Freed,2023_Shao,2023_Bhardwaj} on this topic for more details.

For the invertible global symmetry, which corresponds to the ordinary global symmetry, a standard way to describe it is in terms of a unitary operator $U_g$, also called symmetry operator, which commutes with the Hamiltonian $H$ as 
\be
[U_g, \,H]=0.
\ee
Here $g\in G$ is the element of the symmetry group $G$.
In the Euclidean path integral formalism, the above equation should be generalized to the invariance under deformation in spacetime. In other words, the operator $U_g$ should be topological: 
\be
\label{Topological_U}
U_g(\Sigma_{1})=U_g(\Sigma'_{1}),
\ee
where $\Sigma_1$ ($\Sigma'_1$) is a co-dimension 1 manifold, which is a one dimensional line here, and $\Sigma_1$ and $\Sigma'_1$ can be continuously deformed to each other.
Depending on where we define the one dimensional manifold $\Sigma_1$ in the two dimensional spacetime, $U_g$ may have different interpretations. Let us consider two cases that will be used later: If we define $\Sigma_1$ with a fixed time, then $U_g$ is the string operator that acts on the Hilbert space $\mathcal H$ of the system.
If $\Sigma_1$ is defined with a fixed spatial coordinate, i.e., it is extended in the time direction, then $U_g$ corresponds to a topological defect line that modifies the quantization, which gives rise to a twisted Hilbert space $\mathcal H_g$. In our later discussion, we will use the terms ``string operators'' and ``defect lines'' interchangeably when we consider the Euclidean path integral. 

Another nice property of $U_g$ in the invertible case is that they can fuse with each other as
\be
\label{group_multiply}
U_g(\Sigma_1)U_{g'}(\Sigma_1)=U_{g g'}(\Sigma_1).
\ee
In addition, each $U_g$ has an inverse $U_{g^{-1}}$ such that $U_gU_{g^{-1}}=U_{g^{-1}} U_g=1$.
This property follows from the fact that when acting on the Hilbert space, the operator $U_g$ is unitary.
Pictorially, one has 
\be
\label{Fusion_inverse}
\begin{tikzpicture}[x=0.75pt,y=0.75pt,yscale=-0.7,xscale=0.7]

\draw [color={rgb, 255:red, 208; green, 2; blue, 27 }  ,draw opacity=1 ][line width=1.0]    (110.5,101) -- (110.5,201) ;

\draw [shift={(110.5,142.7)}, rotate = 90] [fill={rgb, 255:red, 208; green, 2; blue, 27 }  ,fill opacity=1 ][line width=0.08]  [draw opacity=0] (11.61,-5.58) -- (0,0) -- (11.61,5.58) -- cycle    ;

\draw [color={rgb, 255:red, 208; green, 2; blue, 27 }  ,draw opacity=1 ][line width=1.0]    (171.5,100) -- (171.5,200) ;

\draw [shift={(171.5,156.8)}, rotate = 270] [fill={rgb, 255:red, 208; green, 2; blue, 27 }  ,fill opacity=1 ][line width=0.08]  [draw opacity=0] (11.61,-5.58) -- (0,0) -- (11.61,5.58) -- cycle    ;

\draw [color={rgb, 255:red, 208; green, 2; blue, 27 }  ,draw opacity=1 ][line width=1.0]  [dash pattern={on 5.63pt off 4.5pt}]  (289.5,101) -- (289.5,201) ;

\small
\draw (219,136.4) node   {fuse};
\draw (219,155.4) node   {$=$};
\end{tikzpicture}
\ee
where the dashed line denotes the invisible defect line (or identity). Here we label the lines with opposite orientations to indicate they correspond to $U_{g}$ and $U_{g^{-1}}$ respectively.
In our later discussion, we will not show explicitly the orientation of the defect lines in most plots, when there is no potential confusion. 

Now, for a non-invertible symmetry in (1+1)d CFT, it still corresponds to a topological line defect as described above, with the difference that now the fusion of two defect lines no longer obey a group multiplication law in \eqref{group_multiply}. Instead, the fusion now takes the form $\mathcal L_a \times \mathcal L_b=\sum_c N^c_{ab} \mathcal L_c$, where $N^c_{ab}$ is a non-negative integer.
In particular, such topological lines are not invertible, i.e., given $\mathcal L$, there does not exist another topological defect line $\mathcal L^{-1}$ such that 
$\mathcal L\times \mathcal L^{-1}=1$, which is in contrast to the invertible case in \eqref{Fusion_inverse}.
Physically, such non-invertible operators or defect lines can implement duality transformations. For example, they correspond to the Kramers-Wannier duality in the $(1+1)d$ Ising model and three-state Potts model\cite{2004Fuchs_KW} and the $T$-duality (a duality between momentum and winding modes) of the compact free boson CFT\cite{2007_Fuchs_defect_boson,2021_Chang,2021_Thorngren_Wang}.

In the following, let us illustrate these concepts with two simple examples in CFTs as well as their lattice realizations, one is the Ising CFT and the other is the free Dirac fermion CFT.
These two examples will also be discussed in the next part on the symmetry-breaking boundary conditions.
In particular, the free fermion lattice model will be used throughout this work as a numeric check for the CFT results.

\medskip
\textit{-- Ising CFT and lattice model:}

Ising CFT is one of the most simplest CFTs. It has three local Virasoro primary fields, i.e., the identity field $1$, the thermal field $\epsilon$, and the spin field $\sigma$.
The topological defect lines in (1+1)d Ising CFT have been well understood\cite{Oshikawa_1996_defectIsing,2001_Zuber,2004_Fuchs,2019Yin,2023_Shao}. There are in total three topological defect lines, which we denote by $\mathbb I$, $\eta$, and $\mathcal D$, respectively. Here $\mathbb I$ corresponds to the trivial identity line, $\eta$ generates the invertible $\mathbb Z_2$ global symmetry, under which $1$ and $\epsilon$ are even and $\sigma$ is odd, and $\mathcal D$ generates the non-invertible symmetry which corresponds to the Kramers-Wannier duality transformation.

The lattice model that realizes the Ising CFT is the quantum Ising model in the transverse field:
\be
\label{H_Ising}
H_{\text{Ising}}=-
\sum_{i=1}^N (\sigma_i^x\sigma_{i+1}^x+h\sigma_i^z),
\ee
where we consider a 1d chain of $N$ sites with periodic boundary conditions, and $\sigma_i^x$ and $\sigma_i^z$ are Pauli matrices. 
By tuning the transverse field to $h=1$, the system is at the critical point. The (invertible) $\mathbb Z_2$ symmetry is generated by the string operator  
\be
\label{eta_lattice}
\eta=\prod_{i=1}^N\sigma_i^z.
\ee
which acts on the local operator as $\eta \sigma_i^x\eta^{-1}=-\sigma_i^x$ and $\eta\sigma_i^z\eta^{-1}=\sigma_i^z$. The topological nature of $\eta$ on the lattice can be found in, e.g., \cite{2024S_SSS}.

The symmetry operator $\mathcal D$ that generates the non-invertible symmetry (or the Kramers-Wannier self-duality) on the lattice is more nontrivial, and it was recently constructed in \cite{2024_Seiberg_Shao,2024S_SSS}, which we do not review here. 
We give further details in our appendix \ref{Appendix:Ising}.

\medskip
\textit{-- Free Dirac fermion CFT and lattice model:}

Now we consider the example of free fermion Dirac CFT, with the action
\be
\label{S_Dirac}
S_{\text{Dirac}}=i\int dt dx  \, \bar{\psi} \gamma^\mu\partial_\mu \psi,
\ee
where the Gamma matrices are chosen as $\gamma^0=\sigma^x$ and $\gamma^1=-i\sigma_y$, $\psi=(\psi_R, \psi_L)$ and $\bar\psi=\psi^\dag \gamma^0$. With this convention, the action can be rewritten in terms of the left and right moving Weyl fermions as:
\be
\label{S_Dirac}
S_{\text{Dirac}}=i\int dt dx\, [\psi_L^\dag(\partial_t-\partial_x)\psi_L+\psi_R^\dag(\partial_t+\partial_x)\psi_R].
\ee
This system has a vector-$U(1)$ and axial-$U(1)$ symmetry $U(1)_V\times U(1)_A$, i.e., the action in \eqref{S_Dirac} is invariant under the phase rotations of Weyl fermions:
\be
\label{U1_V_A}
\begin{split}
&V:\quad \psi_{R,L}\mapsto e^{-i\theta} \psi_{R,L},\\
&A:\quad \psi_{R,L}\mapsto e^{\mp i\phi}\psi_{R,L}.
\end{split}
\ee
The corresponding symmetry operators are as follows:
\be
\label{UV_UA}
\begin{split}
&U_V(\theta)=e^{i\theta\, Q_V}=e^{i\theta\int dx[\psi^\dag_L(x)\psi_L(x)+\psi^\dag_R(x)\psi_R(x)]},\\
&U_A(\phi)=e^{i\phi\, Q_A}=e^{i\phi\int dx[\psi^\dag_L(x)\psi_L(x)-\psi^\dag_R(x)\psi_R(x)]},\\
\end{split}
\ee
where $\theta,\,\phi\in[0,\,2\pi)$. 
That is, $U_V(\theta)$ defines a family of invertible operators parameterized by $\theta\in S^1$, and similarly for $U_A(\phi)$.

This free fermion CFT can be realized as the low energy limit of the following lattice model at half-filling:
\be
\label{H_Dirac_lattice}
H_{\text{Dirac}}=-\frac{1}{2}\sum_i c_i^\dag c_{i+1} +h.c.
\ee
where $\{c_j,\,c_k\}=\{c_j^\dag,\,c_k^\dag\}=0$ and $\{c_j,c_k^\dag\}=\delta_{j,k}$. As recently studied in \cite{2024_Shao_fermion}, the vector charge $Q_V$ and axial charge $Q_A$ on a lattice are constructed and it is found there is a subtle feature -- they do not commute on a finite lattice, although they do commute in the continuum limit. In this work, the symmetry operator we will use in the numerical calculations is the one corresponding to the vector $U(1)$ symmetry, which is of a simple form:
\be
\label{U_theta}
U(\theta)=e^{i\theta Q_V}=e^{i\theta \sum_j c_j^\dag c_j}, \quad \theta\in \mathbb [0,\,2\pi).
\ee
One can find that it flows to $U_V(\theta)$ in \eqref{UV_UA} in the continuum and low energy limit.

\subsection{Symmetry breaking conformal boundary or interface}
\label{Sec:Sym_break_boundary}

The symmetry operators we have reviewed so far are for a CFT with a global symmetry, either invertible or non-invertible. When the CFT is defined on an open manifold, it has been known that many interesting conformal boundary conditions will (partially) break the global symmetry\cite{Fuchs_1999,2000Fuchs_2,birke1999symmetry,
2001Moore,2001MooreB,2001_Gaberdiel,2002Quella,Quella_2007,2021Tong}.

Here we will not review the history and general mechanism of symmetry-breaking conformal boundary conditions, which can be found in, e.g., Ref.\cite{Fuchs_1999,2000Fuchs_2,birke1999symmetry,
2001Moore,2001MooreB,2001_Gaberdiel,2002Quella,Quella_2007,2021Tong}. Instead, we illustrate with the two simple examples as introduced above in Sec.\ref{Sec:Intro_String},  i.e., the Ising CFT and the Dirac fermion CFT, from the point of views of both field theory and lattice models. 

\medskip
-- \textit{Boundary Ising CFT and lattice model:}

For an Ising CFT, there are in total three conformally invariant boundary conditions, which are characterized by the following three conformal boundary states:\cite{Cardy1989}
\be
\label{3_bdy_Cardy}
|+\rangle,\quad |-\rangle, \quad |\text{free}\rangle,
\ee
where $|\pm\rangle$ represents the $\mathbb Z_2$ symmetry-broken boundary states, and $|\text{free}\rangle$ represents the disordered boundary state where the $\mathbb Z_2$ symmetry is preserved.
To see the symmetry transformation more clearly, it is helpful to write down the conformal boundary states in terms of the Ishibashi states as follows:
\be
\left\{
\begin{split}
&|+\rangle=\frac{1}{\sqrt{2}}|\mathbb I\rangle+\frac{1}{\sqrt 2}|\epsilon\rangle+\frac{1}{2^{1/4}}|\sigma\rangle,\\
&|-\rangle=\frac{1}{\sqrt{2}}|\mathbb I\rangle+\frac{1}{\sqrt 2}|\epsilon\rangle-\frac{1}{2^{1/4}}|\sigma\rangle,\\
&|\text{free}\rangle=|\mathbb I\rangle-|\epsilon\rangle.
\end{split}
\right.
\ee
Here the Ishibashi states $|\mathbb I\rangle$, $|\epsilon\rangle$, and $|\sigma\rangle$ are constructed from the conformal families of the three primary fields $1$, $\epsilon$, and $\sigma$, respectively.
As briefly reviewed in the previous subsection, there are two nontrivial topological defect lines denoted by $\eta$ and $\mathcal D$, which implement the $\mathbb Z_2$ symmetry and non-invertible symmetry transformation respectively. Based on how such topological line defects act on the primary fields\cite{2023_Shao,2019Yin}, one can find that they act on the conformal boundary states as 
\be
\label{Ising_Bdry_transf}
\eta|+\rangle=|-\rangle, \quad \eta|-\rangle=|+\rangle, \quad
\eta|\text{free}\rangle=|\text{free}\rangle,
\ee
and 
\be
\begin{split}
&\mathcal D|+\rangle= |\text{free}\rangle,\quad
\mathcal D|-\rangle= |\text{free}\rangle,\\
&\mathcal D|\text{free}\rangle=  |+\rangle+|-\rangle.
\end{split}
\ee
That is, the boundary states $|+\rangle$ and $|-\rangle$ break the global $\mathbb Z_2$ symmetry and $|\text{free}\rangle$ preserves the $\mathbb Z_2$ symmetry, while all the three conformal boundary states break the non-invertible symmetry. One simple way to understand the breaking of non-invertible symmetry for the three conformal boundary states is based on Cardy's conjecture that 
the ground state of a conformal field theory deformed by relevant bulk operators corresponds to a smeared conformal boundary state\cite{Cardy_2017}. For the Ising CFT, we can consider the following deformation by the thermal operator $\epsilon$ and spin operator $\sigma$\cite{Cardy_2017}:
\be
H_{\text{deform}}=H_{\text{Ising CFT}}+t \int\hat \epsilon \,dx -h \int \hat \sigma\, dx.
\ee
The gapped ground state flows to $|+\rangle$ ($|-\rangle$) for $t=0$ and $h>0$ ($h<0$), and flows to $|\text{free}\rangle$ for $t>0$ and $h=0$.
For such gapped systems, they are not self-dual under a Kramers-Wannier mapping, and therefore do not preserve the non-invertible symmetry which is actually a Kramers-Wannier duality symmetry.

Now let us consider the lattice model in \eqref{H_Ising} defined on the semi-infinite line as:
\be
\begin{split}
\label{H_Ising_open}
H_{\text{Ising}}=&H_{\text{bulk}}+H_{\text{bdry}}
=-
\sum_{i=1}^\infty (\sigma_i^x\sigma_{i+1}^x+h\sigma_i^z)
-h_1\sigma_1^x.
\end{split}
\ee
As we choose $h_1>0$, $h_1<0$, and $h_1=0$, the corresponding boundary conditions will flow to $|+\rangle$, $|-\rangle$, and $|\text{free}\rangle$, respectively.  In particular, for $h_1\neq 0$, we have 
\be
[\eta,H_{\text{bulk}}]=0,\quad [\eta,H_{\text{bdry}}]\neq 0,
\ee
where $\eta=\prod_{i=1}\sigma_i^z$ implements the  $\mathbb Z_2$ transformation. Here the boundary magnetic field breaks the $\mathbb Z_2$ symmetry, and this is why $|+\rangle$ and $|-\rangle$ in \eqref{Ising_Bdry_transf} breaks the $\mathbb Z_2$ symmetry.

On the other hand, for the lattice realization of $\mathcal D$, which implements the non-invertible symmetry transformation and was recently constructed in \cite{2024_Seiberg_Shao,2024S_SSS}, one can find that for an open boundary system in \eqref{H_Ising_open}, one always has $[\mathcal D, H_{\text{Ising}}]\neq 0$, which is due to the breaking of translation symmetry.

\medskip
\textit{-- Boundary Dirac fermion CFT and lattice model:}

Now, let us consider the conformal boundary conditions for a Dirac fermion CFT in \eqref{S_Dirac}. One can find that each conformal boundary condition will break one of the $U(1)$ symmetries in \eqref{U1_V_A}.

Let us define the CFT on the semi-infinite line $[0,+\infty)$, i.e., the boundary is introduced at $x=0$.
One can find there are two interesting types of boundary conditions, which we call $B_\beta$ and $A_\alpha$ boundary conditions as follows:\cite{hori2003mirror}
\be
\label{CBC:fermion}
\begin{split}
&B_\beta: \quad \psi_L=e^{-i\beta} \psi_R,\quad \psi_L^\dag=e^{i\beta} \psi_R^\dag,\\
&A_\alpha: \quad \psi_L=e^{-i\alpha}\psi_R^\dag, \quad \psi_L^\dag=e^{i\alpha} \psi_R.
\end{split}
\ee
Considering the $U(1)_V\times U(1)_A$ symmetry transformation in \eqref{U1_V_A}, one can find that the $B_\beta$ boundary condition preserves the $U(1)_V$ symmetry but breaks the $U(1)_A$ symmetry, while the $A_\alpha$ boundary condition preserves the $U(1)_A$ symmetry but breaks the $U(1)_V$ symmetry.
One can further write down the conformal boundary states corresponding to the boundary conditions in \eqref{CBC:fermion} \cite{hori2003mirror}, which we don't repeat here.  

In terms of the Hamiltonian, the bulk Hamiltonian which preserves the $U(1)_V\times U(1)_A$ symmetry has the form
\be
H_{\text{bulk}}=i\int_{0}^{+\infty} \left( \psi_L^\dag \partial_x\psi_L-\psi_R^\dag \partial_x\psi_R
\right),
\ee
The boundary Hamiltonians that can give rise to the two different types of boundary condition in \eqref{CBC:fermion} are
\be
\begin{split}
&H_{\text{bdry}}^{B}\propto m \, e^{i\beta} \psi_L^\dag \psi_R+ m \, e^{-i\beta} \psi_R^\dag \psi_L,\\
&H_{\text{bdry}}^{A}\propto h \, e^{i\alpha} \psi_L^\dag \psi_R^\dag +h \, e^{-i\alpha} \psi_R\psi_L,\\
\end{split}
\ee
where $m,\,h\in \mathbb R$, and $\beta,\,\alpha\in[0,2\pi)$.
One can find that 
\be
[U_V, H_{\text{bdry}}^{B}]=0, \quad [U_A, H_{\text{bdry}}^{B}]\neq 0,
\ee
and
\be
[U_V, H_{\text{bdry}}^{A}]\neq 0, \quad [U_A, H_{\text{bdry}}^{A}]= 0,
\ee
where $U_V$ and $U_A$ are the symmetry operators given in \eqref{UV_UA}.

Now, let us consider the free fermion lattice model in \eqref{H_Dirac_lattice}. To introduce the boundary that breaks the $U(1)_V$ symmetry, one can consider the following Hamiltonian 
\be
\label{HU1_Z2}
H_{\text{Dirac}}^A=-\frac{1}{2}\sum_{i=1}^{\infty} c_i^\dag c_{i+1} +h\, c_1^\dag c_2^\dag+h.c., \quad 
h\in\mathbb{R}.
\ee
 Physically, this corresponds to connecting the end of the $1d$ chain to a superconductor with pairings. One can check explicitly that $[U(\theta), H_{\text{Dirac}}^A]\neq 0$ where $U(\theta)$ is given in \eqref{U_theta}. Similarly, to generate the boundary condition $B_\beta$ in \eqref{CBC:fermion}, one can consider a boundary Hamiltonian proportional to $mc_1^\dag c_2+h.c.$.
 Certainly, one can add boundary terms that break both $U(1)_V$ and $U(1)_A$ and then one can obtain the more general family of symmetry-breaking conformal boundary states.

In addition to the symmetry-breaking conformal boundaries, one can of course consider the more general situation of symmetry-breaking conformal interfaces \cite{2002Quella}. 
Comparing to a boundary, interface objects can be (partially) transmissible and allow for nontrivial correlations across them to host richer phenomena\cite{Oshikawa_1996_defectIsing, 2002Quella, Quella_2007, Sakai_2008_EE_interface, Eisler_2012_EE_defect, Tang_2023_interface, Cogburn_2023_lattice_RGdomain, Karch_2024_bound_interface}. 
In our interested case of inserting an interface that breaks global symmetry, one can convert the problem to the boundary case by using a folding trick\cite{Wong_1994, Oshikawa_1996_defectIsing, Bachas_2001}. 
Based on this consideration, we would expect some universal features that are common for both symmetry-breaking boundaries and interfaces. 
In fact, this is what we observed in our calculations.

Before we move on to the next step, let us comment on one more useful picture. Without boundary symmetry breaking, the symmetry operator is a topological line that can move freely in the spacetime path-integral. Now, by introducing a symmetry breaking at the boundary, the defect line in general can no longer move along the boundary, although it can still move freely and is topological in the bulk:
\be
\label{BSB_picture1}
\begin{tikzpicture}[x=0.75pt,y=0.75pt,yscale=-0.7,xscale=0.7]

\draw [color={rgb, 255:red, 0; green, 0; blue, 0 }  ,draw opacity=1 ][line width=0.75]    (110.5,101) -- (110.5,201) ;
\draw [color={rgb, 255:red, 0; green, 0; blue, 0 }  ,draw opacity=1 ][line width=0.75]    (110.5,101) -- (100.33,111) ;
\draw [color={rgb, 255:red, 0; green, 0; blue, 0 }  ,draw opacity=1 ][line width=0.75]    (110.5,111) -- (100.33,121) ;
\draw [color={rgb, 255:red, 0; green, 0; blue, 0 }  ,draw opacity=1 ][line width=0.75]    (110.17,120.33) -- (100,130.33) ;
\draw [color={rgb, 255:red, 0; green, 0; blue, 0 }  ,draw opacity=1 ][line width=0.75]    (110.5,131) -- (100.33,141) ;
\draw [color={rgb, 255:red, 0; green, 0; blue, 0 }  ,draw opacity=1 ][line width=0.75]    (110.5,141) -- (100.33,151) ;
\draw [color={rgb, 255:red, 0; green, 0; blue, 0 }  ,draw opacity=1 ][line width=0.75]    (110.17,150.33) -- (100,160.33) ;
\draw [color={rgb, 255:red, 0; green, 0; blue, 0 }  ,draw opacity=1 ][line width=0.75]    (110.5,160) -- (100.33,170) ;
\draw [color={rgb, 255:red, 0; green, 0; blue, 0 }  ,draw opacity=1 ][line width=0.75]    (110.5,170) -- (100.33,180) ;
\draw [color={rgb, 255:red, 0; green, 0; blue, 0 }  ,draw opacity=1 ][line width=0.75]    (110.17,179.33) -- (100,189.33) ;
\draw [color={rgb, 255:red, 0; green, 0; blue, 0 }  ,draw opacity=1 ][line width=0.75]    (110.5,190) -- (100.33,200) ;
\draw [color={rgb, 255:red, 208; green, 2; blue, 27 }  ,draw opacity=1 ][line width=1.0]    (111.33,146) .. controls (151.33,116) and (149.67,172.33) .. (189.67,142.33) ;
\draw [shift={(156.33,147.94)}, rotate = 212.65] [fill={rgb, 255:red, 208; green, 2; blue, 27 }  ,fill opacity=1 ][line width=0.08]  [draw opacity=0] (11.61,-5.58) -- (0,0) -- (11.61,5.58) -- cycle    ;
\draw [color={rgb, 255:red, 0; green, 0; blue, 0 }  ,draw opacity=1 ][line width=0.75]    (260.5,98.67) -- (260.5,198.67) ;
\draw [color={rgb, 255:red, 0; green, 0; blue, 0 }  ,draw opacity=1 ][line width=0.75]    (260.5,98.67) -- (250.33,108.67) ;
\draw [color={rgb, 255:red, 0; green, 0; blue, 0 }  ,draw opacity=1 ][line width=0.75]    (260.5,108.67) -- (250.33,118.67) ;
\draw [color={rgb, 255:red, 0; green, 0; blue, 0 }  ,draw opacity=1 ][line width=0.75]    (260.17,118) -- (250,128) ;
\draw [color={rgb, 255:red, 0; green, 0; blue, 0 }  ,draw opacity=1 ][line width=0.75]    (260.5,128.67) -- (250.33,138.67) ;
\draw [color={rgb, 255:red, 0; green, 0; blue, 0 }  ,draw opacity=1 ][line width=0.75]    (260.5,138.67) -- (250.33,148.67) ;
\draw [color={rgb, 255:red, 0; green, 0; blue, 0 }  ,draw opacity=1 ][line width=0.75]    (260.17,148) -- (250,158) ;
\draw [color={rgb, 255:red, 0; green, 0; blue, 0 }  ,draw opacity=1 ][line width=0.75]    (260.5,157.67) -- (250.33,167.67) ;
\draw [color={rgb, 255:red, 0; green, 0; blue, 0 }  ,draw opacity=1 ][line width=0.75]    (260.5,167.67) -- (250.33,177.67) ;
\draw [color={rgb, 255:red, 0; green, 0; blue, 0 }  ,draw opacity=1 ][line width=0.75]    (260.17,177) -- (250,187) ;
\draw [color={rgb, 255:red, 0; green, 0; blue, 0 }  ,draw opacity=1 ][line width=0.75]    (260.5,187.67) -- (250.33,197.67) ;
\draw [color={rgb, 255:red, 208; green, 2; blue, 27 }  ,draw opacity=1 ][line width=1.0]    (261.33,143.67) .. controls (296.33,172.33) and (300.33,125.67) .. (336.33,134.33) ;
\draw [shift={(304.15,142.35)}, rotate = 143.84] [fill={rgb, 255:red, 208; green, 2; blue, 27 }  ,fill opacity=1 ][line width=0.08]  [draw opacity=0] (11.61,-5.58) -- (0,0) -- (11.61,5.58) -- cycle    ;

\draw (204.67,139.73) node [anchor=north west][inner sep=0.75pt]    {$=$};

\begin{scope}[xshift=-15pt,yshift=60pt]
\node at (50pt,35pt){\textcolor[rgb]{0.29,0.56,0.89}{Boundary}};
\node at (50pt,49pt){\textcolor[rgb]{0.29,0.56,0.89}{symmetry}};
\node at (50pt,63pt){\textcolor[rgb]{0.29,0.56,0.89}{broken}};
\end{scope}

\end{tikzpicture}
\ee
Another deformation we will use frequently in our later study is the following one:
\be
\label{BSB_picture2}
\begin{tikzpicture}[x=0.75pt,y=0.75pt,yscale=-0.7,xscale=0.7]

\draw [color={rgb, 255:red, 0; green, 0; blue, 0 }  ,draw opacity=1 ][line width=0.75]    (110.5,101) -- (110.5,201) ;
\draw [color={rgb, 255:red, 0; green, 0; blue, 0 }  ,draw opacity=1 ][line width=0.75]    (110.5,101) -- (100.33,111) ;
\draw [color={rgb, 255:red, 0; green, 0; blue, 0 }  ,draw opacity=1 ][line width=0.75]    (110.5,111) -- (100.33,121) ;
\draw [color={rgb, 255:red, 0; green, 0; blue, 0 }  ,draw opacity=1 ][line width=0.75]    (110.17,120.33) -- (100,130.33) ;
\draw [color={rgb, 255:red, 0; green, 0; blue, 0 }  ,draw opacity=1 ][line width=0.75]    (110.5,131) -- (100.33,141) ;
\draw [color={rgb, 255:red, 0; green, 0; blue, 0 }  ,draw opacity=1 ][line width=0.75]    (110.5,141) -- (100.33,151) ;
\draw [color={rgb, 255:red, 0; green, 0; blue, 0 }  ,draw opacity=1 ][line width=0.75]    (110.17,150.33) -- (100,160.33) ;
\draw [color={rgb, 255:red, 0; green, 0; blue, 0 }  ,draw opacity=1 ][line width=0.75]    (110.5,160) -- (100.33,170) ;
\draw [color={rgb, 255:red, 0; green, 0; blue, 0 }  ,draw opacity=1 ][line width=0.75]    (110.5,170) -- (100.33,180) ;
\draw [color={rgb, 255:red, 0; green, 0; blue, 0 }  ,draw opacity=1 ][line width=0.75]    (110.17,179.33) -- (100,189.33) ;
\draw [color={rgb, 255:red, 0; green, 0; blue, 0 }  ,draw opacity=1 ][line width=0.75]    (110.5,190) -- (100.33,200) ;

\draw [color={rgb, 255:red, 0; green, 0; blue, 0 }  ,draw opacity=1 ][line width=0.75]    (260.5,98.67) -- (260.5,198.67) ;
\draw [color={rgb, 255:red, 0; green, 0; blue, 0 }  ,draw opacity=1 ][line width=0.75]    (260.5,98.67) -- (250.33,108.67) ;
\draw [color={rgb, 255:red, 0; green, 0; blue, 0 }  ,draw opacity=1 ][line width=0.75]    (260.5,108.67) -- (250.33,118.67) ;
\draw [color={rgb, 255:red, 0; green, 0; blue, 0 }  ,draw opacity=1 ][line width=0.75]    (260.17,118) -- (250,128) ;
\draw [color={rgb, 255:red, 0; green, 0; blue, 0 }  ,draw opacity=1 ][line width=0.75]    (260.5,128.67) -- (250.33,138.67) ;
\draw [color={rgb, 255:red, 0; green, 0; blue, 0 }  ,draw opacity=1 ][line width=0.75]    (260.5,138.67) -- (250.33,148.67) ;
\draw [color={rgb, 255:red, 0; green, 0; blue, 0 }  ,draw opacity=1 ][line width=0.75]    (260.17,148) -- (250,158) ;
\draw [color={rgb, 255:red, 0; green, 0; blue, 0 }  ,draw opacity=1 ][line width=0.75]    (260.5,157.67) -- (250.33,167.67) ;
\draw [color={rgb, 255:red, 0; green, 0; blue, 0 }  ,draw opacity=1 ][line width=0.75]    (260.5,167.67) -- (250.33,177.67) ;
\draw [color={rgb, 255:red, 0; green, 0; blue, 0 }  ,draw opacity=1 ][line width=0.75]    (260.17,177) -- (250,187) ;
\draw [color={rgb, 255:red, 0; green, 0; blue, 0 }  ,draw opacity=1 ][line width=0.75]    (260.5,187.67) -- (250.33,197.67) ;

\draw (204.67,139.73) node [anchor=north west][inner sep=0.75pt]    {$=$};

\begin{scope}[xshift=-15pt,yshift=60pt]
\node at (50pt,35pt){\textcolor[rgb]{0.29,0.56,0.89}{Boundary}};
\node at (50pt,49pt){\textcolor[rgb]{0.29,0.56,0.89}{symmetry}};
\node at (50pt,63pt){\textcolor[rgb]{0.29,0.56,0.89}{broken}};
\end{scope}

\begin{scope}[xshift=-12pt,yshift=40pt]
\draw [color={rgb, 255:red, 208; green, 2; blue, 27 }  ,draw opacity=1 ][line width=1.0]    (170,140) .. controls (193,107) and (157,90) .. (172,51) ;
\draw [shift={(171.32,89.26)}, rotate = 72.64] [fill={rgb, 255:red, 208; green, 2; blue, 27 }  ,fill opacity=1 ][line width=0.08]  [draw opacity=0] (11.61,-5.58) -- (0,0) -- (11.61,5.58) -- cycle    ;
\end{scope}

\begin{scope}[xshift=-5pt,yshift=40pt]
\draw [color={rgb, 255:red, 208; green, 2; blue, 27 }  ,draw opacity=1 ][line width=1.0]    (318,139) .. controls (305,100) and (336,96) .. (319,52) ;
\draw [shift={(322.63,88.95)}, rotate = 106.57] [fill={rgb, 255:red, 208; green, 2; blue, 27 }  ,fill opacity=1 ][line width=0.08]  [draw opacity=0] (11.61,-5.58) -- (0,0) -- (11.61,5.58) -- cycle    ;

\end{scope}

\end{tikzpicture}
\ee
That is, the topological defect line can move freely in the bulk, as long as it does not touch the symmetry-breaking boundary.
A picture similar to \eqref{BSB_picture1} and \eqref{BSB_picture2} also works for the symmetry-breaking interface.

\subsection{Setup and main results}

Having introduced the string operator as well as the boundary symmetry-breaking CFT, now we are ready to introduce our setup.
The goal is to detect the boundary or interface symmetry-breaking when the system is out of equilibrium.

More concretely, given a string operator which corresponds to the (either invertible or non-invertible) symmetry operator $\mathcal L$ in a CFT, we are interested in the time evolution $\langle \psi(t)|\mathcal L|\psi(t)\rangle$, where $|\psi(t)\rangle=e^{-iHt}|\psi_0\rangle$. If the symmetry is preserved over the whole system, then one has $[\mathcal L, H]=0$ and therefore 
\be
\label{L_trivial}
\langle \psi(t)|\mathcal L|\psi(t)\rangle=\langle \psi_0|\mathcal L|\psi_0\rangle.
\ee
That is, $\langle \psi(t)|\mathcal L|\psi(t)\rangle$ is independent of time. However, if we introduce a symmetry-breaking boundary or interface, then $[\mathcal L,H]\neq 0$ and $\langle \psi(t)|\mathcal L|\psi(t)\rangle$ will in general depend on time. 

In this work, we study the universal feature in $\langle \psi(t)|\mathcal L|\psi(t)\rangle$ for a general CFT when there is a symmetry breaking at the boundary or interface. For the time dependent states $|\psi(t)\rangle$,  we consider the two canonical setups of quantum quenches in (1+1) dimensional CFTs, i.e., the global quench and the local quench \cite{CC_Global,CC2007_local}.

We will give more details on the setups of quantum quenches in Sec.\ref{Sec:BSB}. Briefly, for the global quantum quench, we start from a short-range entangled state, which may be considered as the ground state of a gapped Hamiltonian. Then at $t=0$, we evolve this initial state with a CFT Hamiltonian where the boundary or interface is symmetry-breaking \cite{CC_Global}. For the local quantum quench, we start from two decoupled CFTs with each CFT in its ground state. Then at $t=0$, we connect the two CFTs at their ends, i.e., we change the Hamiltonian locally in space \cite{CC2007_local}. A symmetry breaking is introduced at the interface between the two CFTs when they are connected.

\medskip
For the string operator $\mathcal L$, although we are mainly interested in the case that $\mathcal L$ is defined over the whole system, we will also study the case that $\mathcal L$ is defined only within a subsystem. This later case is related to the so-called full counting statistics, 
which characterizes the charge fluctuations in a subsystem,
as extensively studied in literaturev\cite{1996_Levitov,2001_BN,2009_Klich,2014_fermion_Klich,2023_Sarang_Romain,2023_FCS_nonequilibrium}. See also Sec.\ref{Sec:Partial_String} for more discussions.

\begin{figure}[t]
\centering
\begin{tikzpicture}[x=0.75pt,y=0.75pt,yscale=-0.9,xscale=0.9]

\draw [color={rgb, 255:red, 74; green, 144; blue, 226 }  ,draw opacity=1 ][line width=1.5]    (100.33,121.33) -- (100.33,40) ;
\draw [line width=1.0]    (170.33,121) -- (170.33,39.67) ;
\draw [color={rgb, 255:red, 208; green, 2; blue, 27 }  ,draw opacity=1 ][line width=1.2]    (170.33,80.33) -- (100.33,80.67) ;
\draw [line width=1.0]    (221,120.67) -- (221,39.33) ;
\draw [line width=1.0]    (291,120.33) -- (291,39) ;
\draw [color={rgb, 255:red, 208; green, 2; blue, 27 }  ,draw opacity=1 ][line width=1.2]    (291,79.67) -- (221,80) ;

\draw [color={rgb, 255:red, 74; green, 144; blue, 226 }  ,draw opacity=1 ][line width=1.5]  [dash pattern={on 5.63pt off 4.5pt}]  (240.33,120) -- (240.33,38.67) ;

\node at (62pt,15pt){Type I:};

\begin{scope}[yshift=20pt]

\node at (60pt,95pt){Type II:};

\draw [color={rgb, 255:red, 74; green, 144; blue, 226 }  ,draw opacity=1 ][line width=1.5]    (100.33,221.33) -- (100.33,140) ;

\draw [color={rgb, 255:red, 74; green, 144; blue, 226 }  ,draw opacity=1 ][line width=1.5]  [dash pattern={on 5.63pt off 4.5pt}]  (255,221) -- (255,140) ;

\draw [line width=1.0]    (170.33,221) -- (170.33,139.67) ;
\draw [color={rgb, 255:red, 208; green, 2; blue, 27 }  ,draw opacity=1 ][line width=1.2]    (139.33,180.33) -- (100.33,180.67) ;
\draw [line width=1.0]    (221,220.67) -- (221,139.33) ;
\draw [line width=1.0]    (291,220.33) -- (291,139) ;
\draw [color={rgb, 255:red, 208; green, 2; blue, 27 }  ,draw opacity=1 ][line width=1.2]    (271,179.67) -- (240,180) ;
\end{scope}

\small
\node at (102pt,69pt){\textcolor{red}{string OP}};

\node at (50pt,49pt){\textcolor[rgb]{0.29,0.56,0.89}{symmetry}};
\node at (50pt,59pt){\textcolor[rgb]{0.29,0.56,0.89}{broken}};

\begin{scope}[xshift=148pt,yshift=-31pt]
\node at (50pt,49pt){\textcolor[rgb]{0.29,0.56,0.89}{symmetry}};
\node at (50pt,59pt){\textcolor[rgb]{0.29,0.56,0.89}{broken}};
\end{scope}

\small
\node at (100pt,165pt){\textcolor{red}{string OP}};

\end{tikzpicture}
\caption{Two types of string operators, which are illustrated here in the ground state of a finite system with open boundary conditions in the two dimensional Euclidean spacetime $z=x+i\tau$. The physical boundaries are along vertical solid lines. The global symmetry is (partially) broken along the boundary (blue solid lines) or along the conformal interface (blue dashed lines). 
}
\label{Fig:SringOP_illustration}
\end{figure}
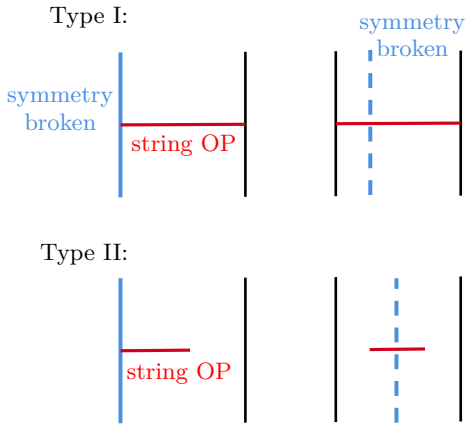

More explicitly, in this work, we study the following two types of string operators under different conditions (See also Fig.\ref{Fig:SringOP_illustration}):
\begin{enumerate}
    \item Type I: The string operator $\mathcal L$ is defined over the whole system, and the symmetry is broken along the boundary or interface. 

\item Type II: The string operator is defined only within a subsystem $A$. The symmetry is either (i) preserved over the whole system, or (ii) broken along the boundary or interface.
To distinguish from the type-I case, we will use $\mathcal L_A$ to label the string operator defined in a subsystem $A$.

\end{enumerate}

As an illustration in the path integral, for a finite system with open boundary conditions in the ground state $|G\rangle$, the expectation values of these two types of string operators, which we call type-I and type-II string order parameters,\footnote{We admit that the name of ``string order parameter'' is more appropriate for type-I string operators, which can distinguish whether there is a symmetry breaking or not at the boundary/interface. For type-II string operators, however, the leading term of the corresponding string order parameter evolution cannot detect whether the boundary/interface is symmetry-breaking or not.}
are represented in Fig.\ref{Fig:SringOP_illustration} for different cases.

\medskip

\textit{-- Main results}
\medskip

In this work, we are interested in the scaling behavior of these two types of string order parameters in non-equilibrium time evolutions.
In the following, we summarize the main results of our work.
These results hold for the string operators constructed from \textit{both invertible and non-invertible} symmetry operators.
Based on both the field theory approach and the lattice model calculations, we obtain the following main results.

-- As studied in Sec.\ref{Sec:BSB}, in the presence of boundary/interface symmetry breaking, there are universal features in the time evolution of type-I string order parameters after a quantum quench:
 
\begin{enumerate}

\item Global quench. For either an infinite system or a semi-infinite system after a global quantum quench, the string order parameter decays exponentially in time, i.e., 
\be
\label{MainResult_global1}
\log |\langle \psi(t)|\mathcal L |\psi(t)\rangle|\simeq - \kappa\, t,\quad \kappa >0,
\ee
where $\kappa$ depends on the details of the CFT and the choices of string operators.

\item Local quench. For an infinite system after a local quench, the string order parameter decays as a power-law in time, i.e., 
\be
\label{MainResult_local}
\log |\langle \psi(t)|\mathcal L|\psi(t)\rangle|\simeq - \kappa\,\log t,
\quad \kappa >0.
\ee

\end{enumerate}

-- As studied in Sec.\ref{Sec:Partial_String}, for type-II string order parameters that are defined over a finite subsystem $A$, no matter the boundary/interface breaks the symmetry or not, one can always observe a universal behavior in the time evolution of type-II string order parameters:

\begin{enumerate}

\item Semi-infinite subsystem after a global quench. When the total system is infinite and the string operator is defined over a semi-infinite subsystem, the type-II string order parameter decays exponentially in time, as described by \eqref{MainResult_global1}. But it is emphasized that the coefficients $\kappa$ for type-I and type-II string order parameters are in general different (for both global and local quantum quenches), as discussed in Sec.\ref{Period_kappa}.

\item Finite subsystem after global quench.
When the total system is semi-infinite, and the string operator is defined over a finite interval of length $l$ at the end of the system, the type-II string order parameter evolves in time as
\be
\label{Finite_typeII_Global2_main}
\log |\langle \psi(t)|\mathcal L_A|\psi(t)\rangle| \simeq
\left\{
\begin{split}
&-\kappa \, t, \quad t<l,\\
&-\kappa \, l, \quad t>l,
\end{split}
\right.
\ee
where $\kappa>0$. That is, the string order parameter decays exponentially in time first, and then saturates for $t>l$.

\item Semi-infinite subsystem after a local quench. When the total system is infinite and the string operator is defined over the semi-infinite line, one can find the type-II string order parameter decays as a power-law in time, as described by \eqref{MainResult_local}.

\end{enumerate}

In addition to the non-equilibrium cases, we also present results of type-I and type-II string order parameters in the ground state $|G\rangle$ for various interesting configurations in the appendices. In particular, in appendix \ref{Appendix:Ising}, we  
give a concrete calculation on the string order parameter that is constructed from a non-invertible symmetry operator in quantum Ising model.

\medskip
The structure of the rest of this work is organized as follows.
In Sec.\ref{Sec:BSB}, we study the universal feature in the time evolution of type-I string order parameter after a quantum quench, including both global and local quenches.
Then in Sec.\ref{Sec:Partial_String}, we study the universal time evolution of type-II string order parameter in the same setup in Sec.\ref{Sec:BSB}, by considering both symmetry-breaking and symmetry preserved boundaries/interfaces. Then we discuss the generalization of our study as well as some interesting future directions in Sec.\ref{Sec:Discuss}. There are also several appendices:
In Appendix \ref{Appendix:Ising}, we study the scaling behavior of string order parameters in the ground state of quantum Ising model,
where the string operators are constructed from both invertible and non-invertible symmetries. 
Then we study the scaling behavior of type-I and type-II string order parameters in the ground state of a general CFT with various configurations in Appendix \ref{Appendix:Ground}. In Appendix \ref{Appendix:TypeI/II}, we give more details on the features of type-I and type-II string order parameters after a quantum quench. The details of our lattice calculations are given in Appendix \ref{Appendix:Lattice}.

\section{ String order parameter over the whole system with boundary symmetry breaking}
\label{Sec:BSB}

In this section, we consider the time evolution of type-I string order parameter after a global/local quantum quench when there is a symmetry breaking at the boundary or interface.
It is noted that there are several setups of quantum quenches where the time evolution can be analytically studied \cite{CC_Global,CC2007_local,2013_Nozaki,2014_Nozaki,2014_He}. 
In this work, we consider the setups as proposed in \cite{CC_Global,CC2007_local}, which will be briefly reviewed as follows.

\subsection{Global quench: a semi-infinite system}
\label{Sec:Global1}

Let us first consider a semi-infinite system defined on $[0,+\infty)$ after a global quench. For $t=0$, the system is initially prepared in a short-range entangled state $|\psi_0\rangle$, which may be viewed as the ground state of a massive theory (or a CFT perturbed by a relevant operator).
Then at $t=0$ we switch to the critical point where the Hamiltonian is denoted as $H_{\text{CFT}}$, and let the system evolve according to $|\psi(t)\rangle=e^{-i H_{\text{CFT}}t}|\psi_0\rangle$. For the boundary condition at $x=0$, we choose the one that breaks the global symmetry. 
In the following, we study $\langle \psi(t)|\mathcal L|\psi(t)\rangle$ by using both the BCFT and lattice-model calculations. 

\subsubsection{BCFT calculation}
\label{Sec:Global1_BCFT}

Our setup for a BCFT calculation follows that proposed in Ref.\,\cite{CC_Global}. In the context of CFT, 
the initial state $|\psi_0\rangle$ can be approximated by the regularized conformal boundary state:
\be
\label{Psi0_global1}
|\psi_0\rangle=e^{-\frac{\beta}{4} H_{\text{CFT}}} |B\rangle,
\ee
where $H_{\text{CFT}}$ denotes the boundary CFT Hamiltonian and $|B\rangle$ is a conformal boundary state. Here $|B\rangle$ may be chosen to preserve the symmetry. That is, we have  
$\mathcal L |B\rangle=\langle \mathcal L\rangle |B\rangle$\footnote{Note that one has $|\langle \mathcal L\rangle|=1$ if the symmetry is invertible, which is in general not true for non-invertible symmetries. For example, as shown in \cite{2019Yin}, one has $|\langle \mathcal L\rangle|>1$ for the symmetry invariant ground state of a unitary and compact CFT.}. 
It is noted that there is no length scale in the conformal boundary state, and $|B\rangle$ itself is not normalizable and has zero real-space entanglement \cite{2015Miyaji}.
By introducing the regularization factor $e^{-\frac{\beta}{4} H_{\text{CFT}}}$ in \eqref{Psi0_global1}, $|\psi_0\rangle$ becomes normalizable and has a finite real-space entanglement~\footnote{The factor of $\frac{\beta}{4}$ is chosen to let the normalized state $| \psi_0 \rangle$ have the same expectation value of free energy as a thermal state at finite temperature $\beta^{-1}$.}. Physically, the parameter $\beta$ in \eqref{Psi0_global1} characterizes the correlation length of the initial state. A larger $\beta$ corresponds to a larger correlation length.

\begin{figure}
\centering
\begin{tikzpicture}[x=0.75pt,y=0.75pt,yscale=-0.65,xscale=0.65]


\begin{scope}[xshift=50pt, yshift=5pt]
\draw (140pt,40pt)--(155pt,40pt);
\draw (140pt,40pt)--(140pt,25pt);
\node at (147pt,30pt){$z$};

\end{scope}

\node at (10pt,100pt){$|b\rangle$};
\node at (100pt,40pt){$\langle B|$};
\node at (105pt,160pt){$|B\rangle$};

\node at (10pt,55pt){$\beta/4$};
\node at (10pt,145pt){$-\beta/4$};

\draw [color={rgb, 255:red, 74; green, 144; blue, 226 }  ,draw opacity=1 ][line width=1.5]    (43,73) -- (43,191) ;
\draw [color={rgb, 255:red, 128; green, 128; blue, 128 }  ,draw opacity=1 ][line width=1.5]    (43,73) -- (269,73) ;
\draw [color={rgb, 255:red, 128; green, 128; blue, 128 }  ,draw opacity=1 ][line width=1.5]    (43,191) -- (269,191) ;
\draw [color={rgb, 255:red, 208; green, 2; blue, 27 }  ,draw opacity=1 ][line width=1.5]    (43.32,112.53) -- (269.32,112.53) ;
\draw  [draw opacity=0][dash pattern={on 4.5pt off 4.5pt}] (44.84,98) .. controls (44.84,98) and (44.84,98) .. (44.84,98) .. controls (52.58,98) and (58.85,104.27) .. (58.85,112.02) .. controls (58.85,119.76) and (52.58,126.03) .. (44.84,126.03) -- (44.84,112.02) -- cycle ; \draw  [dash pattern={on 4.5pt off 4.5pt}] (44.84,98) .. controls (44.84,98) and (44.84,98) .. (44.84,98) .. controls (52.58,98) and (58.85,104.27) .. (58.85,112.02) .. controls (58.85,119.76) and (52.58,126.03) .. (44.84,126.03) ;  
\end{tikzpicture}
\caption{Type-I string order parameter for a semi-infinite system after a global quench. The gray lines denote the symmetry preserved conformal boundary states that are used to define the initial state. The blue line denote the symmetry-breaking boundary at the left end of the system, and the red line denotes the string operator $\mathcal L$, which is defined along the path $C=\{i\tau+x,x\ge 0\}$.
A small half-disk of radius $\epsilon$ is removed to introduce a UV cutoff, with a conformal boundary condition $|b\rangle$ imposed along its boundary.\
}
\label{Fig:Global1}
\end{figure}
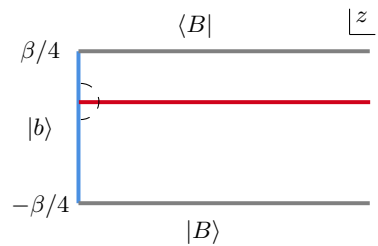

Then from $t=0$, we evolve the initial state with the critical Hamiltonian $H_{\text{CFT}}$. The time dependent wavefucntion after the quantum quench is
\be\label{psi_t_global}
|\psi(t)\rangle=e^{-iH_{\text{CFT}}}|\psi_0\rangle.
\ee
Based on \eqref{Psi0_global1} and \eqref{psi_t_global},
the string order parameter at time $t$ is
\be
\langle \psi(t)|\mathcal L|\psi(t)\rangle=\langle B|e^{-(\frac{\beta}{4}-it)H} \mathcal L\, e^{-(\frac{\beta}{4}+it)H}|B\rangle,
\ee
where we have written $H_{\text{CFT}}$ as $H$ for simplicity. In the path integral, it is convenient to consider the Euclidean spacetime, by taking $\tau=it$. Then we have
\be
\label{Global1_zplane}
\langle \psi(\tau)|\mathcal L|\psi(\tau)\rangle=
\langle B|e^{-(\frac{\beta}{4}-\tau)H} \mathcal L \,e^{-(\frac{\beta}{4}+\tau)H}|B\rangle.
\ee
Pictorially, this corresponds to the configuration in Fig.\ref{Fig:Global1}, which is a semi-infinite strip of width $\beta/2$.
That is, we start from a conformal boundary state $|B\rangle$ defined along $\Im(z)=-\beta/4$.
Then we evolve it for 
an imaginary time interval $\beta/4+\tau$, after which we insert the string operator $\mathcal L$ along $\Im(z)=\tau$. Then we evolve the state in imaginary time direction with interval
$\beta/4-\tau$ to generate another factor $e^{-(\frac{\beta}{4}-\tau)H}$.

To introduce a UV cutoff, we follow the strategy in Ref.\cite{2016CardyTonni} by removing a small half-disc with radius $\epsilon$ around $z=0+i\tau$.
Then we consider the following conformal mapping 
\be
\label{ConformalMap_Global1}
w=f(z)=\log\left(\frac{
\sinh[\pi(z-i\tau)/\beta]
}{
\cosh[\pi (z+i\tau)/\beta]
}\right),
\ee
which maps the configuration in Fig.\ref{Fig:Global1} to a $w$-strip in Fig.\ref{Fig:Global2}.
After this conformal mapping, the string operator (red solid line in Fig.\ref{Fig:Global1}) is mapped to a \textit{curve} in Fig.\ref{Fig:Global2}, and the symmetry-breaking boundary is mapped to the top and bottom edges of the strip. 

As long as the string operator does not hit the symmetry-breaking boundary, one can deform it freely to a straight line, as shown in Fig.\ref{Fig:Global2}. One can find that the width of the $w$-strip is $\pi$, and its length in the $\Re(w)$ direction is (after an analytical continuation $\tau\to it$)
\be
\label{Wt_global1}
W(t)\simeq \log\left[\frac{\cosh (2\pi t/\beta)}{\sin(\pi \epsilon/\beta)}\right]\simeq \frac{2\pi t}{\beta}+\log\frac{\beta }{2\pi  \epsilon}.
\ee

\begin{figure}
\centering
\begin{tikzpicture}[x=0.75pt,y=0.75pt,yscale=-0.75,xscale=0.75]

\begin{scope}[xshift=10pt]
\draw (140pt,40pt)--(155pt,40pt);
\draw (140pt,40pt)--(140pt,25pt);
\node at (147pt,32pt){$w$};
\end{scope}

\begin{scope}[xshift=170pt]
\draw (140pt,40pt)--(155pt,40pt);
\draw (140pt,40pt)--(140pt,25pt);
\node at (147pt,32pt){$w$};
\end{scope}

\begin{scope}[xshift=200pt]
\node at (147pt,47pt){$\pi/2$};
\node at (147pt,105pt){$-\pi/2$};
\end{scope}

\draw [color={rgb, 255:red, 74; green, 144; blue, 226 }  ,draw opacity=1 ][line width=1.0]    (66,141) -- (225.83,141.5) ;
\draw [color={rgb, 255:red, 0; green, 0; blue, 0 }  ,draw opacity=1 ][line width=0.75]  [dash pattern={on 4.5pt off 4.5pt}]  (65,59.5) -- (66,142) ;
\draw [color={rgb, 255:red, 128; green, 128; blue, 128 }  ,draw opacity=1 ][line width=1.5]    (224.83,60) -- (225.83,141.75) ;
\draw [color={rgb, 255:red, 74; green, 144; blue, 226 }  ,draw opacity=1 ][line width=1.0]    (65,60) -- (224.83,60.5) ;
\draw [color={rgb, 255:red, 208; green, 2; blue, 27 }  ,draw opacity=1 ][line width=1.0]    (65.5,100.75) .. controls (122,104) and (190,113) .. (225,124) ;
\draw [color={rgb, 255:red, 74; green, 144; blue, 226 }  ,draw opacity=1 ][line width=1.0]    (281,141) -- (440.83,141.5) ;
\draw [color={rgb, 255:red, 0; green, 0; blue, 0 }  ,draw opacity=1 ][line width=0.75]  [dash pattern={on 4.5pt off 4.5pt}]  (280,60.5) -- (281,143) ;
\draw [color={rgb, 255:red, 128; green, 128; blue, 128 }  ,draw opacity=1 ][line width=1.5]    (439.83,61) -- (440.83,142.75) ;
\draw [color={rgb, 255:red, 74; green, 144; blue, 226 }  ,draw opacity=1 ][line width=1.0]    (280,61) -- (439.83,61.5) ;
\draw [color={rgb, 255:red, 208; green, 2; blue, 27 }  ,draw opacity=1 ][line width=1.0]    (280.5,101.75) -- (440.33,102.25) ;

\draw (250,100) node {$=$};

\end{tikzpicture}
\caption{Type-I string order parameter after the conformal mapping in \eqref{ConformalMap_Global1}.
One can deform the string operator (red line) freely from the curve (left plot) to a straight line (right plot), since only the boundaries at the top and bottom of this $w$-strip break the symmetry.
}
\label{Fig:Global2}
\end{figure}

Next, to evaluate the partition function over the $w$-strip in Fig.\ref{Fig:Global2}, we consider the direction of $\Re w$ as the imaginary time direction, and the Hamiltonian is defined in the direction of $\Im w$ of length $\pi$.
For the configuration in Fig.\ref{Fig:Global2} (right plot), the effect of string operator is to introduce a defect in the Hamiltonian. 
Let us denote the Hamiltonians with/without defect as $H_{\text{defect}}$/$H_0$, then the string order parameter has the expression
\be
\langle \psi(t)|\mathcal L|\psi(t)\rangle=\frac{Z_{\text{defect}}}{Z_0}
=\frac{\langle b|e^{-H_{\text{defect}} W(t)}|B\rangle}{\langle b|e^{-H_{0} W(t)}|B\rangle}.
\label{SOP_global1}
\ee
Pictorially, we have defined
\be
\label{Z_defect_global1}
\begin{tikzpicture}[x=0.75pt,y=0.75pt,yscale=-0.6,xscale=0.6]
\draw [color={rgb, 255:red, 74; green, 144; blue, 226 }  ,draw opacity=1 ][line width=0.75]    (100.3,39.71) -- (250.3,39.71) ;
\draw [color={rgb, 255:red, 74; green, 144; blue, 226 }  ,draw opacity=1 ][line width=0.75]    (100.3,100.71) -- (250.3,100.71) ;
\draw [color={rgb, 255:red, 208; green, 2; blue, 27 }  ,draw opacity=1 ][line width=0.75]    (100.3,71.71) -- (250.3,71.71) ;
\draw  [dash pattern={on 4.5pt off 4.5pt}]  (100.3,39.71) -- (100.3,100.71) ;
\draw [color={rgb, 255:red, 155; green, 155; blue, 155 }  ,draw opacity=1 ][line width=1.5]    (250.3,40.21) -- (250.3,101.21) ;
\draw [color={rgb, 255:red, 245; green, 166; blue, 35 }  ,draw opacity=1 ]   (141.3,39.71) -- (141.3,100.71) ;

\begin{scope}[xshift=0pt,yshift=29pt]
\draw [line width=0.5] [>=stealth,<->]   (100.3,170) -- (250.3,170) ;
\end{scope}

\draw [color={rgb, 255:red, 74; green, 144; blue, 226 }  ,draw opacity=1 ][line width=0.75]    (101.3,132.71) -- (251.3,132.71) ;
\draw [color={rgb, 255:red, 74; green, 144; blue, 226 }  ,draw opacity=1 ][line width=0.75]    (101.3,193.71) -- (251.3,193.71) ;
\draw  [dash pattern={on 4.5pt off 4.5pt}]  (101.3,132.71) -- (101.3,193.71) ;
\draw [color={rgb, 255:red, 155; green, 155; blue, 155 }  ,draw opacity=1 ][line width=1.5]    (251.3,133.21) -- (251.3,194.21) ;
\draw [color={rgb, 255:red, 245; green, 166; blue, 35 }  ,draw opacity=1 ]   (142.3,132.71) -- (142.3,193.71) ;

\node at (10pt,52pt){$Z_{\text{defect}}=$};

\node at (25pt,122pt){$Z_{0}=$};

\small

\node at (58pt,52pt){$\langle b|$};
\node at (58pt,122pt){$\langle b|$};

\node at (208pt,52pt){$|B\rangle$};
\node at (208pt,122pt){$|B\rangle$};

\node at (127pt,170pt){$W(t)$};
\node at (117pt,135pt){\textcolor{orange}{$H_0$}};
\node at (131pt,65pt){\textcolor{orange}{$H_{\text{defect}}$}};

\end{tikzpicture}
\ee
where $H_{\text{defect}}$/$H_0$ are defined along the vertical orange lines.
Since we are interested in the long time limit $t\gg \beta$ in \eqref{Wt_global1}, or equivalently $W(t)\gg 1$, the string order parameter in \eqref{SOP_global1} is mainly determined by the ground-state energies of $H_{\text{defect}}$ and $H_0$, which we denote as $E_{\text{defect}}^0$ and $E^0$ respectively. Then one can find
\be
\begin{split}
\langle \psi(t)|\mathcal L|\psi(t)\rangle
\simeq & \frac{\langle b|G_d\rangle \langle G_d|B\rangle}{\langle b|G_0\rangle \langle G_0|B\rangle}
\frac{e^{-E^0_{\text{defect}} W(t)}}{e^{-E^0 W(t)}}\\
\simeq & \frac{\langle b|G_d\rangle \langle G_d|B\rangle}{\langle b|G_0\rangle \langle G_0|B\rangle}  e^{-(E_{\text{defect}}^0-E^0)\cdot \frac{2\pi t}{\beta} },
\end{split}
\label{SOP_global1_b}
\ee
where$|G_d\rangle$ and $|G_0\rangle$ are the ground states of $H_{\text{defect}}$ and $H_0$ respectively.

From \eqref{SOP_global1_b}, one can find the string order parameter decays exponentially in time after a global quench, with
\be
\label{Eq:Global_semi}
\log |\langle \psi(t)|\mathcal L|\psi(t)\rangle| \simeq - \kappa \, t,
\ee
up to a constant shift. Here the decaying rate is
\be
\label{kappa_global1}
\kappa=\frac{2\pi}{\beta}\cdot(E_{\text{defect}}^0-E^0).
\ee
That is, the exponential decaying rate is proportional to $1/\beta$ that characterizes the (inverse of) correlation length of the initial state. The factor $(E_{\text{defect}}^0-E^0)$ in \eqref{kappa_global1} is not universal and depends on the details of the string operator
\footnote{ We would like to point out that, although the factor is generally not universal, it is potentially able to reveal important information about the concrete model.  
See also the discussion in Sec.\ref{Period_kappa}. }.

\subsubsection{Lattice calculation}

To verify the CFT results, we consider a free fermion lattice model after a global quench. The initial state $|\psi_0\rangle$ is chosen as the ground state of the 
gapped Hamiltonian
\be
\label{H0_globalQuench}
H_0=-\frac{1}{2}\sum_{1\le j\le L-1} c_j^\dag c_{j+1}+h.c.
+m \sum_j (-1)^j c_j^\dag c_j,
\ee
with half filling, and $m>0$ is the mass term. 
The relation between $m$ in \eqref{H0_globalQuench} and the parameter $\beta$ in \eqref{Psi0_global1} was discussed in, e.g., Ref.\cite{2018_Wen_Wang_Ryu}.
Briefly, a larger mass $m$ corresponds to a smaller $\beta$, and therefore a smaller correlation length in the initial state $|\psi_0\rangle$.

\begin{figure}[t]
\centering
\begin{tikzpicture}

    \node[inner sep=0pt] (russell) at (15pt,-85pt)
    {\includegraphics[width=.24\textwidth]{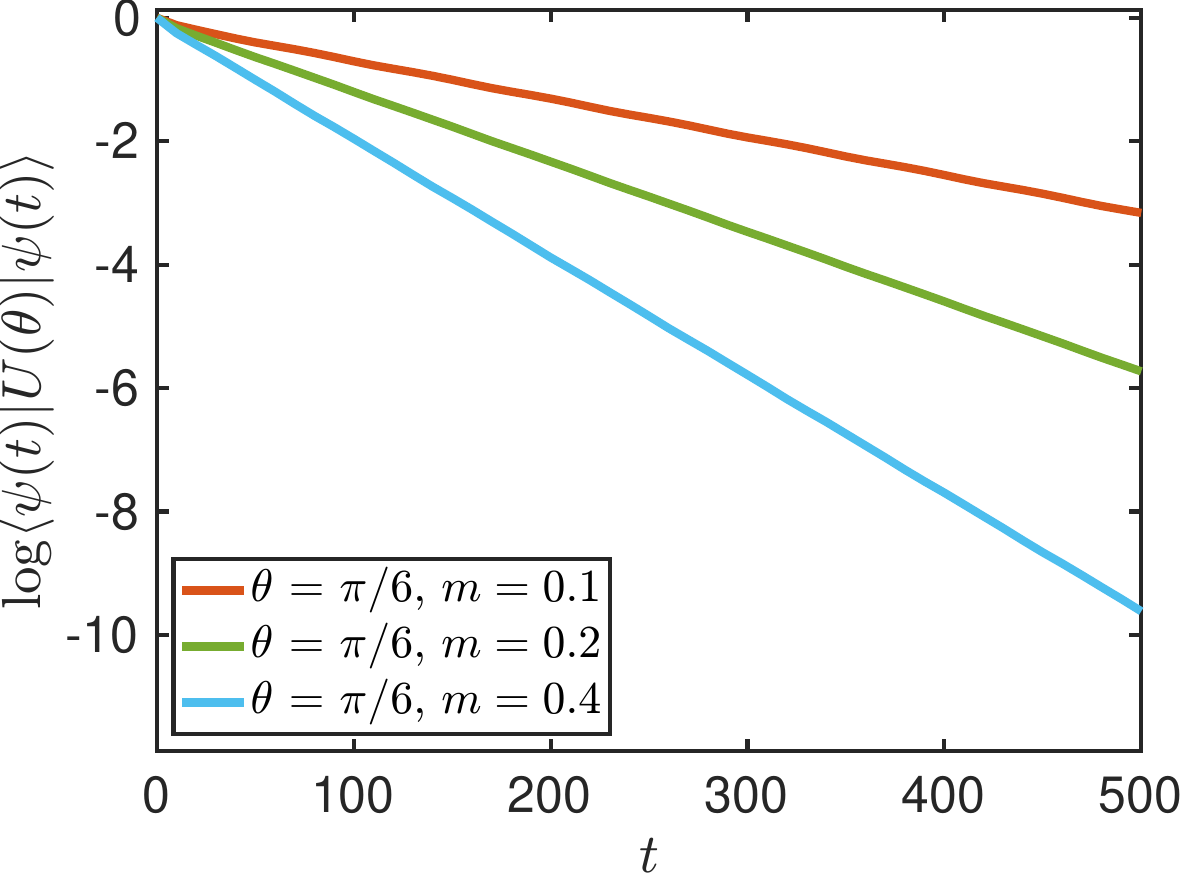}};
        \node[inner sep=0pt] (russell) at (140pt,-85pt)
    {\includegraphics[width=.24\textwidth]{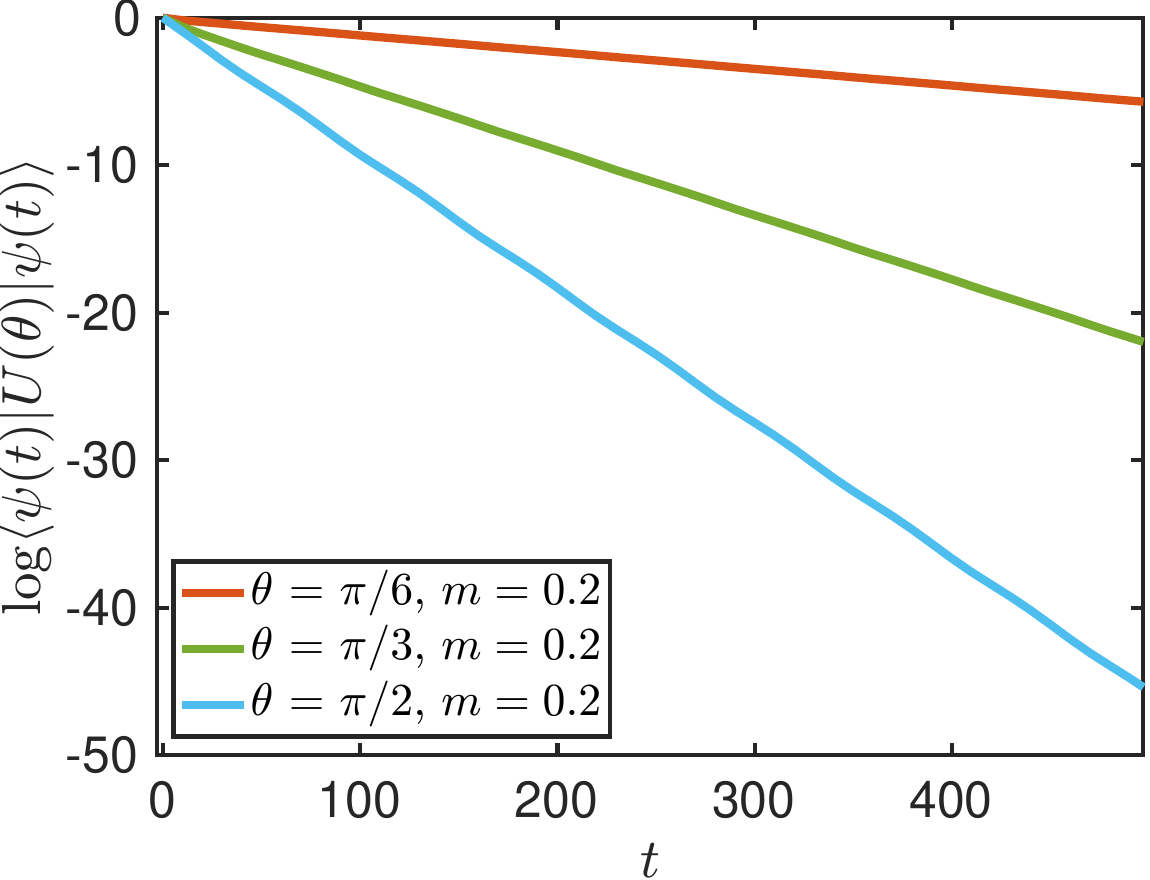}};

             \node at (52pt, -48pt){(a)};
            \node at (175pt, -110pt){(b)};

    \end{tikzpicture}
\caption{Lattice calculation of type-I string order parameter evolution after a global quantum quench.
We choose the total system size $L=600$, and the strength of symmetry breaking $h = 0.5$ in \eqref{H1_Global_semi}. (a) String order parameter evolution for different mass terms $m$ in \eqref{H0_globalQuench} with a fixed $\theta$ in $U(\theta)$. (b) String order parameter evolution for different $\theta$ with a fixed mass term $m$. 
}
\label{Fig:SymBreak_Global_finite}
\end{figure}

At $t=0$, we switch the Hamiltonian to the critical point with a boundary symmetry breaking as follows: 
\be\label{H1_Global_semi}
H_1=
-\frac{1}{2}\sum_{1\le j\le L-1} c_j^\dag c_{j+1}+
h \, c_1^\dag c_2^\dag +h.c., 
\ee
where $h>0$ characterizes the strength of symmetry breaking of $U(1)$ at the left boundary. For $t<L$, during which the effect of symmetry breaking at the left boundary does not propagate to the right boundary, we can use this finite-length lattice to simulate the non-equilibrium dynamics in a semi-infinite system.
Then we study the time evolution $\langle \psi(t)|U(\theta)|\psi(t)\rangle$, with the string operator $U(\theta)$ defined in \eqref{U_theta}.

As shown in Fig.\ref{Fig:SymBreak_Global_finite}, one can see clearly that the type-I string order parameter decays exponentially in time, which agrees with the result in \eqref{Eq:Global_semi}. In addition, the decay rates $\kappa$ depend on both the mass term (or equivalently  the correlation length in the initial state) and the parameter $\theta$ in $U(\theta)$. As we increases $m$,
which corresponds to a smaller $\beta$ in the initial state in \eqref{Psi0_global1}, one can find the decaying rate in Fig.\ref{Fig:SymBreak_Global_finite} increases accordingly.
This agrees with the CFT result in \eqref{kappa_global1}.
See Appendix \ref{Appendix:Mass_kappa} for a more detailed analysis on this $m$-dependence. For the $\theta$ dependence, we find the decay rate has a period of $\pi$, which results from $\langle U(\theta)\rangle=\langle U(\theta+\pi)\rangle$. This is because the symmetry operator $U(\theta)$ at $\theta=\pi$ measures the $\mathbb Z_2$ fermion parity, which is preserved by the boundary term in \eqref{H1_Global_semi}.
Therefore, one has $|\langle U(\theta=\pi)\rangle|=|\langle U(\theta=0)\rangle|=1$, which results in the $\pi$-period.
See also Sec.\ref{Period_kappa} for more discussions on this $\theta$ dependence.

\subsection{Global quench: an infinite system}
\label{Sec:Global2}

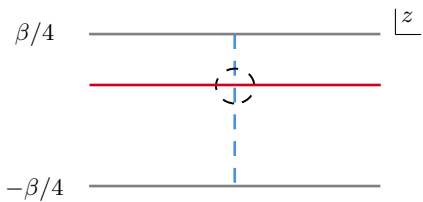
\begin{figure}
\centering
\begin{tikzpicture}[x=0.75pt,y=0.75pt,yscale=-0.65,xscale=0.65]

\begin{scope}[xshift=60pt, yshift=15pt]
\draw (140pt,40pt)--(155pt,40pt);
\draw (140pt,40pt)--(140pt,25pt);
\node at (147pt,30pt){$z$};
\end{scope}

\draw [color={rgb, 255:red, 74; green, 144; blue, 226 }  ,draw opacity=1 ][line width=1.0]  [dash pattern={on 5.63pt off 4.5pt}]  (142,73) -- (142,191) ;
\draw  [draw opacity=0][dash pattern={on 4.5pt off 4.5pt}][line width=0.75]  (141.51,126.87) .. controls (133.59,126.48) and (127.34,120.46) .. (127.44,113.19) .. controls (127.53,105.73) and (134.26,99.76) .. (142.48,99.87) .. controls (150.69,99.97) and (157.27,106.11) .. (157.17,113.57) .. controls (157.08,120.84) and (150.68,126.7) .. (142.75,126.89) -- (142.3,113.38) -- cycle ; \draw  [dash pattern={on 4.5pt off 4.5pt}][line width=0.75]  (141.51,126.87) .. controls (133.59,126.48) and (127.34,120.46) .. (127.44,113.19) .. controls (127.53,105.73) and (134.26,99.76) .. (142.48,99.87) .. controls (150.69,99.97) and (157.27,106.11) .. (157.17,113.57) .. controls (157.08,120.84) and (150.68,126.7) .. (142.75,126.89) ;  
\draw [color={rgb, 255:red, 128; green, 128; blue, 128 }  ,draw opacity=1 ][line width=1.0]    (29,73) -- (255,73) ;
\draw [color={rgb, 255:red, 128; green, 128; blue, 128 }  ,draw opacity=1 ][line width=1.0]    (29,191) -- (255,191) ;
\draw [color={rgb, 255:red, 208; green, 2; blue, 27 }  ,draw opacity=1 ][line width=1.0]    (29.32,112.53) -- (255.32,112.53) ;

\node at (-10pt,55pt){$\beta/4$};

\node at (-10pt,145pt){$-\beta/4$};

\end{tikzpicture}
\caption{Type-I string order parameter for an infinite system after a global quench.
The dashed blue line corresponds to the conformal interface that breaks the global symmetry, and the red line corresponds to the string operator. A small disk of radius $\epsilon$ is removed to introduce a UV cutoff, with a conformal boundary condition $|b\rangle$ imposed along its boundary.
}
\label{Fig:GlobalQuench2}
\end{figure}

Now let us consider an infinite system after a global quench, where a conformal interface is inserted in the middle of the system. The global symmetry is broken only along this conformal interface, as shown in Fig.\ref{Fig:GlobalQuench2}.

\subsubsection{BCFT calculation}

In the BCFT description, the initial state $|\psi_0\rangle$, which preserves the global symmetry, is again chosen as a regularized conformal boundary state in \eqref{Psi0_global1}, where $|B\rangle$ is now defined over the infinite line. At $t=0$, we evolve the state with a CFT Hamiltonian $H_{\text{CFT}}$ where the global symmetry is broken only at the interface, say, along $x=0$.

As shown in Fig.\ref{Fig:GlobalQuench2}, the string order parameter $\langle \psi(t)|\mathcal L|\psi(t)\rangle$ corresponds to a path integral defined over an infinitely long strip of width $\beta/2$. The interpretation of this path integral is similar to that in Fig.\ref{Fig:Global1}, which we don't repeat here.

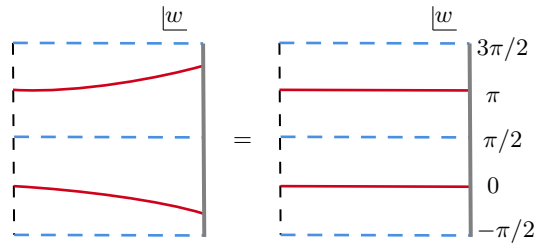
\begin{figure}
\centering
\begin{tikzpicture}[x=0.75pt,y=0.75pt,yscale=-0.6,xscale=0.6]

\begin{scope}[xshift=95pt,yshift=-20pt]
\draw (140pt,40pt)--(155pt,40pt);
\draw (140pt,40pt)--(140pt,25pt);
\node at (147pt,32pt){$w$};
\end{scope}

\begin{scope}[xshift=265pt,yshift=-20pt]
\draw (140pt,40pt)--(155pt,40pt);
\draw (140pt,40pt)--(140pt,25pt);
\node at (147pt,32pt){$w$};
\end{scope}

\draw [color={rgb, 255:red, 0; green, 0; blue, 0 }  ,draw opacity=1 ][line width=0.75]  [dash pattern={on 4.5pt off 4.5pt}]  (187,37.5) -- (188,199) ;
\draw [color={rgb, 255:red, 74; green, 144; blue, 226 }  ,draw opacity=1 ][line width=1.0]  [dash pattern={on 5.63pt off 4.5pt}]  (188,199) -- (347.83,199.5) ;
\draw [color={rgb, 255:red, 74; green, 144; blue, 226 }  ,draw opacity=1 ][line width=1.0]  [dash pattern={on 5.63pt off 4.5pt}]  (188,116.5) -- (347.83,117) ;
\draw [color={rgb, 255:red, 74; green, 144; blue, 226 }  ,draw opacity=1 ][line width=1.0]  [dash pattern={on 5.63pt off 4.5pt}]  (187,37.5) -- (346.83,38) ;
\draw [color={rgb, 255:red, 208; green, 2; blue, 27 }  ,draw opacity=1 ][line width=1.0]    (188,158) .. controls (244.5,161.25) and (312.5,170.25) .. (347.5,181.25) ;
\draw [color={rgb, 255:red, 208; green, 2; blue, 27 }  ,draw opacity=1 ][line width=1.0]    (187,77) .. controls (243.5,80.25) and (314,66) .. (346,57) ;
\draw [color={rgb, 255:red, 0; green, 0; blue, 0 }  ,draw opacity=1 ][line width=0.75]  [dash pattern={on 4.5pt off 4.5pt}]  (411,37.5) -- (412,199) ;
\draw [color={rgb, 255:red, 74; green, 144; blue, 226 }  ,draw opacity=1 ][line width=1.0]  [dash pattern={on 5.63pt off 4.5pt}]  (412,199) -- (571.83,199.5) ;
\draw [color={rgb, 255:red, 74; green, 144; blue, 226 }  ,draw opacity=1 ][line width=1.0]  [dash pattern={on 5.63pt off 4.5pt}]  (412,116.5) -- (571.83,117) ;
\draw [color={rgb, 255:red, 208; green, 2; blue, 27 }  ,draw opacity=1 ][line width=1.0]    (411,77) -- (570.83,77.5) ;
\draw [color={rgb, 255:red, 74; green, 144; blue, 226 }  ,draw opacity=1 ][line width=1.0]  [dash pattern={on 5.63pt off 4.5pt}]  (411,37.5) -- (570.83,38) ;
\draw [color={rgb, 255:red, 208; green, 2; blue, 27 }  ,draw opacity=1 ][line width=1.0]    (412,158) -- (571.83,158.5) ;
\draw [color={rgb, 255:red, 128; green, 128; blue, 128 }  ,draw opacity=1 ][line width=1.5]    (346.83,38) -- (347.83,200.75) ;
\draw [color={rgb, 255:red, 128; green, 128; blue, 128 }  ,draw opacity=1 ][line width=1.5]    (570.83,38) -- (571.83,200.75) ;

\begin{scope}[xshift=87pt]
\node at (362pt,32pt){$3\pi/2$};
\node at (356pt,62pt){$\pi$};
\node at (362pt,90pt){$\pi/2$};
\node at (356pt,118pt){$0$};
\node at (364pt,147pt){$-\pi/2$};
\end{scope}

\draw (380,120) node {$=$};

\end{tikzpicture}
\caption{Type-I string order parameter after a conformal mapping from Fig.\ref{Fig:GlobalQuench2}.
The string operator (defined along the red solid lines) can be freely deformed from curves (left plot) to straight lines (right plot) as long as they do not touch the symmetry-breaking interface (dashed blue lines). The width of this $w$-cylinder in $\Im(w)$ direction is $2\pi$, and its length in $\Re(w)$ direction is approximately $2\pi t/\beta$.
}
\label{Global2_cylinder}
\end{figure}

To evaluate the string order parameter, we consider the same conformal mapping in \eqref{ConformalMap_Global1}, except that now we have $\Re (z)\in (-\infty,+\infty)$. Then the strip configuration in Fig.\ref{Fig:GlobalQuench2} is mapped to a $w$-cylinder in Fig.\ref{Global2_cylinder}, where the top and bottom edges are identified. 
Note there is a subtle difference from the case of a semi-infinite system in Fig.\ref{Fig:Global2}. Here in Fig.\ref{Global2_cylinder}, 
the length of this $w$-cylinder has the same expression as \eqref{Wt_global1}, but the width in the $\Im(w)$ direction is $2\pi$. In addition, there are now two defect lines inserted in the $w$-cylinder.

Then, we consider the procedure similar to that in Sec.\ref{Sec:Global1_BCFT}. 
That is, we consider the defect Hamiltonian $H_{\text{defect}}$ defined in the $\Im (w)$ direction, and the imaginary time in $\Re (w)$ direction.
Different from the configuration in \eqref{SOP_global1}, here $H_{\text{defect}}$ is defined on a circle of length $2\pi$ (see \eqref{Eq:H_defect2} below), with two conformal interfaces and two string operators inserted.

Then, one can find the time-dependent string order parameter $\langle \psi(t)|\mathcal L|\psi(t)\rangle$ has the same expression as \eqref{SOP_global1}, except that now $Z_{\text{defect}}$ and $Z_0$ are defined on cylinders as follows:
\be
\label{Eq:H_defect2}
\begin{tikzpicture}[x=0.75pt,y=0.75pt,yscale=-0.7,xscale=0.7]

\draw  [dash pattern={on 4.5pt off 4.5pt}] (100,147.5) .. controls (100,130.1) and (106.16,116) .. (113.75,116) .. controls (121.34,116) and (127.5,130.1) .. (127.5,147.5) .. controls (127.5,164.9) and (121.34,179) .. (113.75,179) .. controls (106.16,179) and (100,164.9) .. (100,147.5) -- cycle ;
\draw  [color={rgb, 255:red, 155; green, 155; blue, 155 }  ,draw opacity=1 ][line width=1.5]  (246,147.5) .. controls (246,130.1) and (252.16,116) .. (259.75,116) .. controls (267.34,116) and (273.5,130.1) .. (273.5,147.5) .. controls (273.5,164.9) and (267.34,179) .. (259.75,179) .. controls (252.16,179) and (246,164.9) .. (246,147.5) -- cycle ;
\draw [color={rgb, 255:red, 74; green, 144; blue, 226 }  ,draw opacity=1 ][line width=1.5]  [dash pattern={on 5.63pt off 4.5pt}]  (113.75,116) -- (259.75,116) ;
\draw [color={rgb, 255:red, 74; green, 144; blue, 226 }  ,draw opacity=1 ][line width=1.5]  [dash pattern={on 5.63pt off 4.5pt}]  (113.75,179) -- (259.75,179) ;
\draw  [color={rgb, 255:red, 245; green, 166; blue, 35 }  ,draw opacity=1 ] (143,147.5) .. controls (143,130.1) and (149.16,116) .. (156.75,116) .. controls (164.34,116) and (170.5,130.1) .. (170.5,147.5) .. controls (170.5,164.9) and (164.34,179) .. (156.75,179) .. controls (149.16,179) and (143,164.9) .. (143,147.5) -- cycle ;
\draw [color={rgb, 255:red, 208; green, 2; blue, 27 }  ,draw opacity=1 ][line width=0.75]    (127.75,151) -- (273.75,151) ;
\draw [color={rgb, 255:red, 208; green, 2; blue, 27 }  ,draw opacity=1 ][line width=0.75]    (100,143.5) -- (246,143.5) ;

\begin{scope}[xshift=0pt,yshift=69pt]
\draw  [dash pattern={on 4.5pt off 4.5pt}] (100,147.5) .. controls (100,130.1) and (106.16,116) .. (113.75,116) .. controls (121.34,116) and (127.5,130.1) .. (127.5,147.5) .. controls (127.5,164.9) and (121.34,179) .. (113.75,179) .. controls (106.16,179) and (100,164.9) .. (100,147.5) -- cycle ;
\draw  [color={rgb, 255:red, 155; green, 155; blue, 155 }  ,draw opacity=1 ][line width=1.5]  (246,147.5) .. controls (246,130.1) and (252.16,116) .. (259.75,116) .. controls (267.34,116) and (273.5,130.1) .. (273.5,147.5) .. controls (273.5,164.9) and (267.34,179) .. (259.75,179) .. controls (252.16,179) and (246,164.9) .. (246,147.5) -- cycle ;
\draw [color={rgb, 255:red, 74; green, 144; blue, 226 }  ,draw opacity=1 ][line width=1.5]  [dash pattern={on 5.63pt off 4.5pt}]  (113.75,116) -- (259.75,116) ;
\draw [color={rgb, 255:red, 74; green, 144; blue, 226 }  ,draw opacity=1 ][line width=1.5]  [dash pattern={on 5.63pt off 4.5pt}]  (113.75,179) -- (259.75,179) ;
\draw  [color={rgb, 255:red, 245; green, 166; blue, 35 }  ,draw opacity=1 ] (143,147.5) .. controls (143,130.1) and (149.16,116) .. (156.75,116) .. controls (164.34,116) and (170.5,130.1) .. (170.5,147.5) .. controls (170.5,164.9) and (164.34,179) .. (156.75,179) .. controls (149.16,179) and (143,164.9) .. (143,147.5) -- cycle ;

\end{scope}

\node at (10pt,109pt){$Z_{\text{defect}}=$};
\node at (25pt,182pt){$Z_{0}=$};

\small

\node at (58pt,109pt){$\langle b|$};
\node at (222pt,109pt){$|B\rangle$};

\node at (58pt,180pt){$\langle b|$};
\node at (222pt,180pt){$|B\rangle$};

\node at (137pt,185pt){\textcolor{orange}{$H_0$}};
\node at (143pt,95pt){\textcolor{orange}{$H_{\text{defect}}$}};

\end{tikzpicture}
\ee
where $H_{\text{defect}}$/$H_0$ are defined along the vertical orange circles.
It is helpful to compare the above configurations with \eqref{Z_defect_global1}, which are defined on strips.
The length of the cylinder in \eqref{Eq:H_defect2} is $\frac{2\pi t}{\beta}$ in the limit $t\gg \beta$. By using a similar argument near \eqref{SOP_global1_b}, one can find that the type-I string order parameter decays exponentially in time as:
\be
\label{String_Global2}
\log |\langle \psi(t)|\mathcal L|\psi(t)\rangle| \simeq -\kappa\,t,
\ee
where $\kappa$ has the same expression as \eqref{kappa_global1}, except that now the values of $E^0_{\text{defect}}$ and $E_0$ are different from those in \eqref{SOP_global1_b}, since the parent Hamiltonians are now different. 
Similar to the conclusion in Sec.\ref{Sec:Global1}, here the decaying rate $\kappa$ is proportional to $1/\beta$, and the other factor $(E_{\text{defect}}^0-E^0)$ is not universal and depends on the details of the string operators.

\subsubsection{Lattice calculation}
\label{Sec:Numerics_typeI_Global2}

\begin{figure}[t]
\centering
\begin{tikzpicture}

    \node[inner sep=0pt] (russell) at (15pt,-85pt)
    {\includegraphics[width=.24\textwidth]{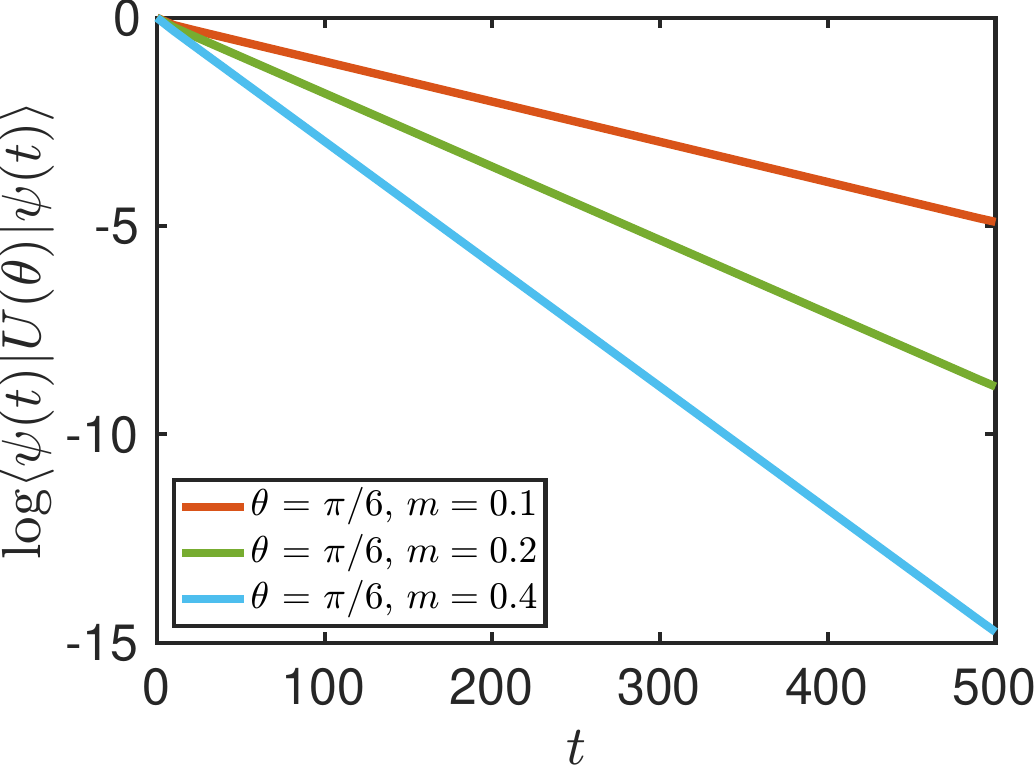}};
        \node[inner sep=0pt] (russell) at (140pt,-85pt)
    {\includegraphics[width=.24\textwidth]{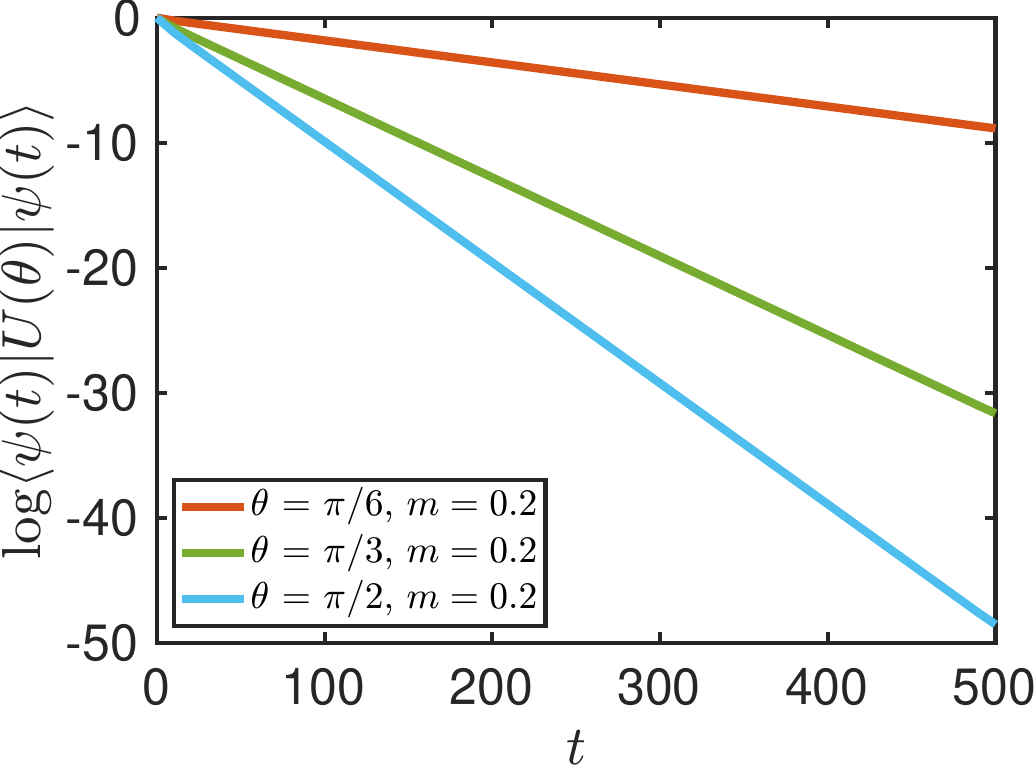}};

             \node at (52pt, -50pt){(a)};
            \node at (109pt, -70pt){(b)};

    \end{tikzpicture}
\caption{Lattice calculation of type-I string order parameter evolution after a global quantum quench. We choose the total system size $L=1000$, and the strength of symmetry breaking at the interface is $h = 0.5$ in \eqref{Global_defect_middle}. (a) Type-I string order parameter evolution for different mass terms $m$ in \eqref{H0_globalQuench} with a fixed $\theta$ in $U(\theta)$. (b) Type-I string order parameter evolution for different $\theta$ with a fixed mass term $m$. 
}
\label{Fig:SymBreak_Global_infi}
\end{figure}

To simulate the string operator evolution on a free fermion lattice, we prepare the initial state as the ground state of the gapped Hamiltonian in \eqref{H0_globalQuench} with a length of $L$ and half-filling. 
Starting from $t=0$, we evolve the initial state with the following Hamiltonian:
\be
\label{Global_defect_middle}
H_1=
-\frac{1}{2}\sum_{1\le j\le L-1} c_j^\dag c_{j+1}+
h c_{L/2}^\dag c_{L/2+1}^\dag +h.c., 
\ee
where the second term breaks the $U(1)$ symmetry at the interface, which is now located in the middle of the system.
For $t<L/2$, during which the effect of the symmetry breaking interface does not propagate to the boundary, this lattice system can be used to simulate the infinite system with a symmetry breaking interface. Then we study $\langle \psi(t)|U(\theta)|\psi(t)\rangle$, with the string operator $U(\theta)$ defined in \eqref{U_theta}.

The lattice results are shown in Fig.\ref{Fig:SymBreak_Global_infi}, where one can clearly observe an exponential decay in the time evolution of the string order parameter. The dependence of decaying rates $\kappa$ on $m$ and $\theta$ is similar to  Fig.\ref{Fig:SymBreak_Global_finite} with the same underlying reason.

\subsection{Local quench}
\label{Sec:Local}

For the local quantum quench, here we consider the setup proposed in Ref.\onlinecite{CC2007_local}. Briefly, we consider two decoupled CFTs, and join them locally at their ends suddenly, and let the system evolve in time.

\subsubsection{BCFT calculation}

We consider two CFTs defined on $(-\infty,0)$ and $(0,+\infty)$ that are decoupled from each other. At $x=0$, for simplicity we impose the same conformal boundary condition for both CFTs at their ends \footnote{In general, one can consider different conformal boundary conditions, in particular when the two CFTs are different.}.
Each CFT is prepared in their ground state, and the initial state is 
\be
|\psi_0\rangle=e^{-\lambda H_{\text{CFT}}} (|G_L\rangle\otimes |G_R\rangle),
\ee
where the factor $e^{-\lambda H_{\text{CFT}}}$ is introduced as a regularization with $\lambda>0$. Note that this initial state is not translationally invariant.
Then at $t=0$, we join the two ends of the CFTs at $x=0$ suddenly, and let the system evolve under the CFT Hamiltonian $H_{\text{CFT}}$, where a conformal interface at $x=0$ is introduced. At this conformal interface, the global symmetry is explicitly broken. Then the time dependent state is $|\psi(t)\rangle=e^{-i H_{\text{CFT}}t}|\psi_0\rangle$.
Similar to the case of a global quantum quench, it is convenient to consider the imaginary time evolution $|\psi(\tau)\rangle=e^{- H_{\text{CFT}}\tau}|\psi_0\rangle$ first, and then do the analytical continuation $\tau\to it$ in the last step.
Note that a similar setup in the presence of a conformal interface was previously considered in \cite{2018_Wen_Wang_Ryu}, but for other motivations.

\begin{figure}
\centering
\begin{tikzpicture}[x=0.75pt,y=0.75pt,yscale=-0.65,xscale=0.65]

\draw (140pt,40pt)--(155pt,40pt);
\draw (140pt,40pt)--(140pt,25pt);
\node at (147pt,30pt){$z$};

\node at (97pt,57pt){$+i\lambda$};
\node at (97pt,108pt){$-i\lambda$};

\draw [color={rgb, 255:red, 74; green, 144; blue, 226 }  ,draw opacity=1 ][line width=1.5]  [dash pattern={on 5.63pt off 4.5pt}]  (108.32,79.03) -- (108.32,146.03) ;
\draw [color={rgb, 255:red, 208; green, 2; blue, 27 }  ,draw opacity=1 ][line width=1.0]    (21.32,112.53) -- (195.32,112.53) ;
\draw  [draw opacity=0][dash pattern={on 4.5pt off 4.5pt}] (107.59,126.53) .. controls (100.19,126.15) and (94.31,120.03) .. (94.31,112.53) .. controls (94.31,104.79) and (100.58,98.52) .. (108.32,98.52) .. controls (116.06,98.52) and (122.34,104.79) .. (122.34,112.53) .. controls (122.34,120.27) and (116.06,126.55) .. (108.32,126.55) -- (108.32,112.53) -- cycle ; 
\draw  [dash pattern={on 4.5pt off 4.5pt}] (107.59,126.53) .. controls (100.19,126.15) and (94.31,120.03) .. (94.31,112.53) .. controls (94.31,104.79) and (100.58,98.52) .. (108.32,98.52) .. controls (116.06,98.52) and (122.34,104.79) .. (122.34,112.53) .. controls (122.34,120.27) and (116.06,126.55) .. (108.32,126.55) ;  
\draw [color={rgb, 255:red, 155; green, 155; blue, 155 }  ,draw opacity=1 ][line width=3]    (108.32,29.03) -- (108.32,79.03) ;
\draw [color={rgb, 255:red, 155; green, 155; blue, 155 }  ,draw opacity=1 ][line width=3]    (108.32,146.03) -- (108.32,194.03) ;

\end{tikzpicture}
\caption{Type-I string order parameter
after a local quantum quench. 
The symmetry is broken along the conformal interface (blue dashed line). The string operator (red solid line) is defined along $C=\{i\tau+x, \, -\infty<x<+\infty\}$.
A small disc of radius $\epsilon$ is removed at $z_0=i\tau$ to introduce a UV cutoff, with a conformal boundary condition $|b\rangle$ imposed along its boundary.
}
\label{Fig_local}
\end{figure}
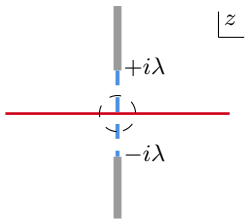

In the path integral, as shown in Fig.\ref{Fig_local}, we consider a plane with two slits along the imaginary axis: one slit goes from $-i\lambda$ to $-i\infty$ and the other one goes from $i\lambda$ to $+i\infty$. 
A conformal interface, which breaks the global symmetry, runs from $-i\lambda$ to $i\lambda$.
Then the string operator $\mathcal L$ is inserted
along $C=\{i\tau+x, \, -\infty<x<+\infty\}$.

To evaluate the time-dependent string order parameter $\langle \psi(t)|\mathcal L|\psi(t)\rangle$, we first remove a small disc of radius $\epsilon$ at $z=i\tau$ to introduce a UV cutoff.
Then we consider the following conformal mapping
\be
\label{LocalQuench_conformalMap}
w=\log\left(
\frac{\sqrt{(\lambda^2-\tau^2)(\lambda^2+z^2)}-i\tau z-\lambda^2}{\lambda(z-i\tau)}
\right),
\ee
which maps the $z$-plane in Fig.\ref{Fig_local} to a $w$-cylinder similar to that in Fig.\ref{Global2_cylinder}. Note that the boundary of the removed disc is mapped to one boundary of the $w$-cylinder, and the two slits in Fig.\ref{Fig_local} are mapped to the other boundary of the $w$-cylinder.
Then by deforming the string operator freely to straight lines, the $w$-cylinder looks exactly the same as Fig.\ref{Global2_cylinder} (right plot).
The only difference is that now the 
the length of the $w$-cylinder depends on time as
(after an analytical continuation $\tau\to it$):
\be
\label{local_W}
W(t)\simeq \log\frac{2(\lambda^2+t^2)}{\epsilon \lambda}.
\ee
Based on the same analysis as in Sec.\ref{Sec:Global1} and Sec.\ref{Sec:Global2},
one can find that in the long time limit $t\gg \lambda$ and $t\gg \epsilon$ such that $W(t)\simeq 2\log t$, and the string order parameter depends on time as
\be
\label{BCFT_local_typeI}
\log |\langle \psi(t)|\mathcal L|\psi(t)\rangle| \simeq
-\kappa \log t,
\ee
up to a constant shift and $\kappa=2(E_{\text{defect}}^0-E^0)$. That is, the order parameter evolution decays as a \textit{power-law} in time. Here $E_{\text{defect}}^0$ and $E^0$ are the ground-state energies of the Hamiltonians $H_\text{defect}$ and $H_0$ respectively, which have the same definitions as those in \eqref{Eq:H_defect2}. With the similar reason as before, here the coefficient $\kappa$ is not universal and it depends on the details of the string operators.

\subsubsection{Lattice calculation}

\begin{figure}
\centering
{\includegraphics[width=.32\textwidth]{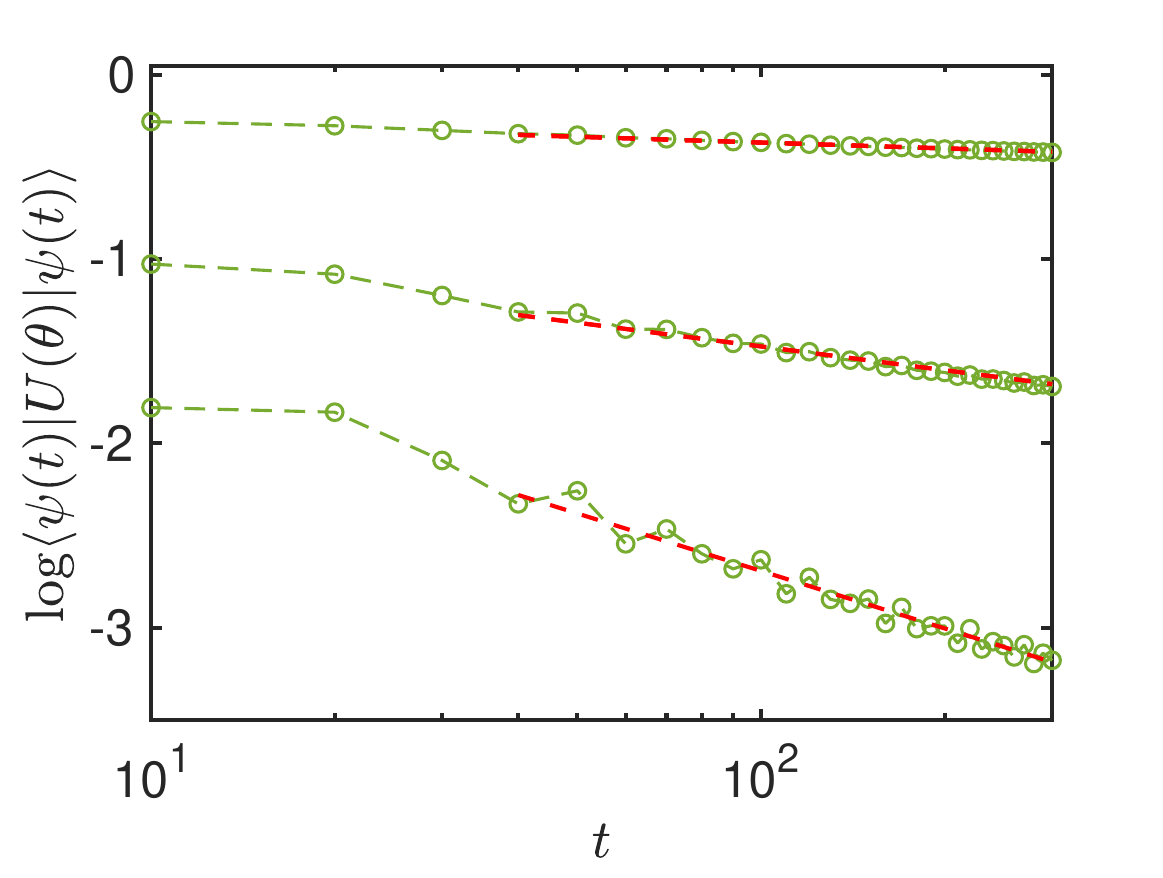}}
    \caption{
    Lattice calculation of type-I string order parameter evolution after a local quantum quench.
    The total system is of length $L=1000$. We choose $h=1/2$ in the symmetry breaking term in \eqref{Local_defect_middle}.
    From top to bottom, we take  $\theta=\pi/6$, $\pi/3$, and $\pi/2$.
    The red dashed lines are fitting lines of the form $y=-a \log t+b$.}
    \label{Fig:SymBreak_Local} 
\end{figure}

To simulate the local quench on a lattice, 
we start from two critical free-fermion chains that are decoupled from each other, with the Hamiltonian
\be
\label{H0_local_quench}
H_0=-\frac{1}{2}\sum_{1\le j\le L/2-1} c_j^\dag c_{j+1}
-\frac{1}{2}\sum_{L/2+1\le j\le L-1} c_j^\dag c_{j+1}+h.c.
\nonumber
\ee
Within each chain, we prepare the system in its ground state with a half-filling. Then from $t=0$, we evolve this initial state with the 
new Hamiltonian $H_1$ in \eqref{Global_defect_middle}, which we rewrite here
\be
\label{Local_defect_middle}
H_1=
-\frac{1}{2}\sum_{1\le j\le L-1} c_j^\dag c_{j+1}+
h \, c_{L/2}^\dag c_{L/2+1}^\dag +h.c., 
\ee
where the $U(1)$ symmetry is broken at the interface in the middle of the total system. Then we calculate $\langle \psi(t)|U(\theta)|\psi(t)\rangle$ with the string operator $U(\theta)$ given in \eqref{U_theta}.
The lattice results are shown in Fig.\ref{Fig:SymBreak_Local}. One can find that the scaling behaviors of the string order parameter evolution agree with the CFT result in \eqref{BCFT_local_typeI}.


\section{
String order parameter over a subsystem with or without boundary symmetry breaking}
\label{Sec:Partial_String}

For the type-I string order parameter, if the symmetry is preserved over the whole system, then it becomes a constant with no interesting scaling behaviors.
However, if the string operator is defined only over a subsystem $A$, since the total charge in a subsystem is in general not conserved, one can still expect a universal scaling behavior of $\langle \psi(t)| \mathcal L_A|\psi(t) \rangle$ even if the symmetry is preserved everywhere. For this reason, we will study the time evolution of type-II string order parameter (where the string operator is defined in a subsystem) with or without boundary/interface symmetry breaking.

It is noted that this type-II string order parameter is closely related to the so-called full-counting statistics (FCS), which characterizes the charge fluctuations in a subsystem \cite{1996_Levitov,2001_BN,2009_Klich,2014_fermion_Klich,2023_Sarang_Romain,2023_FCS_nonequilibrium}. For example, let us consider the string operator in \eqref{U_theta} defined over a subsystem $A$. Then one has 
\be
\label{U_theta_A}
U(\theta)=e^{i\theta Q_V(A)}=e^{i\theta \sum_{j\in A} c_j^\dag c_j}, \quad \theta\in \mathbb [0,\,2\pi).
\ee
Then the expectation value $\langle  e^{i\theta \sum_{j\in A} c_j^\dag c_j} \rangle$ is nothing but the FCS in a free fermion system. While the FCS discussed in literature is usually related to an invertible symmetry and therefore corresponds to the expectation value of an invertible operator defined in a subregion, our discussion here works for both invertible and non-invertible operators.

The method we use in this section is based on Ref.\onlinecite{2016CardyTonni}, where one can map the path integral of the reduced density matrix $\rho_A$ for subsystem $A$ to a cylinder (or strip) with a branch cut. Since our string operator is defined only within the subsystem $A$, once we know how to obtain $\rho_A$, we can evaluate the type-II string order parameters as
\be
\langle \psi(t)| \mathcal L_A|\psi(t)\rangle =\text{Tr}[\rho_A(t) \mathcal L_A],
\ee
where $|\psi(t)\rangle$ is a time-dependent state in general.

\subsection{Global quench}

Now let us consider the type-II string order parameter evolution after a global quantum quench. The setup is the same as those in Sec.\ref{Sec:Global1} and Sec.\ref{Sec:Global2}, with the difference that now the string operator is defined in a \textit{subsystem} rather than over the whole system. More concretely, we will consider the following two cases:
\begin{enumerate}
\item An infinite system over $(-\infty,+\infty)$ after a global quench, with the string operator defined over the subsystem $[0,+\infty)$.

\item A semi-infinite system over $[0,+\infty)$ after a global quench, with the string operator is defined over a subsystem $[0,l]$.

\end{enumerate}
For such type-II string operators, no matter there is a symmetry-breaking interface/boundary or not, one will always have universal features in the time evolution. This is in contrast to type-I string order parameters, which become trivial if the symmetry is preserved over the whole system.

\begin{figure}
\centering
\begin{tikzpicture}[x=0.75pt,y=0.75pt,yscale=-0.6,xscale=0.6]

\draw [color={rgb, 255:red, 128; green, 128; blue, 128 }  ,draw opacity=1 ][line width=1.5]    (57,73) -- (258,73) ;
\draw [color={rgb, 255:red, 128; green, 128; blue, 128 }  ,draw opacity=1 ][line width=1.5]    (57,172) -- (259,172) ;

\begin{scope}[xshift=30pt]
\draw [color={rgb, 255:red, 208; green, 2; blue, 27 }  ,draw opacity=1 ][line width=1.5]    (112.32,112.53) -- (217.32,112.53) ;

\draw  [dash pattern={on 4.5pt off 4.5pt}] (111.59,126.53) .. controls (104.19,126.15) and (98.31,120.03) .. (98.31,112.53) .. controls (98.31,104.79) and (104.58,98.52) .. (112.32,98.52) .. controls (120.06,98.52) and (126.34,104.79) .. (126.34,112.53) .. controls (126.34,120.27) and (120.06,126.55) .. (112.32,126.55) ; 
\end{scope}

\begin{scope}[xshift=-53.5pt]
\draw [color={rgb, 255:red, 74; green, 144; blue, 226 }  ,draw opacity=1 ][line width=1.0][dashed]    (224,72) -- (224,171) ;
\end{scope}

\draw [color={rgb, 255:red, 74; green, 144; blue, 226 }  ,draw opacity=1 ][line width=1.2]    (324,72) -- (324,171) ;

\draw [color={rgb, 255:red, 128; green, 128; blue, 128 }  ,draw opacity=1 ][line width=1.5]    (324,72) -- (525,72) ;
\draw [color={rgb, 255:red, 128; green, 128; blue, 128 }  ,draw opacity=1 ][line width=1.5]    (324,170) -- (526,170) ;
\draw [color={rgb, 255:red, 208; green, 2; blue, 27 }  ,draw opacity=1 ][line width=1.5]    (324.32,111.53) -- (454.32,111.53) ;


\draw  [dash pattern={on 4.5pt off 4.5pt}] (453.59,125.53) .. controls (446.19,125.15) and (440.31,119.03) .. (440.31,111.53) .. controls (440.31,103.79) and (446.58,97.52) .. (454.32,97.52) .. controls (462.06,97.52) and (468.34,103.79) .. (468.34,111.53) .. controls (468.34,119.27) and (462.06,125.55) .. (454.32,125.55) ;  

\node at (15pt,55pt){$\beta/4$};

\node at (15pt,125pt){$-\beta/4$};

\end{tikzpicture}
\caption{
The expectation value of the type-II string operator  over a subsystem after a global quantum quench.
Left: A semi-infinite subsystem in an infinite system. Right: A finite-length subsystem at the end of a semi-infinite system.
There is a symmetry breaking at the interface (blue dashed line) or at the boundary (blue solid line). A small disc (black dashed line) of radius $\epsilon$ is removed at the end of the string operator, with a conformal boundary condition $|b\rangle$ imposed.
}
\label{Fig:PartialString_Global}
\end{figure}
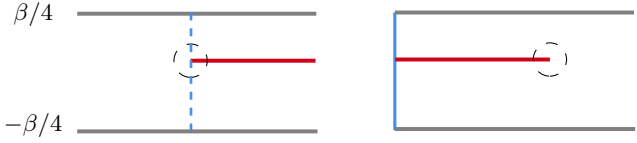

\subsubsection{Semi-infinite interval in an infinite line}

Let us first consider an infinite system over $(-\infty,+\infty)$ after a global quench, 
which is the same as the setup in Sec.\ref{Sec:Global2}.
The difference is that now the string operator is inserted only in the semi-infinite subsystem $[0,+\infty)$.
To be general, we will consider a symmetry-breaking interface at $x=0$. \footnote{One can simply set this interface to be identity and then one can reproduce the case with no symmetry-breaking interface.}

\begin{figure}[h]
\begin{tikzpicture}[x=0.75pt,y=0.75pt,yscale=-0.5,xscale=0.5]

\begin{scope}[xshift=95pt,yshift=-20pt]
\draw (140pt,40pt)--(155pt,40pt);
\draw (140pt,40pt)--(140pt,25pt);
\small\node at (149pt,30pt){$w$};
\end{scope}

\begin{scope}[xshift=265pt,yshift=-20pt]
\draw (140pt,40pt)--(155pt,40pt);
\draw (140pt,40pt)--(140pt,25pt);
\small\node at (149pt,30pt){$w$};
\end{scope}

\draw [color={rgb, 255:red, 0; green, 0; blue, 0 }  ,draw opacity=1 ][line width=0.75]  [dash pattern={on 4.5pt off 4.5pt}]  (187,37.5) -- (188,199) ;
\draw [color={rgb, 255:red, 74; green, 144; blue, 226 }  ,draw opacity=1 ][line width=1.0]  [dash pattern={on 5.63pt off 4.5pt}]  (188,199) -- (347.83,199.5) ;
\draw [color={rgb, 255:red, 74; green, 144; blue, 226 }  ,draw opacity=1 ][line width=1.0]  [dash pattern={on 5.63pt off 4.5pt}]  (188,116.5) -- (347.83,117) ;
\draw [color={rgb, 255:red, 74; green, 144; blue, 226 }  ,draw opacity=1 ][line width=1.0]  [dash pattern={on 5.63pt off 4.5pt}]  (187,37.5) -- (346.83,38) ;
\draw [color={rgb, 255:red, 208; green, 2; blue, 27 }  ,draw opacity=1 ][line width=1.0]    (188,158) .. controls (244.5,161.25) and (312.5,170.25) .. (347.5,181.25) ;
\draw [color={rgb, 255:red, 0; green, 0; blue, 0 }  ,draw opacity=1 ][line width=0.75]  [dash pattern={on 4.5pt off 4.5pt}]  (411,37.5) -- (412,199) ;
\draw [color={rgb, 255:red, 74; green, 144; blue, 226 }  ,draw opacity=1 ][line width=1.0]  [dash pattern={on 5.63pt off 4.5pt}]  (412,199) -- (571.83,199.5) ;
\draw [color={rgb, 255:red, 74; green, 144; blue, 226 }  ,draw opacity=1 ][line width=1.0]  [dash pattern={on 5.63pt off 4.5pt}]  (412,116.5) -- (571.83,117) ;
\draw [color={rgb, 255:red, 74; green, 144; blue, 226 }  ,draw opacity=1 ][line width=1.0]  [dash pattern={on 5.63pt off 4.5pt}]  (411,37.5) -- (570.83,38) ;
\draw [color={rgb, 255:red, 208; green, 2; blue, 27 }  ,draw opacity=1 ][line width=1.0]    (412,158) -- (571.83,158.5) ;
\draw [color={rgb, 255:red, 128; green, 128; blue, 128 }  ,draw opacity=1 ][line width=1.5]    (346.83,38) -- (347.83,200.75) ;
\draw [color={rgb, 255:red, 128; green, 128; blue, 128 }  ,draw opacity=1 ][line width=1.5]    (570.83,38) -- (571.83,200.75) ;

\small
\begin{scope}[xshift=87pt]
\node at (372pt,32pt){$3\pi/2$};
\node at (366pt,62pt){$\pi$};
\node at (372pt,90pt){$\pi/2$};
 \node at (366pt,118pt){$0$};
\node at (374pt,147pt){$-\pi/2$};
\end{scope}

\draw (380,120) node {$=$};

\end{tikzpicture}
\caption{
The expectation value of the type-II string operator  over a sub-region after a global quantum quench, which is obtained from Fig.\ref{Fig:PartialString_Global} (left) after the conformal mapping in \eqref{ConformalMap_Global1}.
}
\label{GlobalQuench_typeII_Cylinder}
\end{figure}
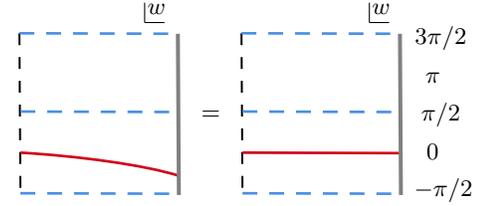

\medskip
The path integral of this setup is shown in Fig.\ref{Fig:PartialString_Global} (See Sec.\ref{Sec:Global2} for more details on this setup), where the string operator is inserted along $C=\{x+i\tau, \,0\le x<+\infty\}$. A small disc of radius $\epsilon$ is removed at $0+i\tau$ to introduce a UV cutoff.
Then, by considering the same conformal mapping in \eqref{ConformalMap_Global1}, one can map the $z$-plane in 
Fig.\ref{Fig:PartialString_Global} (left) to a $w$-cylinder in Fig.\ref{GlobalQuench_typeII_Cylinder}.

In Fig.\ref{GlobalQuench_typeII_Cylinder}, we have deformed the string operator from a curve to a straight line. The width of this cylinder along $\Im(w)$ direction is $2\pi$, and its length along $\Re(w)$ direction is $W\simeq 2\pi t/\beta$, which is the same as that in Fig.\ref{Global2_cylinder}.

It is emphasized that there is a key difference between the configurations in Fig.\ref{GlobalQuench_typeII_Cylinder} and
Fig.\ref{Global2_cylinder}. Let us consider the case when there is no symmetry-breaking interface, i.e., the whole system preserves the global symmetry during the time evolution. It is reminded there is in fact an orientation in the string operator, as shown in \eqref{Fusion_inverse}.
Then the configuration in Fig.\ref{Global2_cylinder} after removing the symmetry-breaking interfaces becomes
\be
\label{Two_string}
\begin{tikzpicture}[x=0.75pt,y=0.75pt,yscale=-0.7,xscale=0.7]

\draw    (99.5,121) -- (220.5,121) ;
\draw    (100.5,200) -- (221.5,200) ;
\draw  [dash pattern={on 4.5pt off 4.5pt}]  (99.5,121) -- (100.5,200) ;
\draw [color={rgb, 255:red, 155; green, 155; blue, 155 }  ,draw opacity=1 ][line width=1.5]    (220.5,121) -- (221.5,200) ;
\draw [color={rgb, 255:red, 208; green, 2; blue, 27 }  ,draw opacity=1 ][line width=1.0]    (99.5,140) -- (220.5,140) ;

\draw [shift={(151.7,140)}, rotate = 0] [fill={rgb, 255:red, 208; green, 2; blue, 27 }  ,fill opacity=1 ][line width=0.08]  [draw opacity=0] (11.61,-5.58) -- (0,0) -- (11.61,5.58) -- cycle    ;
\draw [color={rgb, 255:red, 208; green, 2; blue, 27 }  ,draw opacity=1 ][line width=1.0]    (100.5,181) -- (221.5,181) ;
\draw [shift={(167.8,181)}, rotate = 180] [fill={rgb, 255:red, 208; green, 2; blue, 27 }  ,fill opacity=1 ][line width=0.08]  [draw opacity=0] (11.61,-5.58) -- (0,0) -- (11.61,5.58) -- cycle    ;
\draw    (295.5,121) -- (416.5,121) ;
\draw    (296.5,200) -- (417.5,200) ;
\draw  [dash pattern={on 4.5pt off 4.5pt}]  (295.5,121) -- (296.5,200) ;
\draw [color={rgb, 255:red, 155; green, 155; blue, 155 }  ,draw opacity=1 ][line width=1.5]    (416.5,121) -- (417.5,200) ;

\node at (191pt,111pt){fusing};
\node at (195pt,125pt){$=$};

\end{tikzpicture}
\ee
where we have considered \textit{invertible} topological lines for simplicity. That is, after fusing the two string operators with opposite directions, they annihilate with each other and result in an identity operator. Therefore, the time evolution of type-I string operator becomes trivial when the whole system preserves the symmetry. When the topological lines in \eqref{Two_string} are non-invertible, the fusion results in general will contain more than one defect lines. It is not straightforward to see the configurations in the left of \eqref{Two_string} are independent of time, i.e., independent of the length of the cylinder. In this case, it is more straightforward to understand its time independence based on the argument near \eqref{L_trivial}.

Now, for the configuration in Fig.\ref{GlobalQuench_typeII_Cylinder}, since there is a single string operator living on the cylinder, no matter how we deform it, the string operator cannot be removed, as follows:
\be
\label{One_string}
\begin{tikzpicture}[x=0.75pt,y=0.75pt,yscale=-0.7,xscale=0.7]

\draw    (99.5,121) -- (220.5,121) ;
\draw    (100.5,200) -- (221.5,200) ;
\draw  [dash pattern={on 4.5pt off 4.5pt}]  (99.5,121) -- (100.5,200) ;
\draw [color={rgb, 255:red, 155; green, 155; blue, 155 }  ,draw opacity=1 ][line width=1.5]    (220.5,121) -- (221.5,200) ;
\draw [color={rgb, 255:red, 208; green, 2; blue, 27 }  ,draw opacity=1 ][line width=1.0]    (100.5,181) -- (221.5,181) ;
\draw [shift={(167.8,181)}, rotate = 180] [fill={rgb, 255:red, 208; green, 2; blue, 27 }  ,fill opacity=1 ][line width=0.08]  [draw opacity=0] (11.61,-5.58) -- (0,0) -- (11.61,5.58) -- cycle    ;
\draw    (295.5,121) -- (416.5,121) ;
\draw    (296.5,200) -- (417.5,200) ;
\draw  [dash pattern={on 4.5pt off 4.5pt}]  (295.5,121) -- (296.5,200) ;
\draw [color={rgb, 255:red, 155; green, 155; blue, 155 }  ,draw opacity=1 ][line width=1.5]    (416.5,121) -- (417.5,200) ;
\draw [color={rgb, 255:red, 208; green, 2; blue, 27 }  ,draw opacity=1 ][line width=1.0]    (296.5,181) -- (417.5,181) ;
\draw [shift={(363.8,181)}, rotate = 180] [fill={rgb, 255:red, 208; green, 2; blue, 27 }  ,fill opacity=1 ][line width=0.08]  [draw opacity=0] (11.61,-5.58) -- (0,0) -- (11.61,5.58) -- cycle    ;

\node at (195pt,125pt){$=$};

\end{tikzpicture}
\ee
Physically, this means no matter whether there is a symmetry-breaking interface or not in Fig.\ref{Fig:PartialString_Global} (left), the time evolution of type-II string operator is nontrivial.

\begin{figure}
\centering
\begin{tikzpicture}

    \node[inner sep=0pt] (russell) at (15pt,-85pt)
    {\includegraphics[width=.26\textwidth]{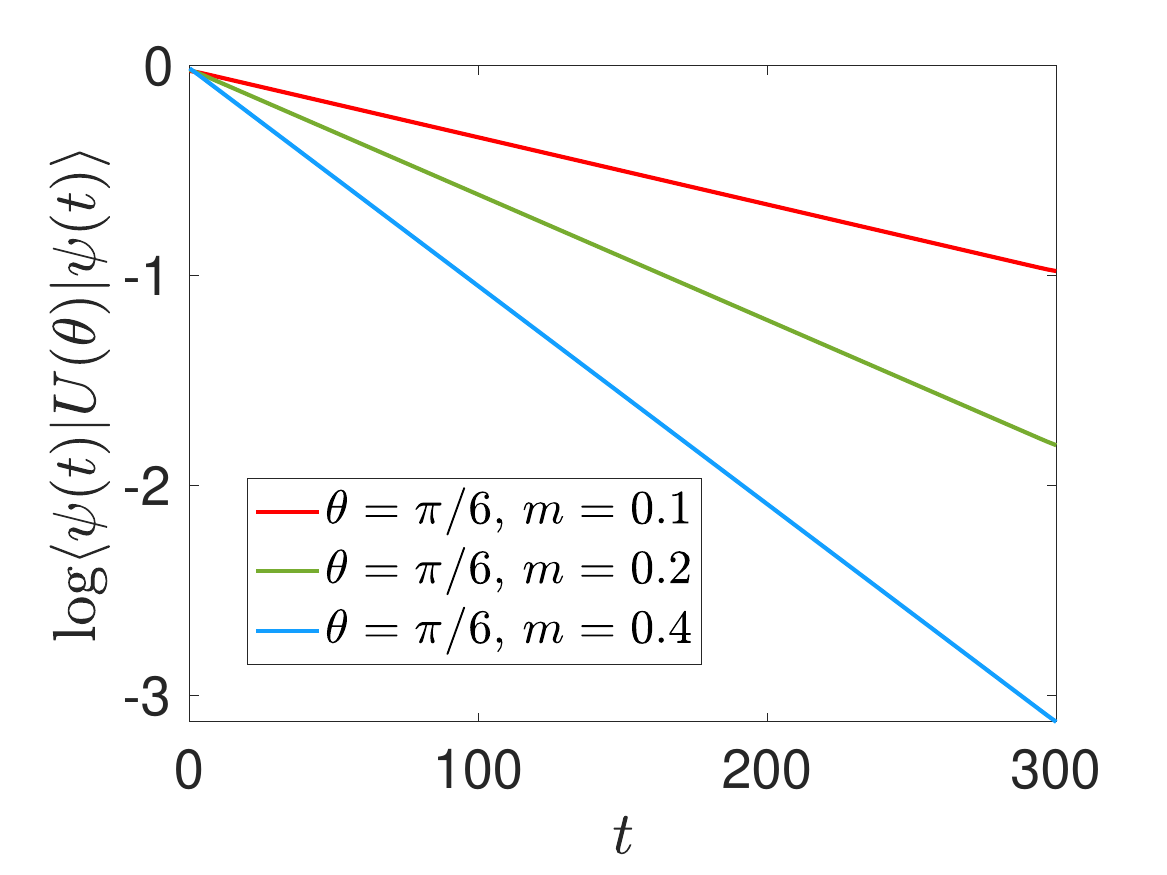}};
    \node[inner sep=0pt] (russell) at (140pt,-85pt)
    {\includegraphics[width=.25\textwidth]{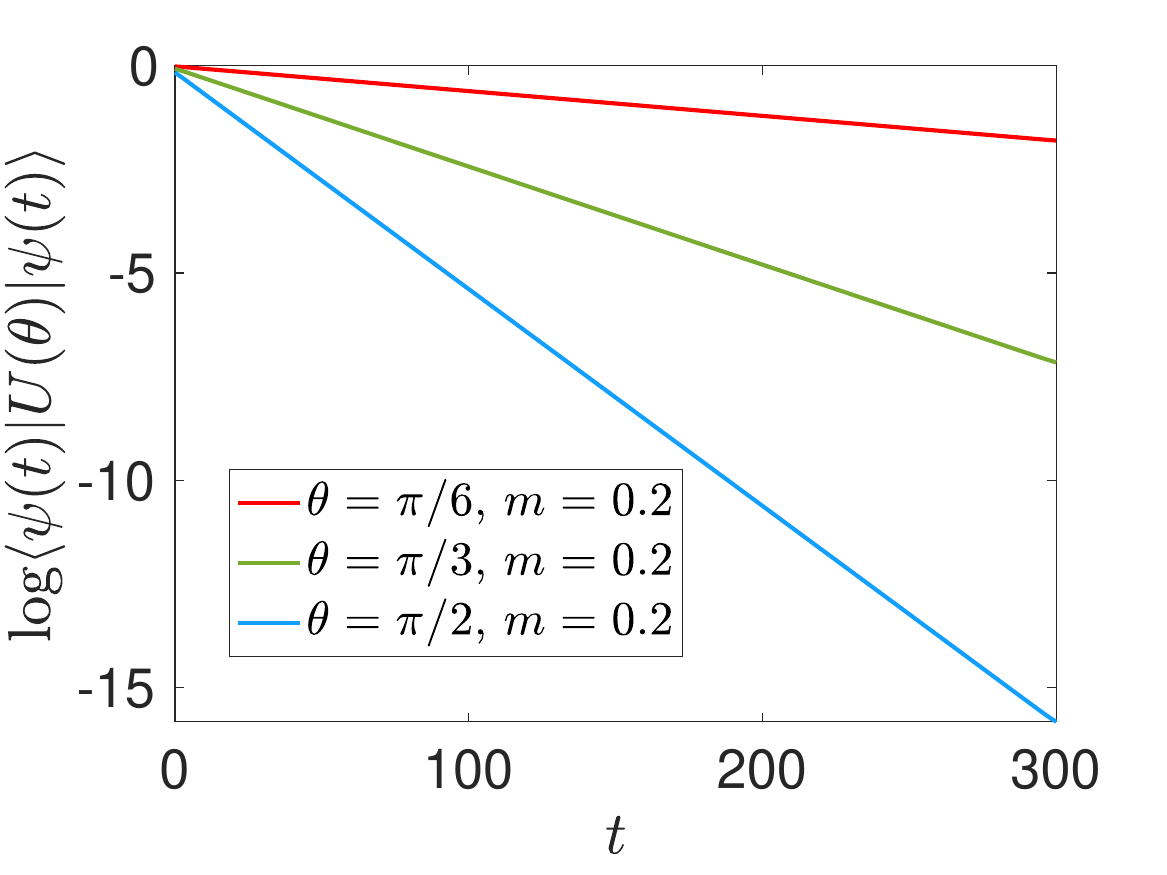}};

    \node[inner sep=0pt] (russell) at (17pt,-185pt)
    {\includegraphics[width=.255\textwidth]{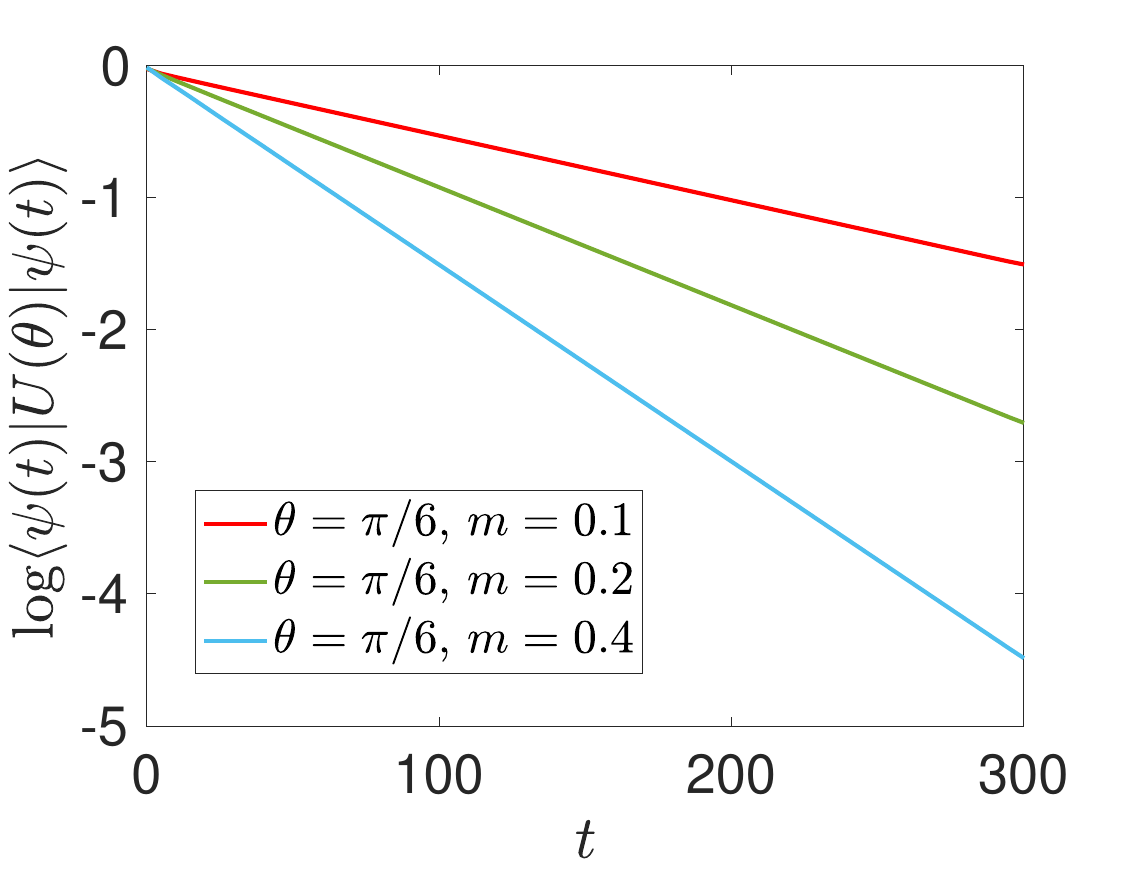}};

      \node[inner sep=0pt] (russell) at (140pt,-185pt)
    {\includegraphics[width=.255\textwidth]{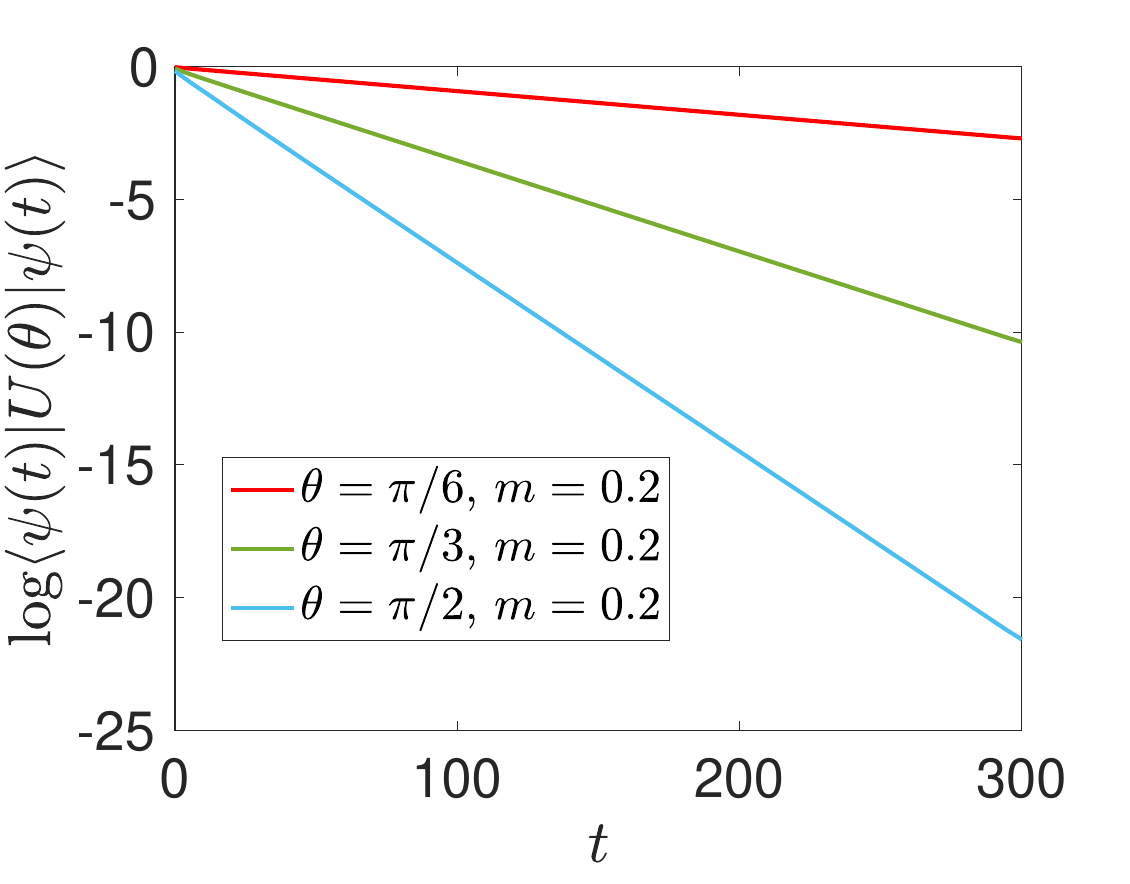}};

             \node at (58pt, -54pt){(a)};
            \node at (180pt, -60pt){(b)};
          
            \node at (58pt, -154pt){(c)};                     \node at (180pt, -160pt){(d)};

    \end{tikzpicture}
\caption{Lattice calculation of type-II string operator evolution in the interval $[0,L]$, where the total system is $[0, 2L]$.
There is no symmetry breaking in (a) and (b), while there is a symmetry-breaking interface in the middle of the system in (c) and (d). 
We consider different mass term $m$ with a fixed $\theta$ in (a) and (c), and different $\theta$ with a fixed mass term $m$ in (b) and (d). For the symmetry-breaking interface introduced in (c) and (d), we choose $h=1/2$ in \eqref{Global_defect_middle}.
}
\label{Fig:TypeII_GlobalQuench1}
\end{figure}
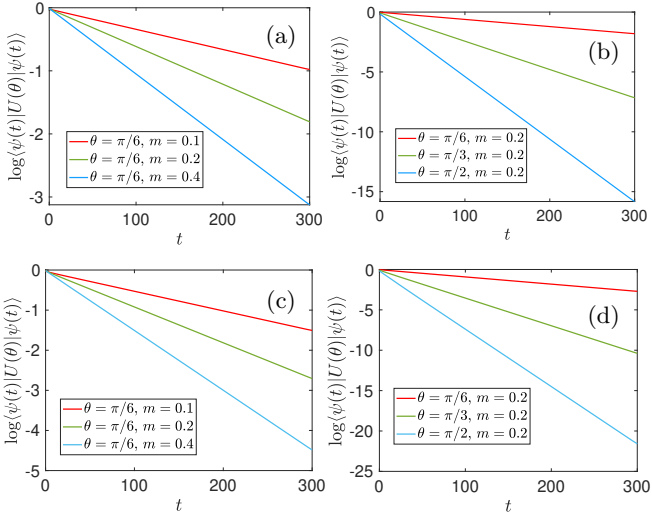

\medskip

Now let us evaluate this type-II string operator evolution explicitly. From Fig.\ref{GlobalQuench_typeII_Cylinder}, one can obtain 
\be
\label{Z_typeII_global_infinite}
\langle \psi(t)|\mathcal L_A|\psi(t)\rangle=\frac{Z_{\text{defect}}}{Z_0}
=\frac{\langle b|e^{-H_{\text{defect}}W(t)}|B\rangle}{
\langle b|e^{-H_0 W(t)}|B\rangle
}
\ee
where $Z_{\text{defect}}$ and $Z_0$ are pictorially shown as follows:
\be
\label{Eq:H_defect2_typeII}
\begin{tikzpicture}[x=0.75pt,y=0.75pt,yscale=-0.7,xscale=0.7]

\draw  [dash pattern={on 4.5pt off 4.5pt}] (100,147.5) .. controls (100,130.1) and (106.16,116) .. (113.75,116) .. controls (121.34,116) and (127.5,130.1) .. (127.5,147.5) .. controls (127.5,164.9) and (121.34,179) .. (113.75,179) .. controls (106.16,179) and (100,164.9) .. (100,147.5) -- cycle ;
\draw  [color={rgb, 255:red, 155; green, 155; blue, 155 }  ,draw opacity=1 ][line width=1.5]  (246,147.5) .. controls (246,130.1) and (252.16,116) .. (259.75,116) .. controls (267.34,116) and (273.5,130.1) .. (273.5,147.5) .. controls (273.5,164.9) and (267.34,179) .. (259.75,179) .. controls (252.16,179) and (246,164.9) .. (246,147.5) -- cycle ;
\draw [color={rgb, 255:red, 74; green, 144; blue, 226 }  ,draw opacity=1 ][line width=1.5]  [dash pattern={on 5.63pt off 4.5pt}]  (113.75,116) -- (259.75,116) ;
\draw [color={rgb, 255:red, 74; green, 144; blue, 226 }  ,draw opacity=1 ][line width=1.5]  [dash pattern={on 5.63pt off 4.5pt}]  (113.75,179) -- (259.75,179) ;
\draw  [color={rgb, 255:red, 245; green, 166; blue, 35 }  ,draw opacity=1 ] (143,147.5) .. controls (143,130.1) and (149.16,116) .. (156.75,116) .. controls (164.34,116) and (170.5,130.1) .. (170.5,147.5) .. controls (170.5,164.9) and (164.34,179) .. (156.75,179) .. controls (149.16,179) and (143,164.9) .. (143,147.5) -- cycle ;
\draw [color={rgb, 255:red, 208; green, 2; blue, 27 }  ,draw opacity=1 ][line width=0.75]    (127.75,151) -- (273.75,151) ;

\begin{scope}[xshift=0pt,yshift=69pt]
\draw  [dash pattern={on 4.5pt off 4.5pt}] (100,147.5) .. controls (100,130.1) and (106.16,116) .. (113.75,116) .. controls (121.34,116) and (127.5,130.1) .. (127.5,147.5) .. controls (127.5,164.9) and (121.34,179) .. (113.75,179) .. controls (106.16,179) and (100,164.9) .. (100,147.5) -- cycle ;
\draw  [color={rgb, 255:red, 155; green, 155; blue, 155 }  ,draw opacity=1 ][line width=1.5]  (246,147.5) .. controls (246,130.1) and (252.16,116) .. (259.75,116) .. controls (267.34,116) and (273.5,130.1) .. (273.5,147.5) .. controls (273.5,164.9) and (267.34,179) .. (259.75,179) .. controls (252.16,179) and (246,164.9) .. (246,147.5) -- cycle ;
\draw [color={rgb, 255:red, 74; green, 144; blue, 226 }  ,draw opacity=1 ][line width=1.5]  [dash pattern={on 5.63pt off 4.5pt}]  (113.75,116) -- (259.75,116) ;
\draw [color={rgb, 255:red, 74; green, 144; blue, 226 }  ,draw opacity=1 ][line width=1.5]  [dash pattern={on 5.63pt off 4.5pt}]  (113.75,179) -- (259.75,179) ;
\draw  [color={rgb, 255:red, 245; green, 166; blue, 35 }  ,draw opacity=1 ] (143,147.5) .. controls (143,130.1) and (149.16,116) .. (156.75,116) .. controls (164.34,116) and (170.5,130.1) .. (170.5,147.5) .. controls (170.5,164.9) and (164.34,179) .. (156.75,179) .. controls (149.16,179) and (143,164.9) .. (143,147.5) -- cycle ;

\end{scope}

\node at (10pt,109pt){$Z_{\text{defect}}=$};
\node at (25pt,182pt){$Z_{0}=$};

\small

\node at (58pt,109pt){$\langle b|$};
\node at (222pt,109pt){$|B\rangle$};

\node at (58pt,180pt){$\langle b|$};
\node at (222pt,180pt){$|B\rangle$};

\node at (137pt,185pt){\textcolor{orange}{$H_0$}};
\node at (143pt,95pt){\textcolor{orange}{$H_{\text{defect}}$}};

\end{tikzpicture}
\ee
Since the length of the cylinder is $W(t)\simeq \frac{2\pi}{\beta}t \gg 1$, $Z_{\text{defect}}$ and $Z_0$ are mainly dominated by the ground states of $H_{\text{defect}}$ and $H_0$, which we denote as $|G_d\rangle$ and $|G_0\rangle$ respectively. Here $H_{\text{defect}}$ and $H_0$ are defined along the orange circles in \eqref{Eq:H_defect2_typeII}. Then we have
\be
\langle \psi(t)|\mathcal L_A|\psi(t)\rangle\simeq \frac{\langle b|G_d\rangle\langle G_d|B\rangle}{\langle b|G_0\rangle\langle G_0|B\rangle}e^{-(E^0_\text{defect}-E^0)W(t)}
\ee
where $E^0_\text{defect}$ and $E^0$ are the ground state energies of the corresponding Hamiltonians. Then one has
\be
\log|\langle \psi(t)|\mathcal L_A|\psi(t)\rangle| \simeq -\kappa \, t,
\ee
up to a constant sub-leading term, and 
\be
\label{kappa_typeII_Global1}
\kappa=\frac{2\pi}{\beta}(E^0_\text{defect}-E^0).
\ee
That is, the expectation value of type-II string operator will decay exponentially in time after a global quench. 

Note that the presence of symmetry-breaking interface or not will in general result in different $\kappa$. This is because the interfaces (blue dashed line) in \eqref{Eq:H_defect2_typeII} will in general modify the corresponding ground state energies of $H_{\text{defect}}$ and $H_0$. 

\medskip

In Fig.\ref{Fig:TypeII_GlobalQuench1}, we study the time evolution  $\langle \psi(t)|\mathcal L_A|\psi(t)\rangle$ in a free-fermion lattice model by following the same procedures in Sec.\ref{Sec:Numerics_typeI_Global2}, with the only difference that now we define the string operator over half of the system $[0,\,L/2]$. In Fig.\ref{Fig:TypeII_GlobalQuench1} (a) and (b), the $U(1)$ symmetry is preserved over the whole system, and in Fig.\ref{Fig:TypeII_GlobalQuench1} (c) and (d) we introduce a symmetry-breaking interface in the middle of the system. One can find that in both cases, $\langle \psi(t)|\mathcal L_A|\psi(t)\rangle$ decays exponentially in time, although the decaying rates are different, as expected from our CFT analysis.

The dependence of $\kappa$ on the mass term in the initial state, as well as its dependence on $\theta$, can be understood in the same way as those in Sec.\ref{Sec:Global1} and Sec.\ref{Sec:Global2}, which we do not repeat here.

\subsubsection{Finite interval at the end of a semi-infinite line}
\label{Sec:TypeII_GlobalQuench}

Now let us turn to the case of a semi-infinite system in $[0,+\infty)$ after a global quantum quench. The setup is the same as that in Sec.\ref{Sec:Global1}, but now the string operator is defined in a subsystem $[0,l]$ near the end of this system. Again, we consider the general case that there is a symmetry breaking at the boundary $x=0$.

\begin{figure}
\begin{tikzpicture}[x=0.75pt,y=0.75pt,yscale=-0.7,xscale=0.7]

\draw  [color={rgb, 255:red, 155; green, 155; blue, 155 }  ,draw opacity=1 ][line width=1.5]  (100,75.5) .. controls (100,58.1) and (106.16,44) .. (113.75,44) .. controls (121.34,44) and (127.5,58.1) .. (127.5,75.5) .. controls (127.5,92.9) and (121.34,107) .. (113.75,107) .. controls (106.16,107) and (100,92.9) .. (100,75.5) -- cycle ;
\draw [color={rgb, 255:red, 0; green, 0; blue, 0 }  ,draw opacity=1 ][line width=0.75]    (113.75,44) -- (240.75,44) ;
\draw [color={rgb, 255:red, 0; green, 0; blue, 0 }  ,draw opacity=1 ][line width=0.75]    (113.75,107) -- (239.75,107) ;
\draw  [dash pattern={on 4.5pt off 4.5pt}] (226,75.5) .. controls (226,58.1) and (232.16,44) .. (239.75,44) .. controls (247.34,44) and (253.5,58.1) .. (253.5,75.5) .. controls (253.5,92.9) and (247.34,107) .. (239.75,107) .. controls (232.16,107) and (226,92.9) .. (226,75.5) -- cycle ;
\draw  [draw opacity=0][line width=2.25]  (126.74,63.04) .. controls (127.55,66.84) and (128,71.06) .. (128,75.5) .. controls (128,79.4) and (127.65,83.13) .. (127.02,86.55) -- (114,75.5) -- cycle ; \draw  [color={rgb, 255:red, 74; green, 144; blue, 226 }  ,draw opacity=1 ][line width=2.25]  (126.74,63.04) .. controls (127.55,66.84) and (128,71.06) .. (128,75.5) .. controls (128,79.4) and (127.65,83.13) .. (127.02,86.55) ;  
\draw    (122.5,64) -- (129.74,61.04) ;
\draw    (123.5,86) -- (130.74,88.04) ;
\draw  [color={rgb, 255:red, 155; green, 155; blue, 155 }  ,draw opacity=1 ][line width=1.5]  (316,75.5) .. controls (316,58.1) and (322.16,44) .. (329.75,44) .. controls (337.34,44) and (343.5,58.1) .. (343.5,75.5) .. controls (343.5,92.9) and (337.34,107) .. (329.75,107) .. controls (322.16,107) and (316,92.9) .. (316,75.5) -- cycle ;
\draw [color={rgb, 255:red, 0; green, 0; blue, 0 }  ,draw opacity=1 ][line width=0.75]    (329.75,44) -- (456.75,44) ;
\draw [color={rgb, 255:red, 0; green, 0; blue, 0 }  ,draw opacity=1 ][line width=0.75]    (329.75,107) -- (455.75,107) ;
\draw  [dash pattern={on 4.5pt off 4.5pt}] (442,75.5) .. controls (442,58.1) and (448.16,44) .. (455.75,44) .. controls (463.34,44) and (469.5,58.1) .. (469.5,75.5) .. controls (469.5,92.9) and (463.34,107) .. (455.75,107) .. controls (448.16,107) and (442,92.9) .. (442,75.5) -- cycle ;
\draw [color={rgb, 255:red, 208; green, 2; blue, 27 }  ,draw opacity=1 ][line width=1.0]    (342.75,77) -- (468.75,77) ;
\draw  [draw opacity=0][line width=2.25]  (342.74,63.04) .. controls (343.55,66.84) and (344,71.06) .. (344,75.5) .. controls (344,79.4) and (343.65,83.13) .. (343.02,86.55) -- (330,75.5) -- cycle ; 

\draw  [color={rgb, 255:red, 74; green, 144; blue, 226 }  ,draw opacity=1 ][line width=2.25]  (342.74,63.04) .. controls (343.55,66.84) and (344,71.06) .. (344,75.5) .. controls (344,79.4) and (343.65,83.13) .. (343.02,86.55) ;  
\draw    (338.5,64) -- (345.74,61.04) ;
\draw    (339.5,86) -- (346.74,88.04) ;
\draw [color={rgb, 255:red, 208; green, 2; blue, 27 }  ,draw opacity=1 ][line width=1.0]    (127.5,75.5) .. controls (172.5,72) and (225.5,79) .. (252.5,89) ;

\node at (210pt,60pt){$=$};

\begin{scope}[xshift=190pt,yshift=-15pt]
\draw (140pt,40pt)--(155pt,40pt);
\draw (140pt,40pt)--(140pt,25pt);
\small\node at (149pt,30pt){$w$};
\end{scope}

\begin{scope}[xshift=30pt,yshift=-15pt]
\draw (140pt,40pt)--(155pt,40pt);
\draw (140pt,40pt)--(140pt,25pt);
\small\node at (149pt,30pt){$w$};
\end{scope}

\end{tikzpicture}
\caption{
The expectation value of the type-II string operator over a subsystem after a global quantum quench, which is obtained from Fig.\ref{Fig:PartialString_Global} (right) after the conformal mapping in \eqref{Map_typeII_global}.
The left boundary of the cylinder is a mixing of two different boundary conditions. The width in the $\Im(w)$ direction is $2\pi$.
}
\label{GlobalQuench2_typeII_Cylinder}
\end{figure}
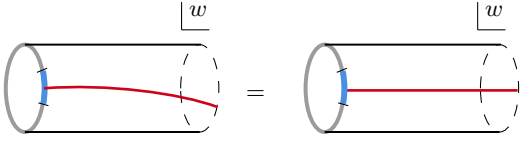

The path integral of $\langle \psi(t)|\mathcal L_A|\psi(t)\rangle$ for this setup is shown in Fig.\ref{Fig:PartialString_Global} (right),
where the string operator is inserted along $C=\{x+i\tau, \,0\le x\le l\}$. A small disc of radius $\epsilon$ is removed at $l+i\tau$ to introduce a UV cutoff, with a conformal boundary condition $|b\rangle$ imposed along the boundary. Then, we consider a two-step conformal mapping \cite{2018Wenb} 
\be
\label{Map_typeII_global}
\left\{
\begin{split}
\xi(z)=&\sinh\left(\frac{2\pi z}{\beta}\right)\\
w(\xi)=&-\log \left(\frac{1+\bar \xi_0}{1+\xi_0}\cdot \frac{\xi-\xi_0}{\xi+\bar \xi_0}\right)
\end{split}
\right.
\ee
where $\xi=\xi(z)$ and $\xi_0=\xi(z_0)$ with $z_0=l+i\tau$.
Then the configuration in Fig.\ref{Fig:PartialString_Global} (right) is mapped to the $w$-cylinder in Fig.\ref{GlobalQuench2_typeII_Cylinder}.
It is noted that the left boundary of the $w$-cylinder is a mixing of two different boundary conditions (with boundary condition changing operator inserted). 
\footnote{The physical meaning of the mixing of boundary conditions is clear by considering the configuration before the conformal mapping. As sketched in Fig.\ref{Fig:PartialString_Global} (right), one boundary condition is imposed along the physical boundary of the system at $x=0$, and the other boundary condition is used to define the initial state. These two boundary conditions are independent of each other.}
This is different from the case of type-I string order parameters, where the string operator is connected to the boundary at infinite.

Before we move on to evaluate $\langle \psi(t)|\mathcal L_A|\psi(t)\rangle$, it is helpful to compare the difference between the type-I string operator in Fig.\ref{Fig:Global1} and the type-II string operator in Fig.\ref{Fig:PartialString_Global} (right), which are in the same setup. 
The string order parameters in these two cases correspond to the partition functions over the strip in Fig.\ref{Fig:Global2} and the cylinder in Fig.\ref{GlobalQuench2_typeII_Cylinder}, respectively.

Let us consider the situation that the symmetry is preserved over the whole system. Then one can find the configuration in Fig.\ref{Fig:Global2} becomes:
\be
\label{Boundary_Absorb}
\begin{tikzpicture}[x=0.75pt,y=0.75pt,yscale=-0.7,xscale=0.7]
\draw [color={rgb, 255:red, 155; green, 155; blue, 155 }  ,draw opacity=1 ][line width=1.5]    (99.5,129) -- (220.5,129) ;
\draw [color={rgb, 255:red, 155; green, 155; blue, 155 }  ,draw opacity=1 ][line width=1.5]    (100.5,191) -- (221.5,191) ;
\draw  [dash pattern={on 4.5pt off 4.5pt}]  (99.5,129) -- (100.5,191) ;
\draw [color={rgb, 255:red, 155; green, 155; blue, 155 }  ,draw opacity=1 ][line width=1.5]    (220.5,128) -- (221.5,191) ;
\draw [color={rgb, 255:red, 208; green, 2; blue, 27 }  ,draw opacity=1 ][line width=1.0]    (100,160.5) -- (221,160.5) ;
\draw [color={rgb, 255:red, 155; green, 155; blue, 155 }  ,draw opacity=1 ][line width=1.5]    (292.5,128) -- (413.5,128) ;
\draw [color={rgb, 255:red, 155; green, 155; blue, 155 }  ,draw opacity=1 ][line width=1.5]    (293.5,190) -- (414.5,190) ;
\draw  [dash pattern={on 4.5pt off 4.5pt}]  (292.5,128) -- (293.5,190) ;
\draw [color={rgb, 255:red, 155; green, 155; blue, 155 }  ,draw opacity=1 ][line width=1.5]    (413.5,127) -- (414.5,190) ;

\draw (227,150.4) node [anchor=north west][inner sep=0.75pt]    {$=\langle \mathcal L\rangle$};

\end{tikzpicture}
\ee
where we have moved the string operator all the way down to the boundary. Since the boundary preserves the symmetry, then one has $\mathcal L|B\rangle=\langle \mathcal L\rangle |B\rangle$. 
That is, the string operator in \eqref{Boundary_Absorb} is absorbed by the boundary. Therefore, the type-I string order parameter is a constant and independent of time when the symmetry is preserved over the whole system.

Now, for the configuration of type-II string order parameters in Fig.\ref{GlobalQuench2_typeII_Cylinder}, even if the symmetry is preserved over the whole system, one can find no matter how we deform the string operator, we cannot remove it, as follows:
\be
\begin{tikzpicture}[x=0.75pt,y=0.75pt,yscale=-0.7,xscale=0.7]

\draw  [color={rgb, 255:red, 155; green, 155; blue, 155 }  ,draw opacity=1 ][line width=1.5]  (63,79.5) .. controls (63,62.1) and (69.16,48) .. (76.75,48) .. controls (84.34,48) and (90.5,62.1) .. (90.5,79.5) .. controls (90.5,96.9) and (84.34,111) .. (76.75,111) .. controls (69.16,111) and (63,96.9) .. (63,79.5) -- cycle ;
\draw [color={rgb, 255:red, 0; green, 0; blue, 0 }  ,draw opacity=1 ][line width=0.75]    (76.75,48) -- (203.75,48) ;
\draw [color={rgb, 255:red, 0; green, 0; blue, 0 }  ,draw opacity=1 ][line width=0.75]    (76.75,111) -- (202.75,111) ;
\draw  [dash pattern={on 4.5pt off 4.5pt}] (189,79.5) .. controls (189,62.1) and (195.16,48) .. (202.75,48) .. controls (210.34,48) and (216.5,62.1) .. (216.5,79.5) .. controls (216.5,96.9) and (210.34,111) .. (202.75,111) .. controls (195.16,111) and (189,96.9) .. (189,79.5) -- cycle ;
\draw [color={rgb, 255:red, 208; green, 2; blue, 27 }  ,draw opacity=1 ][line width=1.0]    (89.75,81) -- (215.75,81) ;
\draw  [color={rgb, 255:red, 155; green, 155; blue, 155 }  ,draw opacity=1 ][line width=1.5]  (272,78.5) .. controls (272,61.1) and (278.16,47) .. (285.75,47) .. controls (293.34,47) and (299.5,61.1) .. (299.5,78.5) .. controls (299.5,95.9) and (293.34,110) .. (285.75,110) .. controls (278.16,110) and (272,95.9) .. (272,78.5) -- cycle ;
\draw [color={rgb, 255:red, 0; green, 0; blue, 0 }  ,draw opacity=1 ][line width=0.75]    (285.75,47) -- (412.75,47) ;
\draw [color={rgb, 255:red, 0; green, 0; blue, 0 }  ,draw opacity=1 ][line width=0.75]    (285.75,110) -- (411.75,110) ;
\draw  [dash pattern={on 4.5pt off 4.5pt}] (398,78.5) .. controls (398,61.1) and (404.16,47) .. (411.75,47) .. controls (419.34,47) and (425.5,61.1) .. (425.5,78.5) .. controls (425.5,95.9) and (419.34,110) .. (411.75,110) .. controls (404.16,110) and (398,95.9) .. (398,78.5) -- cycle ;
\draw [color={rgb, 255:red, 208; green, 2; blue, 27 }  ,draw opacity=1 ][line width=1.0]    (298.75,80) -- (424.75,80) ;

\node at (180pt,60pt){$=$};

\end{tikzpicture}
\ee
In other words, for the type-II string operator, the time evolution $\langle \psi(t)|\mathcal L_A |\psi(t)\rangle$ is nontrivial no matter whether there is a symmetry-breaking boundary or not.

\medskip
Now let us evaluate $\langle \psi(t)|\mathcal L_A|\psi(t)\rangle$ explicitly.
For the $w$-cylinder in Fig.\ref{GlobalQuench2_typeII_Cylinder}, 
one can find its length in the $\Re(w)$ direction depends on whether $t<l$ or $t>l$ as follows \cite{2018Wenb}:
\be
\label{Wt_2scale}
W(t)\simeq \left\{
\begin{split}
&\log\left(\frac{\beta}{2\pi\epsilon}\right)+\frac{2\pi}{\beta}t,\quad t<l,\\
&\log\left(\frac{\beta}{2\pi\epsilon}\right)+\frac{2\pi}{\beta}l,\quad t>l.
\end{split}
\right.
\ee
It is reminded that here $l$ is the length of the subsystem. Then one has
\be
\label{TypeII_Global2}
\langle \psi(t)|\mathcal L_A|\psi(t)\rangle=\frac{Z_{\text{defect}}}{Z_0}=\frac{\langle B_{\text{mix}}|e^{-H_{\text{defect}}W(t)}|b\rangle}{
\langle B_{\text{mix}}|e^{-H_0 W(t)}|b\rangle
}
\ee
Here the state $|B_{\text{mix}}\rangle$ corresponds to the mixing of two boundary conditions in Fig.\eqref{GlobalQuench2_typeII_Cylinder}.
The partition functions $Z_{\text{defect}}$ and $Z_0$ are similarly defined as those in \eqref{Z_typeII_global_infinite} and \eqref{Eq:H_defect2_typeII} by modifying the boundary conditions and removing the conformal interfaces.
For both $t<l$ and $t>l$ in \eqref{Wt_2scale}, we are interested in the limit $W(t)\gg 1$. Then \eqref{TypeII_Global2} is mainly contributed by the ground states of $H_{\text{defect}}$ and $H_0$, which we denote as $|G_d\rangle$ and $|G_0\rangle$ respectively, as follows:
\be
\langle \psi(t)|\mathcal L_A|\psi(t)\rangle\simeq 
\frac{\langle B_{\text{mix}}|G_d\rangle\langle G_d|b\rangle}{
\langle B_{\text{mix}}|G_0\rangle\langle G_0|b\rangle
}e^{-(E^0_\text{defect}-E^0)W(t)}.
\ee
Then one can find 
\be
\label{Decay_typeII_Global2}
\log|\langle \psi(t)|\mathcal L_A|\psi(t)\rangle|\simeq
\left\{
\begin{split}
&-\kappa \, t, \quad t<l,\\
&-\kappa \, l, \quad t>l,
\end{split}
\right.
\ee
up to a constant sub-leading term. Here the coefficient 
$\kappa=\frac{2\pi }{\beta}(E^0_\text{defect}-E^0)$ is again non-universal and depends on both the initial state and the details of the string operators.

\begin{figure}[t]
\centering
\begin{tikzpicture}

    \node[inner sep=0pt] (russell) at (15pt,-85pt)
    {\includegraphics[width=.26\textwidth]{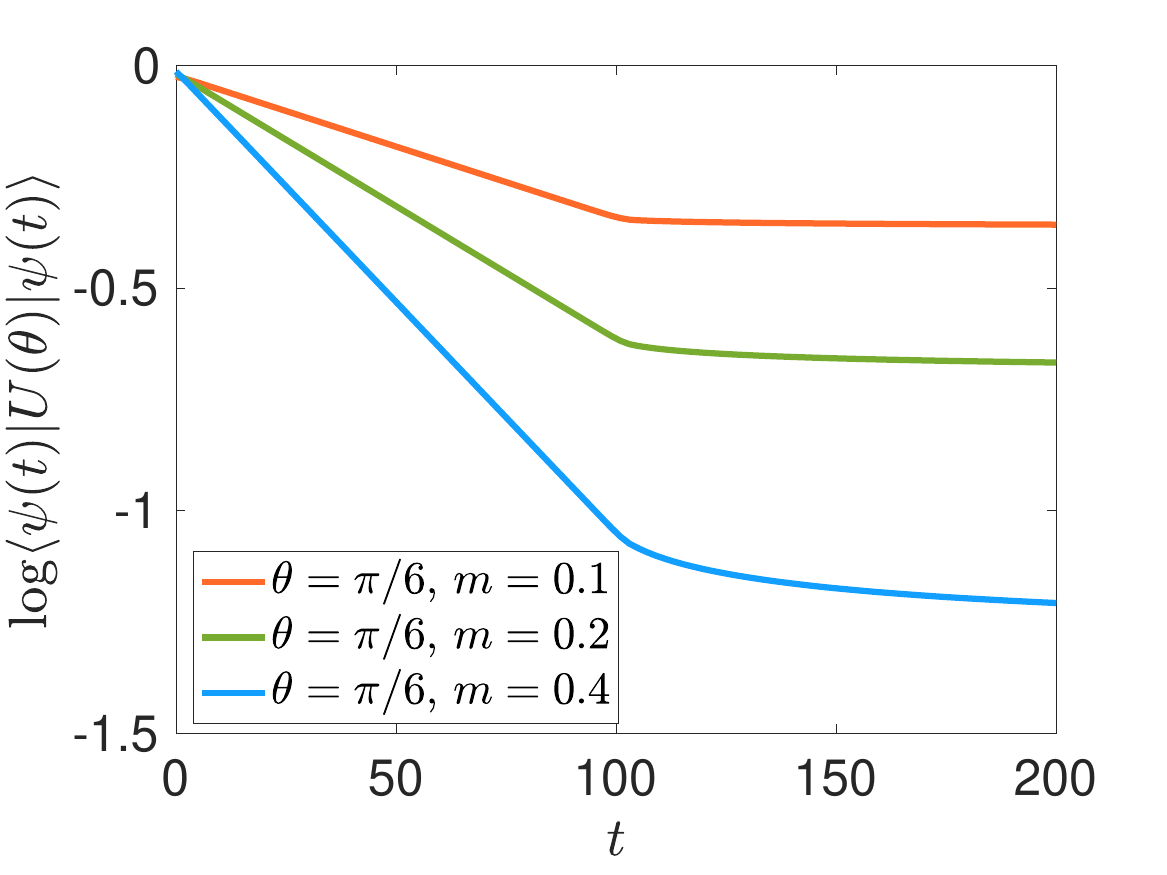}};
        \node[inner sep=0pt] (russell) at (140pt,-85pt)
    {\includegraphics[width=.26\textwidth]{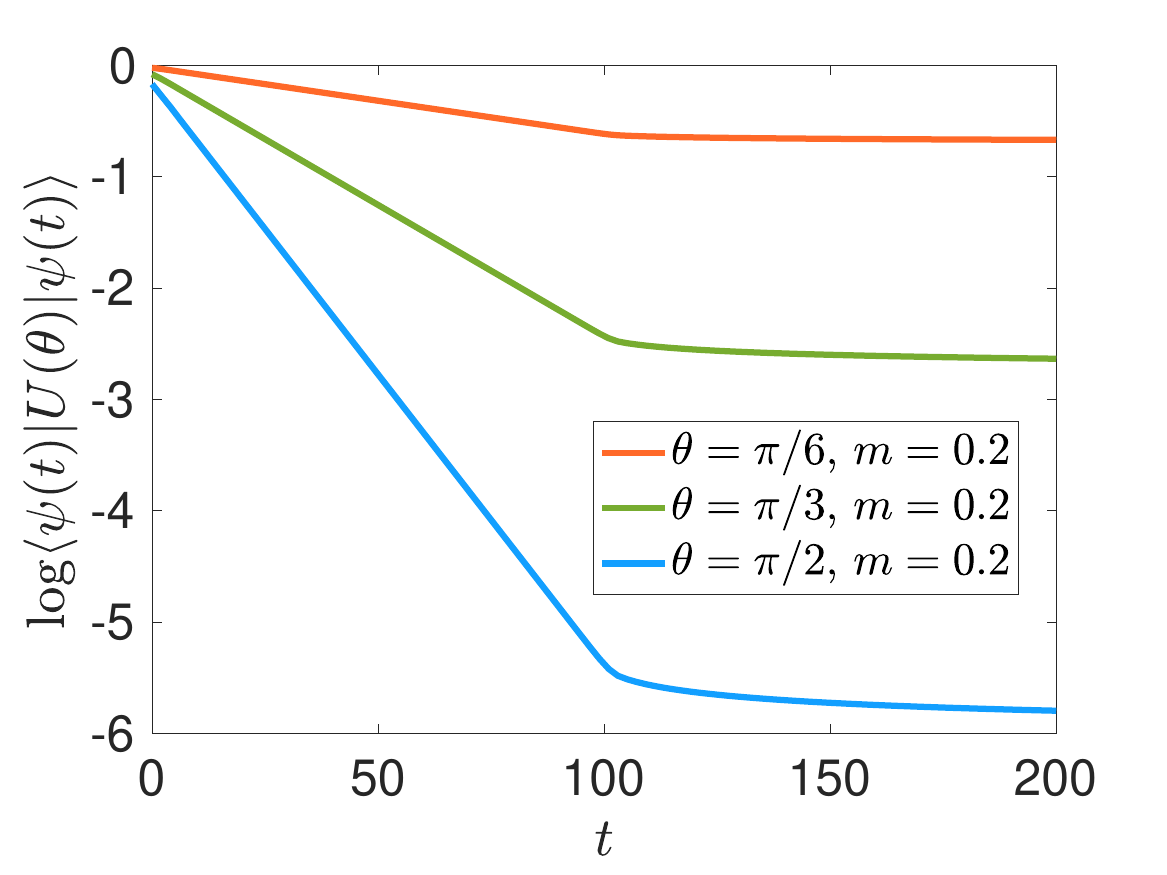}};

    \node[inner sep=0pt] (russell) at (15pt,-185pt)
    {\includegraphics[width=.26\textwidth]{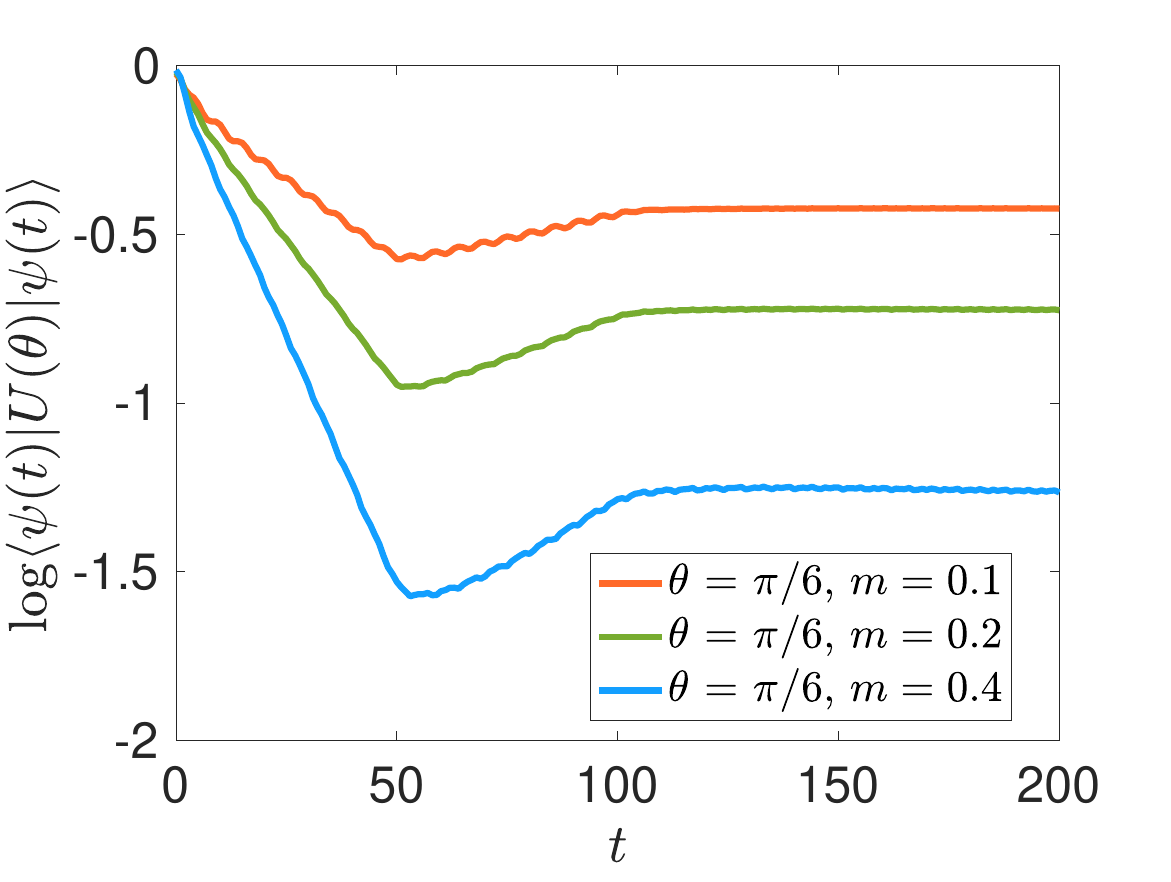}};
    
        \node[inner sep=0pt] (russell) at (140pt,-185pt)
    {\includegraphics[width=.26\textwidth]{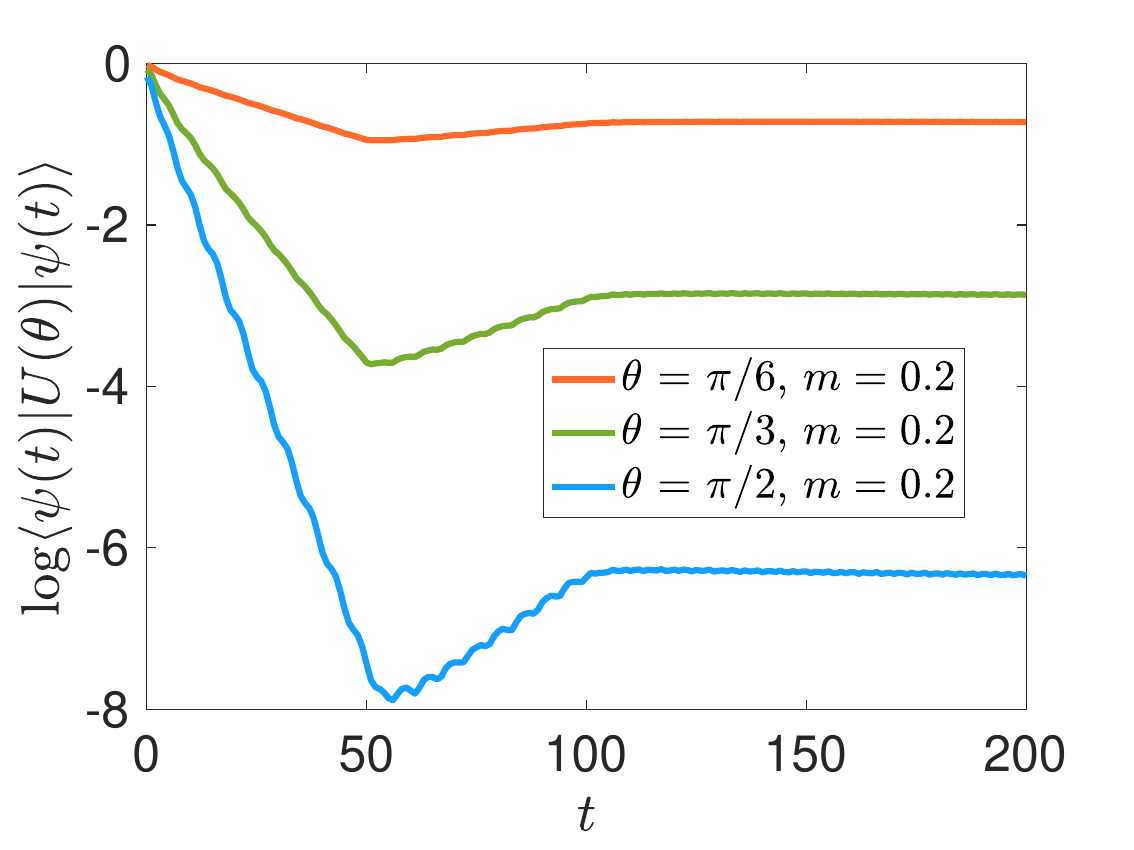}};
  
             \node at (48pt, -52pt){(a)};
            \node at (110pt, -110pt){(b)};
             \node at (48pt, -152pt){(c)};          
             \node at (168pt, -160pt){(d)};                
                          
    \end{tikzpicture}
\caption{ Lattice calculation of type-II string order parameter in the interval $[0,l]$ at the end of a finite system $[0,L]$ after a global quantum quench. We take $l=100$ and $L=400$. 
In (a) and (b), the whole system preserves the global $U(1)$ symmetry. In (c) and (d), we break the symmetry at the boundary, and here we choose $h=1/2$ in \eqref{H1_Global_semi}.
For $t<l$, the string order parameters decay exponentially in time.
Note there is a dip feature in (c) and (d). See the discussion in the main text. From $t=l$, the string order parameters start to saturate.
}
\label{Fig:NoSymBreak_Global}
\end{figure}
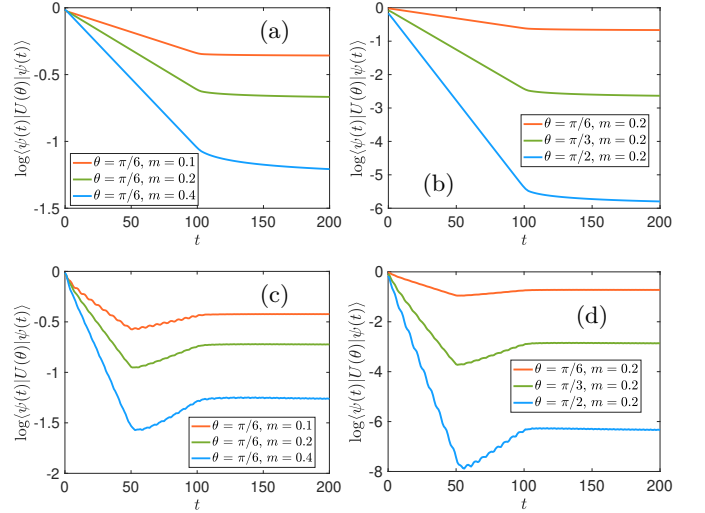

\medskip
A numerical calculation based on the lattice fermion can be found in Fig.\ref{Fig:NoSymBreak_Global}. We consider a finite interval of length $l$ at the end of a lattice of length $L$. In Fig.\ref{Fig:NoSymBreak_Global} (a) and (b), the symmetry is preserved over the whole system. In Fig.\ref{Fig:NoSymBreak_Global} (c) and (d), we introduce a symmetry-breaking boundary at $x=0$ according to \eqref{H1_Global_semi}. One can see the general features of exponential decay for $t<l$, and the saturation for $t>l$, which agrees with our CFT analysis.

One very intriguing feature is the ``dip'' feature in Fig.\ref{Fig:NoSymBreak_Global} (c) and (d), where the boundary at $x=0$ is symmetry-breaking. We give more analysis of this feature in appendix \ref{Appendix:TypeI/II}. It is found this feature depends on the strength of the symmetry-breaking term at the boundary and can vanish for certain parameters (See Fig.\ref{Fig:TypeII_changeH} in the appendix). Briefly, one always sees an exponential decay for $t\in[0,l/2]$ and a saturation for $t>l$. For the time evolution in $t\in (l/2,l]$, currently we do not have a good understanding based on the CFT study, which only gives us the features in \eqref{Decay_typeII_Global2}.
\footnote{We want to mention one possibly related structure:
If one considers the time evolution of the entanglement entropy of subsystem $[0,\,l]$ in Fig.\ref{Fig:PartialString_Global} (right), then the result only depends on whether $t<l$ or $t>l$. However, if one considers the entanglement Hamiltonian of the same subsystem, it has three different features in $t\in[0,\, l/2]$, $t\in(l/2,\,l]$, and $t>l$, respectively. See details in Sec.2.2.1 in  Ref.\cite{2018Wenb}. }
We hope to come back to this feature in the future.

\subsection{Local quench}

\begin{figure}
\centering
\begin{tikzpicture}[x=0.75pt,y=0.75pt,yscale=-0.65,xscale=0.65]

\begin{scope}[yshift=-20pt]
\draw (140pt,40pt)--(155pt,40pt);
\draw (140pt,40pt)--(140pt,25pt);
\node at (147pt,30pt){$z$};
\end{scope}

\begin{scope}[xshift=170pt,yshift=-20pt]
\draw (140pt,40pt)--(155pt,40pt);
\draw (140pt,40pt)--(140pt,25pt);
\node at (147pt,30pt){$w$};
\end{scope}

\node at (97pt,57pt){$+i\lambda$};
\node at (97pt,108pt){$-i\lambda$};

\draw [color={rgb, 255:red, 74; green, 144; blue, 226 }  ,draw opacity=1 ][line width=1.5]  [dash pattern={on 5.63pt off 4.5pt}]  (108.32,79.03) -- (108.32,146.03) ;
\draw [color={rgb, 255:red, 208; green, 2; blue, 27 }  ,draw opacity=1 ][line width=1.0]    (108.32,112.53) -- (195.32,112.53) ;

\draw  [draw opacity=0][dash pattern={on 4.5pt off 4.5pt}] (107.59,126.53) .. controls (100.19,126.15) and (94.31,120.03) .. (94.31,112.53) .. controls (94.31,104.79) and (100.58,98.52) .. (108.32,98.52) .. controls (116.06,98.52) and (122.34,104.79) .. (122.34,112.53) .. controls (122.34,120.27) and (116.06,126.55) .. (108.32,126.55) -- (108.32,112.53) -- cycle ; 

\draw  [dash pattern={on 4.5pt off 4.5pt}] (107.59,126.53) .. controls (100.19,126.15) and (94.31,120.03) .. (94.31,112.53) .. controls (94.31,104.79) and (100.58,98.52) .. (108.32,98.52) .. controls (116.06,98.52) and (122.34,104.79) .. (122.34,112.53) .. controls (122.34,120.27) and (116.06,126.55) .. (108.32,126.55) ;  

\draw [color={rgb, 255:red, 155; green, 155; blue, 155 }  ,draw opacity=1 ][line width=3]    (108.32,29.03) -- (108.32,79.03) ;
\draw [color={rgb, 255:red, 155; green, 155; blue, 155 }  ,draw opacity=1 ][line width=3]    (108.32,146.03) -- (108.32,194.03) ;

\begin{scope}[xshift=-100pt]
\draw [color={rgb, 255:red, 0; green, 0; blue, 0 }  ,draw opacity=1 ][line width=0.75]  [dash pattern={on 4.5pt off 4.5pt}]  (411,37.5) -- (412,199) ;
\draw [color={rgb, 255:red, 74; green, 144; blue, 226 }  ,draw opacity=1 ][line width=1.0]  [dash pattern={on 5.63pt off 4.5pt}]  (412,199) -- (571.83,199.5) ;
\draw [color={rgb, 255:red, 74; green, 144; blue, 226 }  ,draw opacity=1 ][line width=1.0]  [dash pattern={on 5.63pt off 4.5pt}]  (412,116.5) -- (571.83,117) ;
\draw [color={rgb, 255:red, 74; green, 144; blue, 226 }  ,draw opacity=1 ][line width=1.0]  [dash pattern={on 5.63pt off 4.5pt}]  (411,37.5) -- (570.83,38) ;
\draw [color={rgb, 255:red, 208; green, 2; blue, 27 }  ,draw opacity=1 ][line width=1.0]    (412,158) -- (571.83,158.5) ;
\draw [color={rgb, 255:red, 128; green, 128; blue, 128 }  ,draw opacity=1 ][line width=1.5]    (570.83,38) -- (571.83,200.75) ;
\end{scope}

\small
\begin{scope}[xshift=-23pt]
\node at (372pt,32pt){$3\pi/2$};
\node at (366pt,62pt){$\pi$};
\node at (372pt,90pt){$\pi/2$};
 \node at (366pt,118pt){$0$};
\node at (374pt,147pt){$-\pi/2$};
\end{scope}
\end{tikzpicture}
\caption{Left: Type-II string order parameter
after a local quantum quench. 
The symmetry is broken along the conformal interface (blue dashed line). The string operator (red solid line) is defined along $C=\{i\tau+x, \, -\infty<x<+\infty\}$.
A small disc of radius $\epsilon$ is removed at $z_0=i\tau$ to introduce a UV cutoff.
Right: The $z$-plane is mapped to a $w$-cylinder after the conformal mapping in \eqref{LocalQuench_conformalMap}, where we have deformed the string operator (red solid line) to a straight line.
}
\label{Fig_local_typeII}
\end{figure}
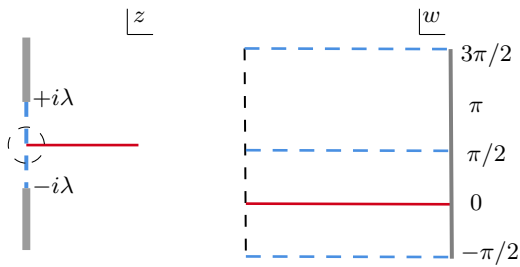

For the local quantum quench, we consider 
the same setup in Sec.\ref{Sec:Local}, but now the string operator  is only defined in the subsystem $[0,+\infty)$, as shown in Fig.\ref{Fig_local_typeII}. We consider the general case that there is a symmetry-breaking interface along $x=0$.

Let us first comment on the key difference between type-I and type-II string order parameter evolution after a local quantum quench, by looking at the path integral of the string order parameters.
For the configuration in Fig.\ref{Fig_local_typeII} (left), we consider the same conformal mapping in \eqref{LocalQuench_conformalMap}, after which the $z$-plane is mapped to a $w$-cylinder in Fig.\ref{Fig_local_typeII} (right). It is noted there is only one string operator inserted on the $w$-cylinder. This is in contrast to the 
type-I string operator evolution in Sec.\ref{Sec:Local}, where there are two string operators on the $w$-cylinder (See Fig.\ref{Global2_cylinder}). This key difference can be seen explicitly when the whole system preserves the global symmetry. For the same reason as illustrated near \eqref{Two_string} and \eqref{One_string}, the type-I string order parameter becomes independent of time if the symmetry is preserved over the whole system, while for type-II string order parameter the time evolution is nontrivial.

\medskip

To evaluate $\langle \psi(t)|\mathcal L_A|\psi(t)\rangle$, one first notices that the path integral in Fig.\ref{Fig_local_typeII} (right) is the same as that in Fig.\ref{GlobalQuench_typeII_Cylinder}, except that now the length of the $w$-cylinder in $\Re(w)$ direction is
$W(t)\simeq \log\frac{2(\lambda^2+t^2)}{\epsilon \lambda}$.
Then by repeating the procedures in Eqs.\eqref{Z_typeII_global_infinite}$\sim$\eqref{kappa_typeII_Global1}, one can find that in the limit $t\gg\lambda$ and $t\gg \epsilon$,
\be
\label{TypeII_local_logt}
\log |\langle \psi(t)|\mathcal L_A|\psi(t)\rangle |\simeq
-\kappa \log t,
\ee
up to a constant sub-leading term. Here the coefficient $\kappa=2(E^0_\text{defect}-E^0)$, where $E^0_\text{defect}$ and $E^0$ are the ground-state energies of $H_{\text{defect}}$ and $H_0$ as depicted in \eqref{Eq:H_defect2_typeII}.

\begin{figure}
\centering
\begin{tikzpicture}

    \node[inner sep=0pt] (russell) at (15pt,-45pt)
    {\includegraphics[width=.32\textwidth]{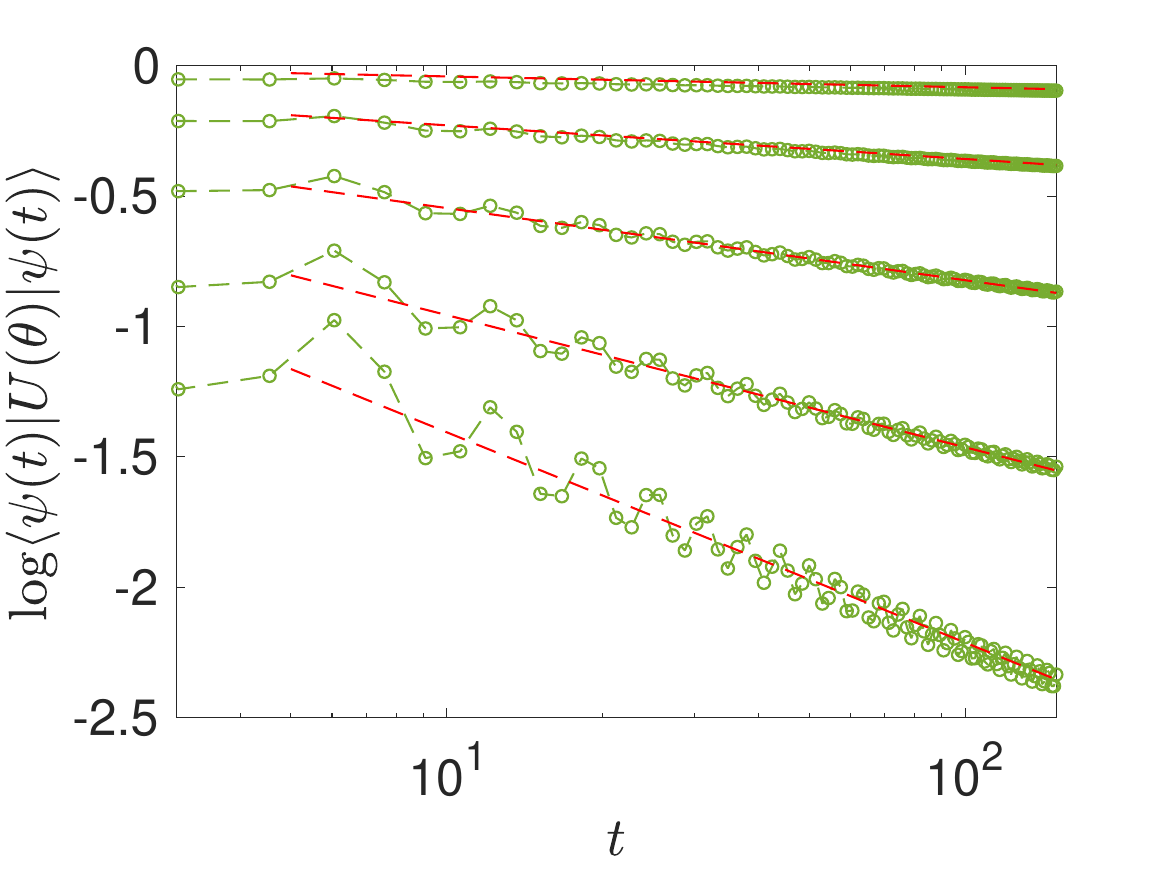}};

    \node[inner sep=0pt] (russell) at (15pt,-165pt)
    {\includegraphics[width=.32\textwidth]{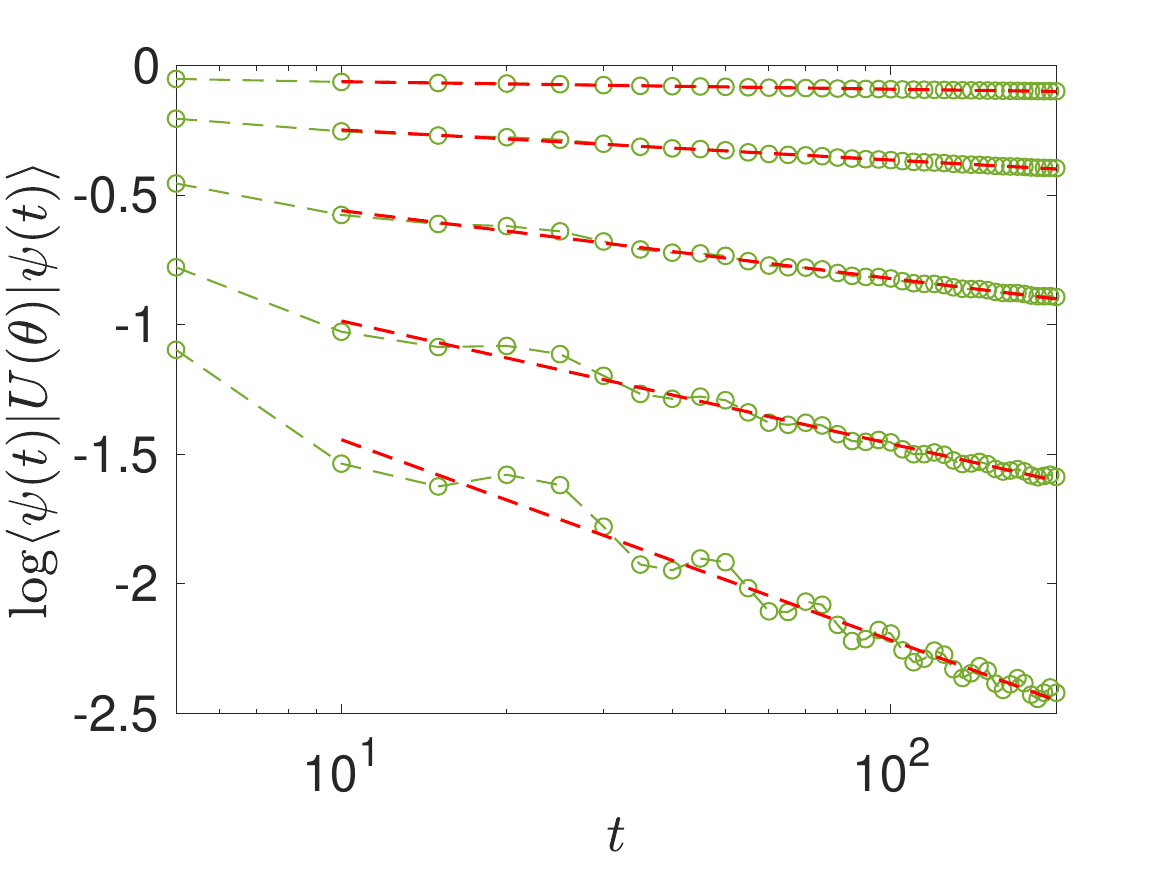}};

             \node at (-26pt, -72pt){(a)};
            \node at (-26pt, -192pt){(b)};

    \end{tikzpicture}
    \caption{Lattice calculation of type-II string order parameter in the subsystem $[0,L]$ after a local quantum quench, where the total system is $[-L,L]$. Here we choose $2L=600$.
    (a) The symmetry is preserved over the whole system.
    (b) We introduce a symmetry-breaking interface in the quenched Hamiltonian in \eqref{Local_defect_middle}, with $h=1/2$.
    From top to bottom in each plot, we take $\theta=\pi/6$, $2\pi/6$, $3\pi/6$, $4\pi/6$, and $5\pi/6$, respectively.
    The red dashed lines are guiding lines of the form $y=-a \log t+b$.}
    \label{Fig:NoSymBreak_Local}
\end{figure}

The time dependence in \eqref{TypeII_local_logt} is verified in a numerical calculation based on the free fermion lattice model, as shown in Fig.\ref{Fig:NoSymBreak_Local}. Here we perform the same numerical calculation as that in Sec.\ref{Sec:Local} but with the string operator defined within the subsystem $[0,L]$. From Fig.\ref{Fig:NoSymBreak_Local}, one can find that the type-II string order parameters have a scaling behavior in \eqref{TypeII_local_logt}, no matter the interface breaks the symmetry or not, as expected.

\section{Discussion and Conclusion}
\label{Sec:Discuss}

\subsection{Periodicity in type-I and type-II string order parameters}
\label{Period_kappa}

For the lattice fermion calculations considered in this work, we used the string operator $U(\theta)$ defined in \eqref{U_theta}.
One can find that both type-I and type-II string order parameters are periodic in $\theta$ in all the setups we have considered in this work.

\begin{figure}[t]
\centering
\begin{tikzpicture}

    \node[inner sep=0pt] (russell) at (15pt,-45pt)
    {\includegraphics[width=.36\textwidth]{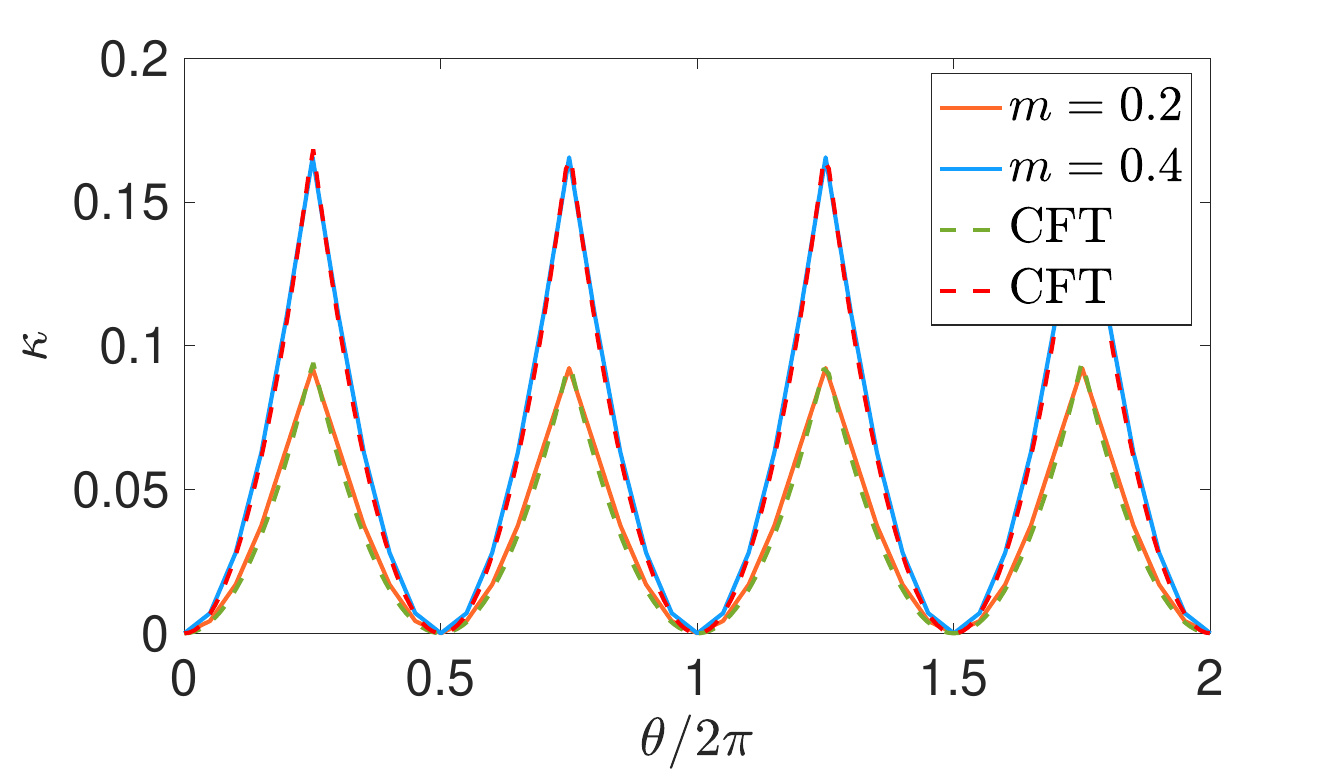}};

    \node[inner sep=0pt] (russell) at (15pt,-150pt)
    {\includegraphics[width=.36\textwidth]{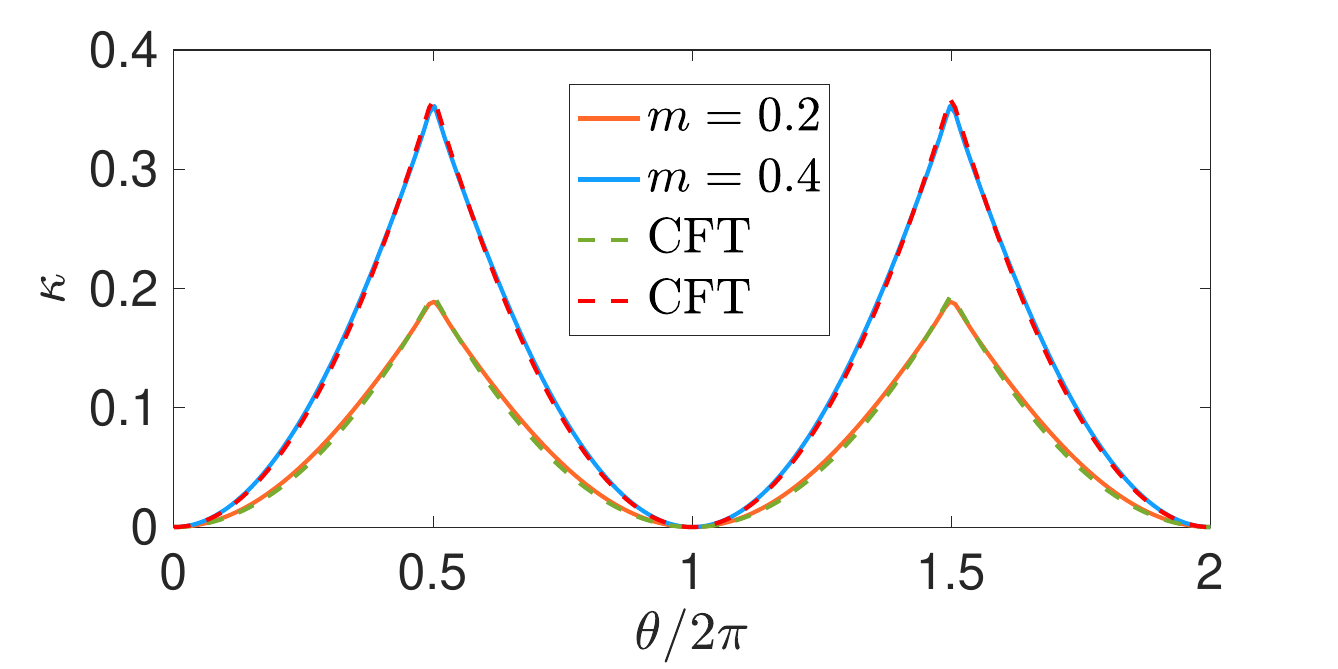}};
    
              
    \end{tikzpicture}
  \caption{The rates of exponential decay of type-I (top) and type-II (bottom) string order parameters after a global quench, which are extracted from the lattice results in Fig.\ref{Fig:SymBreak_Global_finite} and the top plots in  Fig.\ref{Fig:NoSymBreak_Global}, respectively. 
    The coefficients $\kappa$ are defined via \eqref{Eq:Global_semi} and \eqref{Decay_typeII_Global2}, respectively. 
    In each plot, we consider two different mass terms in the definition of initial states.
    The periods of $\kappa$ in $\theta$ are $\pi$
    for type-I string order parameter and $2\pi$ for type-II string order parameter.
    The CFT plots in the bottom plot are obtained by $\kappa=\frac{2\pi}{\beta}(E^0_\text{defect}-E^0)$ with $(E^0_\text{defect}-E^0)$ given in \eqref{CFT_Casimir}, and the CFT plots in the top are obtained from \eqref{kappa_typeI}, with $\beta$ as the only fitting parameter.
    }
    \label{Fig:Period}
\end{figure}

One interesting feature is that the periodicities in the type-I and type-II string order parameters are different from each other.
The period is $\pi$ for type-I string order parameters,
\footnote{Note that this $\pi$-period feature was already obtained in the ground state case in \cite{2023BSB}.}
while it is $2\pi$ for type-II string order parameters. A typical plot is shown in Fig.\ref{Fig:Period} for the case of global quantum quenches.

This difference in periodicity can be understood as follows. Let us consider $U(\theta)$ as defined in \eqref{U_theta}. At $\theta=\pi$, it can be rewritten as
\be
\label{U_Z2}
U_{\mathbb Z_2}=(-1)^{\sum_{i} c_i^\dag c_i},
\ee
which measures the fermion parity in the system (or subsystem).
For type-I string order parameters, $U(\theta)$ with $\theta=\pi$ measures the fermion parity of the total system.
In our setup, the boundary or interface breaks the $U(1)$ symmetry but still preserves the $\mathbb Z_2$ fermion parity. In other words, $U_{\mathbb Z_2}$ in \eqref{U_Z2} is still a good symmetry operator and it commutes with the Hamiltonian. Therefore, $\langle U(\theta)\rangle$ is independent of time at $\theta=\pi$.
This is why we have a $\pi$ period in type-I string order parameters in Fig.\ref{Fig:Period}.
On the other hand, for the type-II string order parameter which is defined in a subsystem $A$, it has a nontrivial time evolution at $\theta=\pi$ (see Fig.\ref{Fig:Period}) because the fermion parity in a subsystem is not conserved, although the fermion parity is conserved in the whole system.

\medskip

Finally, it is also interesting to give a remark on the amplitude of the decaying rate $\kappa$ in Fig.\ref{Fig:Period}.
Let us first consider the simpler case of type-II string order parameter which corresponds to Fig.\ref{Fig:Period} (bottom).
The decaying rate is $\kappa=\frac{2\pi }{\beta}(E^0_\text{defect}-E^0)$, where $E^0_\text{defect}$ and $E^0$ are the ground states of $H_{\text{defect}}$ and $H_0$, respectively. Here $H_{\text{defect}}$ and $H_0$ are defined along the circles in the $\Im(w)$ direction of the cylinder in Fig.\ref{GlobalQuench2_typeII_Cylinder}.
The length of the circle is $2\pi$, and there is a defect line inserted for $H_{\text{defect}}$, while there is no defect line for $H_0$. For $H_{\text{defect}}$, the effect of the topological defect line is to twist the boundary conditions in $\tau=\Im(w)$ direction as 
\be
\psi_{L,R}(x,\tau+2\pi)=(-1)\, e^{i\theta} \, \psi_{L,R}(x,\tau),
\ee
where it is reminded that $\psi_L$ and $\psi_R$
are the left and right moving Weyl fermions in \eqref{S_Dirac}.
Then, one can find that the ground-state energy depends on $\theta$ as follows:\cite{hori2003mirror}
\be
\label{CFT_Casimir}
E^0_\text{defect}-E^0=\left\{\left(\frac{\theta}{2\pi}-\frac{1}{2}\right)-\left[\frac{\theta}{2\pi}-\frac{1}{2}\right]-\frac{1}{2}
\right\}^2,
\ee
where $\theta\in \mathbb R$.
Here $[\alpha]$ denotes the greatest integer less than or equal to $\alpha$. 
Note the function in \eqref{CFT_Casimir} has a period of $2\pi$.
A comparison of this CFT result and the lattice result is shown in Fig.\ref{Fig:Period} (bottom), where one can find an excellent agreement.

For type-I string order parameters, the evaluation of the magnitude of $\kappa$ is more involved, since one needs to consider the symmetry-breaking boundary conditions with the string operator inserted, as seen in Fig.\ref{Fig:Global2}. A detailed calculation of $\kappa$ in the ground-state case has been performed in \cite{2023BSB}. By performing a similar calculation in our setup, one can find that 
\be
\label{kappa_typeI}
\kappa=\left(\frac{\theta}{2\pi}\right)^2\cdot \frac{2\pi}{\beta}, \quad (n-\frac{1}{2})\pi \le \theta\le (n+\frac{1}{2})\pi,\,\, n\in \mathbb Z. 
\ee
Note that the period in \eqref{kappa_typeI} is $\pi$, which is one half of that in \eqref{CFT_Casimir}, with the underlying physics analyzed previously. A comparison of the CFT result and the lattice result is shown in Fig.\ref{Fig:Period} (top), where again one can see an excellent agreement.

\subsection{Non-unitary time evolution}

The time evolution considered in this work is unitary. In fact, our approach can be straightforwardly generalized to the setup of non-unitary time evolutions.
Here we illustrate this generalization based on the setup in Ref.\cite{2024Wen}, where a complex spacetime metric is considered. 
See also the related setups in \cite{2024_Su,2024_Nozaki}.
In Ref.\cite{2024Wen}, it was found that the non-unitary time evolutions of entanglement Hamiltonian and entanglement entropy exhibit qualitatively different scaling behaviors from those in a unitary time evolution. A similar conclusion holds here in the time evolution of string order parameters. 

As an illustration, we give a concrete example on the type-I string order parameter evolution after a global quench.
The same procedure can be generalized to all the cases as discussed in Sec.\ref{Sec:BSB} and Sec.\ref{Sec:Partial_String} of this work.
More explicitly, we consider the following complex time evolution:
\be
\label{Psi_complex}
|\psi(t)\rangle=e^{-iH(1-i\epsilon)t}|\psi_0\rangle, \quad \epsilon>0.
\ee
This non-unitary time evolution corresponds to a complex spacetime metric as discussed in \cite{2024Wen}.
Here the initial state $|\psi_0\rangle$ is chosen as the regularized conformal boundary state, i.e., $|\psi_0\rangle=e^{-\frac{\beta}{4} H_{\text{CFT}}} |B\rangle$. The density matrix corresponding to \eqref{Psi_complex} is
\be
\label{rho_complex}
\rho(t)=e^{-i Ht} e^{-\left(\frac{\beta}{4}+\epsilon t\right) H}|B\rangle\langle B| e^{-\left(\frac{\beta}{4}+\epsilon t\right) H}
e^{i Ht}.
\ee
It is convenient to consider the Euclidean space with the following two-time density matrix
\be
\label{rho_tau1tau2}
\rho(\tau_1,\tau_2)=e^{-\left(\frac{\beta}{4}+\epsilon \tau_1+\tau_2\right) H}|B\rangle \langle B| e^{-\left(\frac{\beta}{4}+\epsilon \tau_1-\tau_2\right) H}.
\ee
Note that by considering the following analytical continuation in $\rho(\tau_1,\tau_2)$:
\be
\label{Analytical_continue}
\tau_1\to t, \quad \tau_2\to it,
\ee
one can obtain $\rho(t)$ in \eqref{rho_complex}.
Then the path integral of the string order parameter evolution $\text{Tr}[U_\theta\, \rho(\tau_1,\tau_2)]$ is shown as follows:
\be
\label{PI_rho_12}
\begin{tikzpicture}[x=0.75pt,y=0.75pt,yscale=-0.6,xscale=0.6]

\begin{scope}[xshift=50pt]
\draw (140pt,40pt)--(155pt,40pt);
\draw (140pt,40pt)--(140pt,25pt);
\node at (147pt,30pt){$z$};
\end{scope}

\draw [color={rgb, 255:red, 74; green, 144; blue, 226 }  ,draw opacity=1 ][line width=1.5]  [dash pattern={on 5.63pt off 4.5pt}]  (142,73) -- (142,191) ;
\draw  [draw opacity=0][dash pattern={on 4.5pt off 4.5pt}][line width=0.75]  (141.51,126.87) .. controls (133.59,126.48) and (127.34,120.46) .. (127.44,113.19) .. controls (127.53,105.73) and (134.26,99.76) .. (142.48,99.87) .. controls (150.69,99.97) and (157.27,106.11) .. (157.17,113.57) .. controls (157.08,120.84) and (150.68,126.7) .. (142.75,126.89) -- (142.3,113.38) -- cycle ; 
\draw  [dash pattern={on 4.5pt off 4.5pt}][line width=0.75]  (141.51,126.87) .. controls (133.59,126.48) and (127.34,120.46) .. (127.44,113.19) .. controls (127.53,105.73) and (134.26,99.76) .. (142.48,99.87) .. controls (150.69,99.97) and (157.27,106.11) .. (157.17,113.57) .. controls (157.08,120.84) and (150.68,126.7) .. (142.75,126.89) ;  
\draw [color={rgb, 255:red, 128; green, 128; blue, 128 }  ,draw opacity=1 ][line width=1.5]    (29,73) -- (255,73) ;
\draw [color={rgb, 255:red, 128; green, 128; blue, 128 }  ,draw opacity=1 ][line width=1.5]    (29,191) -- (255,191) ;
\draw [color={rgb, 255:red, 208; green, 2; blue, 27 }  ,draw opacity=1 ][line width=1.0]    (29.32,112.53) -- (255.32,112.53) ;

\node at (-12pt,55pt){$\beta/4+\epsilon\tau_1$};
\node at (-20pt,85pt){$\tau_2$};
\node at (-17pt,145pt){$-\beta/4-\epsilon\tau_1$};

\end{tikzpicture}
\ee
where the conformal boundaries are imposed along $\Im z=\beta/4+\epsilon\tau_1$ and $\Im z=-\beta/4-\epsilon\tau_1$ respectively.
The string operator (solid red line) is imposed along $\Im z=\tau_2$. A conformal interface (blue dashed line) which breaks the global symmetry is inserted along $\Re(z)=0$.

Next, by considering the following conformal mapping
\be
\label{ConformalMap_Global_complex}
w=f(z)=\log\frac{
\sinh[\pi(z-i\tau_2)/(\beta+4\epsilon\tau_1)]
}{
\cosh[\pi (z+i\tau_2)/(\beta+4\epsilon\tau_1)]
},
\ee
the configuration in \eqref{PI_rho_12} can be mapped to a $w$-cylinder in Fig.\ref{Global2_cylinder}, except that now the length of this cylinder, after the analytical continuation in \eqref{Analytical_continue}, depends on time $t$ as follows:
\be
\small
\label{W_complex}
W(t)=\frac{1}{2} \log\left(\frac{ \cosh[2\pi \epsilon_0/(\beta+4\epsilon t)]+\cosh[ 4\pi t/(\beta+4\epsilon t)]}{2\sinh^2[\pi\epsilon_0/(\beta+4\epsilon t)]}\right).
\ee
If the time evolution is unitary, i.e., $\epsilon=0$, then one has $W(t)\propto \frac{2\pi}{\beta}t$ in the long time limit. Here for $\epsilon>0$, one has $W(t)\propto \log t$.
Therefore, different from the exponential decay in \eqref{String_Global2}, one can find that in the long time limit the type-I string order parameter depends on time in a power-law as:
\be
\log |\langle \psi(t)|\mathcal L|\psi(t)\rangle|\propto
-\log t.
\ee

As a remark, note that there are many other setups of non-unitary time evolutions, such as the ones governed by Lindblad equations. These setups are more relevant when we consider an open system that is coupled to environment. It will be more interesting to study the time evolution of string order parameters in such setups.

\subsection{Conclusion and Outlook}

To conclude, in this work we have investigated the universal time evolution of two different types of string order parameters in (1+1) dimensional quantum critical systems in the presence of boundary/interface symmetry-breaking after quantum quenches.
Besides the interesting scaling behaviors of the string order parameters themselves, our study also provides a method to detect the boundary/interface symmetry breaking in the non-equilibrium dynamics. For the type-I string order parameter that is defined over the whole system with a boundary/interface symmetry breaking, it decays exponentially in time after a global quench, and decays as a power-law in time after a local quench. We also study the type-II string order parameter that is defined in a subsystem (rather than over the whole system). This type of string order parameter is related to the full counting statistics which measures the charge fluctuations in a subsystem. In contrast to the type-I string order parameter, no matter there is a boundary/interface symmetry-breaking or not, the type-II string order parameter always has a nontrivial and universal scaling behavior in time. The underlying physics is that  although the symmetry is preserved in the whole system, there is still a charge fluctuation in the subsystem.
Our conformal field theory results are confirmed by a free-fermion lattice calculation.

\medskip
In the following, let us comment on several possible future problems:


\medskip
-- It is desirable to generalize our study to higher dimensional CFTs, where the boundary or defect could break the global symmetry. As recently studied in \cite{2022_higherD_CFT,2024Herzog}, in higher dimensional CFTs,
there is a large class of boundaries or interfaces that break the global symmetry, with concrete examples on free-field theories fully worked out. Such boundaries or interfaces can be defined in sub-manifolds with different dimensions, e.g., on a line or on a surface. Another bonus in higher dimensions is the higher-form symmetry, which corresponds to topological defects of higher codimensions \cite{2015_Kapustin}. Recall that the string operator in this work is constructed based on the symmetry operator in an ordinary global symmetry which is an invertible 0-form symmetry. 
In general, one can construct the non-local operators based on the symmetry operator of higher form symmetries and break the symmetry in certain sub-manifolds.
It will be interesting to investigate the scaling behavior of non-local order parameters in such setups.

\medskip
-- Another future problem is to consider the string order parameter in a non-unitary CFT. Recently, for the non-Hermitian version of the Su-Schrieffer-Heeger model at the critically, which breaks Hermiticity but preserves a global $U(1)$ symmetry, it was found the critical point is described by the $bc$-ghost CFT with central charge $c=-2$ \cite{NonUnitaryCFT_2020,2023Calabrese},
where the global $U(1)$ symmetry of a $bc$-ghost CFT is associated to the ghost number. It will be interesting to study the feature of string order parameters when the boundary/interface of such non-unitary CFT breaks the $U(1)$ symmetry explicitly. An even more interesting example is the complex CFT that was realized in a (1+1)-dimensional quantum 5-state Potts model recently \cite{2024_Complex_CFT}. Both the central charge and the conformal dimensions of primary operators become complex in this case. It would be interesting to study the behavior of string order parameters in such exotic quantum critical systems.

\medskip
-- It is also interesting to consider the effect of boundary symmetry breaking in a driven system.
In the context of $(1+1)d$ conformal field theories, there are setups where the non-equilibrium dynamics is exactly solvable when the driven Hamiltonians are constructed by deforming the Hamiltonian density in space and time. To mention only a few in the early development in this direction, see, e.g., \cite{2018_WenWu,2020Fan_peak,2020PRR_Lapierre,2020_QuasiPeriodic,2020_Han_Wen,2021part1,2022Spart2,2021_general_Lapierre, Fan_2021}. During such time dependent drivings, depending on the driving parameters, different dynamical phases can emerge during the driving. Physical quantities such as the entanglement entropy and energy density have different universal features in different phases. In the presence of the symmetry-breaking boundary or interface in the driving Hamiltonians, we expect the string order parameters (including both type-I and type-II string operators) may exhibit different universal features in different dynamical phases.
We also hope to mention another setup on the emergent quantum criticality during a periodic driving \cite{2017_Yao,2018_Berdanier,2018_Mitra,2021_You,2024_Zhou_Yu}, where one considers gapped driving Hamiltonians, in contrast to the driven CFTs introduced above, where one considers gapless driving Hamiltonians. It would be interesting to study the time evolution of string order parameters in this case.

-- For the type-II string operator in (1+1) dimensional CFTs, it is also interesting to consider the case where the string operator is defined in disjoint intervals. This problem is difficult for a general CFT, but we expect that it may be studied analytically in a free field theory, by using the approach in \cite{2009Casini}.

-- Besides the string order parameter as studied in this work, we expect there should be other quantities associated to the global symmetries of the system that can be used to detect the boundary or interface symmetry breaking. For example, it was recently proposed in \cite{2023_En_asymmetry} that the so-called entanglement asymmetry can be used to probe symmetry breaking. Some other related quantities that may be used to study the setup in this work include the symmetry-resolved entanglement\cite{2018_Goldstein,2019_Calabrese, 2023_Ooguri} and charged entanglement entropy\cite{2016Matsuura}.

-- Finally, we also noted several recent studies on the interplay between general topological defect lines and the conformal boundary conditions in two dimensional CFTs \cite{2020_Konechny,2023_Huang_Cheng,2024_KV,2024_Choi_A,2024_Choi_long,2024_Das,2024_Quella}. It will be interesting to find the possible connections between our work and these works.

\acknowledgments

This work is supported by a startup at Georgia Institute of Technology (RB, QT, XW).
WZ is supported by the National Key Research and Development Program of China Grant No. 2022YFA1402204 and the National Natural Science Foundation of China Grant No. 12274086.

\appendix

\section{Quantum Ising on a lattice: String operator from non-invertible symmetry}
\label{Appendix:Ising}
As discussed in Sec.\ref{Sec:Sym_break_boundary}, all the three boundary conditions in Ising CFT break the non-invertible symmetry that implements the Kramers-Wannier duality. In this appendix, we give a concrete study of the type-I string order parameter in the ground state of the lattice model.

Let us we consider the transverse-field Ising chain with boundary terms as
\be\label{eq:TFI_ham_mix_bc}
H_{\rm Ising} = - \sum_{j=1}^{N} ( \sigma_j^x \sigma_{j+1}^x + h \sigma_j^z ) + h_b \sigma_N^x \sigma_1^x - h_1 \sigma_1^x .
\ee
When $h_1 = 0$ and $h_b = 0$, the spin chain is under a periodic boundary condition and is self-dual at the critical point $h=1$ under the standard Kramers-Wannier duality transformation
\be
\mu_i^x = \prod_{j \le i} \sigma^z_j , 
\quad
\mu_i^z = \sigma^x_i \sigma^x_{i+1} . 
\ee
Away from the critical point, the Kramers-Wannier duality establishes a mapping between $Z_2$-symmetric ($0 \le h < 1$) and symmetry-breaking ($h > 1$) phases. 
This mapping is 2-to-1 so that it is non-invertible, and its corresponding symmetry operator on lattice was recently introduced~\cite{2024_Seiberg_Shao} in terms of following string operator
\be
\begin{aligned}
\mathrm{D} & = e^{-\frac{2\pi i N}{8}} 
U_{KW}
\frac{1 + \eta}{2} , \\
U_{KW} & = \prod_{j=1}^{N} \left( \frac{1 + i \sigma^z_j}{\sqrt{2}} \frac{1 + i \sigma^x_j \sigma^x_{j+1}}{\sqrt{2}} \right) \left( \frac{1 + i \sigma^z_N}{\sqrt{2}} \right) , 
\end{aligned}
\ee
where $\eta = \prod_{j=1}^{N} \sigma^z_j$ is the parity symmetry operator so that $\frac{1 + \eta}{2}$ is a projector to the parity even sector. 
It is easy to check that the non-invertible string operator $\mathrm{D}$ is translation invariant and commutes with the self-dual Hamiltonian, i.e., the transverse-field Ising chain at critical point with periodic boundary conditions. 
In fact, its square is 
\be
\mathrm{D}^2 = \frac{1 + \eta}{2} \mathbf{T}_{\rm Ising} ,
\ee
where $\mathbf{T}_{\rm Ising}$ is the translation symmetry operator of Ising chain.

Here we are interested in cases of breaking the non-invertible symmetry of Kramers-Wannier duality at critical point $h=1$, by switching the boundary condition of Ising chain to be non-periodic, i.e., imposing nonzero $h_b$ and $h_1$. 
In particular, we consider two boundary conditions as the follows
\begin{enumerate}
\item Both ends of the chain are free to have an open boundary condition, i.e. let $h_b = 1$ and $h_1 = 0$.
\item Adding a boundary field on one end to explicitly break the $Z_2$ parity symmetry and let another end to be free, i.e. let $h_b = 1$ and $h_1$ take nonzero values.
\end{enumerate}
In the first case, the symmetry of Kramers-Wannier duality is broken but the $Z_2$ parity symmetry is preserved, so that we expect a nontrivial scaling behavior for $\langle G | \mathrm{D} | G \rangle$ but a constant for $\langle G | \eta | G \rangle$.
In the second case, both of Kramers-Wannier duality and $Z_2$ parity symmetry are broken by the boundary condition, therefore we expect both $\langle G | \mathrm{D} | G \rangle$ and $\langle G | \eta | G \rangle$ exhibit a scaling on the system size. 

To test these expectations, we perform a numerical calculation by using the density matrix renormalization group algorithm~\cite{white1992dmrg} within a matrix product states representation~\cite{Schollw_ck_2011}. 
The total system size is simulated upto $L = 400$ with a bond dimension $\chi = 200$. 
As both operators $\eta$ and $U_{KW}$ are products of local operators, the evaluation of the parity symmetry operator $\eta$ and the non-invertible string operator $\mathrm{D}$ can be evaluated efficiently in the matrix product states representation.

\begin{figure}\centering
	\includegraphics[width=\columnwidth]{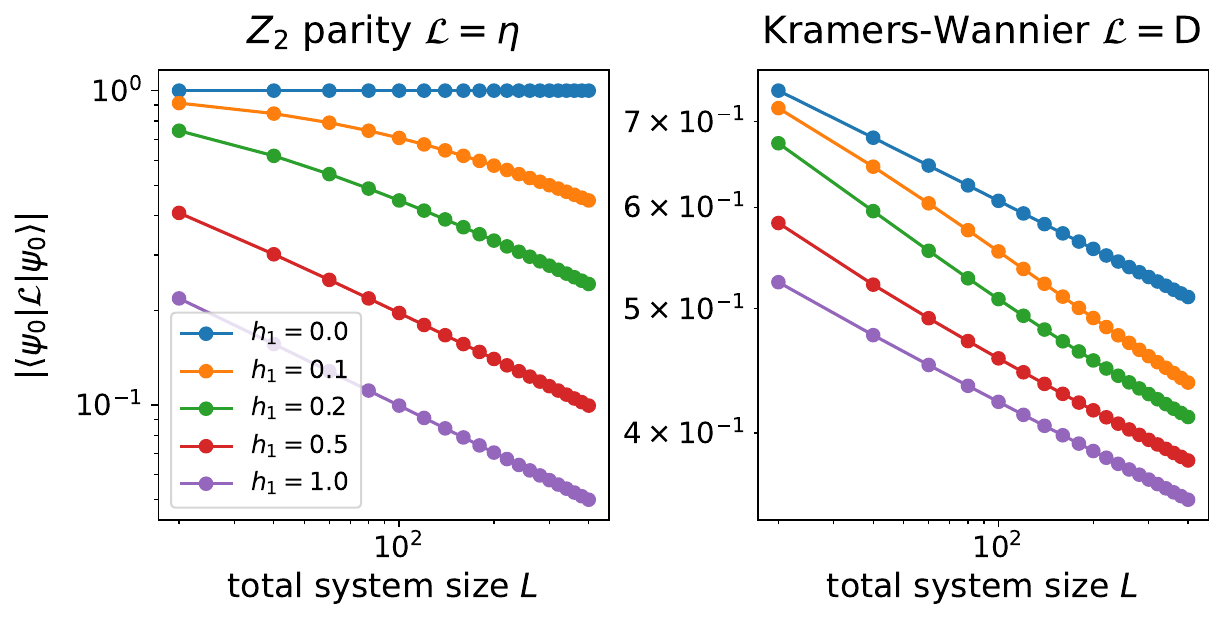}
	\caption{
		\label{fig:TFI_boundary_typeI}
		The expectation value of (left) the $Z_2$ parity symmetry operator $\eta$ and (right) the non-invertible Kramers-Wannier symmetry operator $\mathrm{D}$ for the ground state $|\psi_0 \rangle$ of the transverse-field Ising chain in Eq.~\eqref{eq:TFI_ham_mix_bc}. Here we consider an open boundary condition ($h_b = 1$) with adding a longitudinal field $- h_1 \sigma^x_1$ on one end of the chain to break the $Z_2$ parity symmetry. The total system size is calculated upto $L = 400$. 
	}
\end{figure}

As shown in Fig.~\ref{fig:TFI_boundary_typeI}, if the symmetry is explicitly broken by imposing appreciated boundary condition, both the ordinary invertible symmetry operator $\eta$ and the non-invertible $\mathrm{D}$ exhibit a power-law dependence $|\langle G| \mathcal{L} | G\rangle| \sim L^{-k}$ as predicted by our CFT derivation.
See our discussion on a general CFT in Appendix \ref{Appendix:Ground}.
If the symmetry is not broken, e.g., the free boundary condition $h_b = 1, h_1 = 0$ which preserves $Z_2$ parity symmetry, the expectation value of $\eta$ remains constant and has no scaling with the total system size. 
We have also checked that for periodic boundary conditions ($h_1 = 0, h_b = 0$) the expectation value $\langle G | \mathrm{D} | G \rangle$ is a constant for various total system sizes, as a result of preserving the non-invertible symmetry of the Kramers-Wannier duality.

\section{Ground state}
\label{Appendix:Ground}

In this appendix, we study the scaling behaviors of both type-I and type-II string order parameters in the ground state of a CFT for various configurations.

\subsection{Type-I string order parameter}

We first consider the scaling behavior of type-I string order parameter in the ground state of a CFT where the global symmetry is explicitly broken at the boundary/interface. The boundary symmetry-breaking case has been recently studied in detail in Ref.\cite{2023BSB}. Here we consider the general case where the symmetry is broken at an interface in the middle of the system. Note that by folding the system along this interface, one can obtain the case of symmetry-breaking at the boundary, with a doubled central charge in the bulk. Later in this appendix, we also consider the scaling behavior of type-I string order parameter in the ground state of an off-critical system.

\subsubsection{Boundary symmetry breaking}

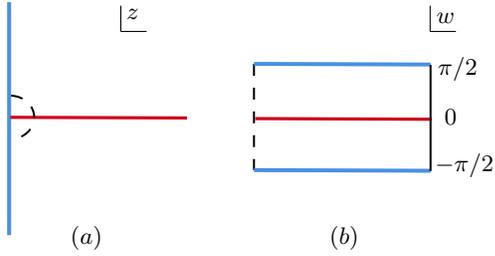
\begin{figure}
\centering
\begin{tikzpicture}[x=0.75pt,y=0.75pt,yscale=-0.65,xscale=0.65]

\draw (140pt,40pt)--(155pt,40pt);
\draw (140pt,40pt)--(140pt,25pt);
\node at (147pt,30pt){$z$};

\begin{scope}[xshift=180pt]
\draw (140pt,40pt)--(155pt,40pt);
\draw (140pt,40pt)--(140pt,25pt);
\node at (149pt,32pt){$w$};
\end{scope}

\node at (336pt,62pt){$\pi/2$};
\node at (332pt,90pt){$0$};
\node at (340pt,118pt){$-\pi/2$};

\node at (120pt,160pt){$(a)$};
\node at (270pt,160pt){$(b)$};

\draw [color={rgb, 255:red, 74; green, 144; blue, 226 }  ,draw opacity=1 ][line width=1.5]    (100,30.5) -- (100,211.5) ;
\draw [color={rgb, 255:red, 208; green, 2; blue, 27 }  ,draw opacity=1 ][line width=1.2]    (101,120) -- (238,120.5) ;
\draw  [draw opacity=0][dash pattern={on 4.5pt off 4.5pt}][line width=0.75]  (101.53,103.21) .. controls (111.59,103.5) and (119.59,111.06) .. (119.48,120.24) .. controls (119.36,129.28) and (111.41,136.55) .. (101.55,136.79) -- (101,120) -- cycle ; 
\draw  [dash pattern={on 4.5pt off 4.5pt}][line width=0.75]  (101.53,103.21) .. controls (111.59,103.5) and (119.59,111.06) .. (119.48,120.24) .. controls (119.36,129.28) and (111.41,136.55) .. (101.55,136.79) ;  
\draw [color={rgb, 255:red, 208; green, 2; blue, 27 }  ,draw opacity=1 ][line width=1.2]    (291,121) -- (428,121.5) ;
\draw [color={rgb, 255:red, 0; green, 0; blue, 0 }  ,draw opacity=1 ][line width=0.75]  [dash pattern={on 4.5pt off 4.5pt}]  (290,78.5) -- (290,161) ;
\draw [color={rgb, 255:red, 74; green, 144; blue, 226 }  ,draw opacity=1 ][line width=1.5]    (290,161) -- (427,161.5) ;
\draw [color={rgb, 255:red, 74; green, 144; blue, 226 }  ,draw opacity=1 ][line width=1.5]    (290,78.5) -- (427,79) ;
\draw [color={rgb, 255:red, 0; green, 0; blue, 0 }  ,draw opacity=1 ][line width=0.75]    (427,79.25) -- (427,161.75) ;
\end{tikzpicture}
\caption{(a) Type-I string order parameter $\langle G|\mathcal L|G\rangle$ in the ground state of a CFT of a finite length $L$. The global symmetry is broken along the left boundary (solid blue line). 
The red solid line corresponds to the string operator which is inserted over the whole system.
(b)The $w$-strip which is obtained from the $z$ half-plane in (a) based on a conformal mapping in \eqref{ConformalMap_ground}.
}
\label{Ground_bdry}
\end{figure}

We consider a CFT of a finite length $L$, with the global symmetry broken only at one boundary, say the left boundary. The ground state $|G\rangle$ of this finite system is represented by the Euclidean path integral that runs from $\tau=-\infty$ to $\tau=0$.

The type-I string order parameter $\langle G|\mathcal L|G\rangle$ corresponds to the configuration in Fig.\ref{Ground_bdry} (a). To calculate $\langle G|\mathcal L|G\rangle$, we first introduce a UV cutoff around $z=0$, which is realized by removing a half-disk of radius $\epsilon$ around $z=0$. We impose a conformal boundary condition $|b_1\rangle$ along the boundary of this half-disk.  Then we consider a conformal mapping
\be
\label{ConformalMap_ground}
w=\log z,
\ee
which maps the configuration in $z$-plane to the $w$-strip in Fig.\ref{Ground_bdry} (b). The physical boundary which breaks the global symmetry is mapped to the two edges of this $w$-strip of length $W=\log(L/\epsilon)$. It is noted that one can move the string operator freely in the system, but it cannot be absorbed by the boundary (solid blue line) which breaks the symmetry explicitly.

To evaluate the string order parameter, we consider the $\Re(w)$ direction as the Euclidean time, and the defect Hamiltonian $H_{\text{defect}}$ is defined along $\Im(w)$. Then we have
\be
\langle G|\mathcal L|G\rangle
=\frac{Z_{\text{defect}}}{Z_0}
=
\frac{\langle b_1| e^{-H_{\text{defect}}W} |b_2\rangle}{ \langle b_1| e^{-H_{0}W} |b_2\rangle},
\label{eq:U_ground_1}
\ee
where $|b_2\rangle$ is the symmetry-perserving conformal boundary condition imposed at the right end of the system.
Here $H_0$ denotes the CFT Hamiltonian without inserting a defect. 
Note that the length of the $w$-cylinder is $W=\log(L/\epsilon)\gg 1$. Then $\langle G|\mathcal L|G\rangle$ in \eqref{eq:U_ground_1} is dominated by the ground energies $E^0_{\text{defect}}$ and $E^0$,
as well as ground states $|G_d\rangle$ and $|G_0\rangle$ of the Hamiltonians $H_{\text{defect}}$ and $H_0$ respectively, as follows
\be
\langle G|\mathcal L|G\rangle \simeq \frac{\langle b_1|G_d\rangle\langle G_d|b_2\rangle}{\langle b_1|G_0\rangle\langle G_0|b_2\rangle} e^{-(E^0_{\text{defect}}-E^0) \log \frac{L}{\epsilon}}.
\ee
based on which we have
\be
\log |\langle G|\mathcal L|G\rangle | \propto \log \frac{L}{\epsilon}.
\ee

\subsubsection{Interface symmetry breaking}

\begin{figure}
\centering
\begin{tikzpicture}[x=0.75pt,y=0.75pt,yscale=-0.65,xscale=0.65]

\begin{scope}[yshift=-20pt]
\draw (140pt,40pt)--(155pt,40pt);
\draw (140pt,40pt)--(140pt,25pt);
\node at (147pt,30pt){$z$};
\end{scope}

\begin{scope}[xshift=180pt,yshift=-20pt]
\draw (140pt,40pt)--(155pt,40pt);
\draw (140pt,40pt)--(140pt,25pt);
\node at (149pt,32pt){$w$};
\end{scope}

\node at (362pt,32pt){$3\pi/2$};
\node at (356pt,62pt){$\pi$};
\node at (362pt,90pt){$\pi/2$};
\node at (356pt,118pt){$0$};
\node at (364pt,147pt){$-\pi/2$};

\node at (110pt,175pt){$(a)$};
\node at (290pt,175pt){$(b)$};

\draw [color={rgb, 255:red, 208; green, 2; blue, 27 }  ,draw opacity=1 ][line width=1.2]    (301,160) -- (460.83,160.5) ;
\draw [color={rgb, 255:red, 0; green, 0; blue, 0 }  ,draw opacity=1 ][line width=0.75]  [dash pattern={on 4.5pt off 4.5pt}]  (300,39.5) -- (301,201) ;
\draw [color={rgb, 255:red, 74; green, 144; blue, 226 }  ,draw opacity=1 ][line width=1.5]  [dash pattern={on 5.63pt off 4.5pt}]  (301,201) -- (460.83,201.5) ;
\draw [color={rgb, 255:red, 74; green, 144; blue, 226 }  ,draw opacity=1 ][line width=1.5]  [dash pattern={on 5.63pt off 4.5pt}]  (301,118.5) -- (460.83,119) ;
\draw [color={rgb, 255:red, 0; green, 0; blue, 0 }  ,draw opacity=1 ][line width=0.75]    (459.83,40) -- (460.83,201.75) ;
\draw [color={rgb, 255:red, 74; green, 144; blue, 226 }  ,draw opacity=1 ][line width=1.5]  [dash pattern={on 5.63pt off 4.5pt}]  (142,31) -- (142,212) ;
\draw [color={rgb, 255:red, 208; green, 2; blue, 27 }  ,draw opacity=1 ][line width=1.2]    (53,120.5) -- (243,120.5) ;
\draw  [draw opacity=0][dash pattern={on 4.5pt off 4.5pt}][line width=0.75]  (142.01,137.27) .. controls (132.17,136.77) and (124.41,129.3) .. (124.52,120.26) .. controls (124.64,110.99) and (133.01,103.58) .. (143.21,103.71) .. controls (153.42,103.84) and (161.59,111.46) .. (161.48,120.74) .. controls (161.36,129.78) and (153.41,137.05) .. (143.55,137.29) -- (143,120.5) -- cycle ; 
\draw  [dash pattern={on 4.5pt off 4.5pt}][line width=0.75]  (142.01,137.27) .. controls (132.17,136.77) and (124.41,129.3) .. (124.52,120.26) .. controls (124.64,110.99) and (133.01,103.58) .. (143.21,103.71) .. controls (153.42,103.84) and (161.59,111.46) .. (161.48,120.74) .. controls (161.36,129.78) and (153.41,137.05) .. (143.55,137.29) ;  
\draw [color={rgb, 255:red, 208; green, 2; blue, 27 }  ,draw opacity=1 ][line width=1.2]    (300,79) -- (459.83,79.5) ;
\draw [color={rgb, 255:red, 74; green, 144; blue, 226 }  ,draw opacity=1 ][line width=1.5]  [dash pattern={on 5.63pt off 4.5pt}]  (300,39.5) -- (459.83,40) ;
\end{tikzpicture}
\caption{(a) Type-I string order parameter $\langle G|\mathcal L|G\rangle$ in the ground state $|G\rangle$ of a CFT with a conformal interface inserted in the middle of the system, where the global symmetry is broken only along this interface. The red solid line corresponds to the string order parameter. (b) The $w$-cylinder which is obtained from the $z$-plane in (a) based on a conformal mapping in \eqref{ConformalMap_ground}.
}
\label{Ground_interface}
\end{figure}
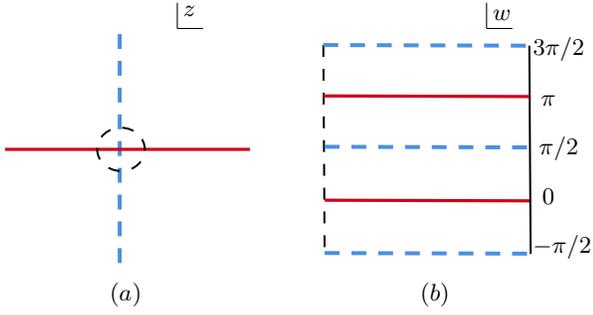

Now let us consider the case where an interface operator is inserted in the middle of the system, and the global symmetry is broken along this interface (See Fig.\ref{Ground_interface}).
The system is of length $L$ defined over $[-L/2, \,L/2]$, with the interface at $x=0$.

In the specific case when the interface corresponds to a conformal boundary, i.e., the two halves of CFTs defined on $[-L/2,0]$ and $[0, L/2]$ are decoupled from each other, the discussion is the same as in the previous part. 
For a general conformal interface inserted in the middle of the system, we can use the folding-trick to map the conformal interface problem to the conformal boundary problem. Here, let us consider the conformal interface case directly without using folding trick.

As shown in Fig.\ref{Ground_interface} (a), we introduce a UV cutoff by removing a small disk of radius $\epsilon$ at $x=0$.
Then we can use the same conformal mapping in \eqref{ConformalMap_ground} to map the $z$-plane to the $w$-cylinder in Fig.\ref{Ground_interface} (b).
Different from Fig.\ref{Ground_bdry} where we have a $w$-strip, now we have a cylinder. The string operator is mapped to the solid lines along $\Im( w)=0$ and $\pi$, and the conformal interface is mapped to $\Im (w)=\pi/2$ and $3\pi/2$, respectively.

\begin{figure}
\centering
{\includegraphics[width=.32\textwidth]{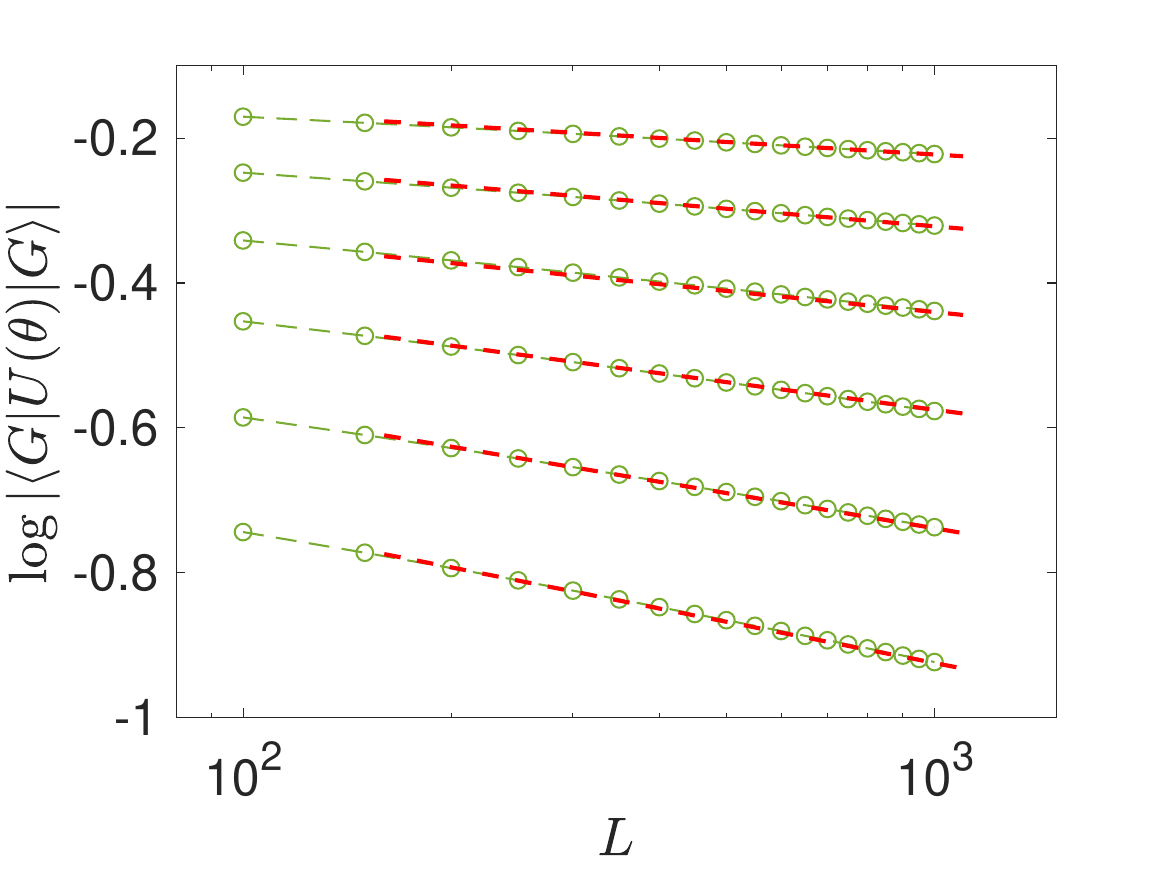}}
    \caption{
    Lattice calculation of type-I string order parameter 
    in the ground state of \eqref{Global_defect_middle}, where we choose $h=0.6$ in the symmetry breaking term in the middle of the chain, and the length of the total system is $L$.
    From top to bottom, we take  $\theta=0.5$, $0.6$, $0.7$, $0.8$, $0.9$, and $1$.
    The red dashed lines are fitting lines of the form $y=-a \log L+b$.}
    \label{Fig:Ground_PBC} 
\end{figure}

Then the expectation value of the type-I string order parameter has the same expression as \eqref{eq:U_ground_1}, except that now the Hamiltonian is defined along $\Im (w)\in [0, 2\pi]$ and the concrete form of $H_\text{defect}$ is different. With a similar procedure in the previous subsection, one can finally arrive at
\be
\label{Ground_interface_typeI}
\log | \langle G|\mathcal L|G\rangle | \propto -\log \frac{L}{\epsilon}.
\ee
To verify this scaling behavior, we consider the lattice model with the Hamiltonian in \eqref{Global_defect_middle}. The results of the type-I string order parameter $\langle G|U(\theta)|G\rangle$ are  shown in Fig.\ref{Fig:Ground_PBC}, where one can find the scaling behavior that is in agreement with \eqref{Ground_interface_typeI}.

\subsubsection{Ground state of off-critical systems}
\label{Appendix:off-critical}

When the system is off the critical point, a new length scale, i.e., the correlation length $\xi$, will be introduced. In this case, it is expected that the type-I string order parameter will saturate when the total length $L$ of the system is much larger than $\xi$.

More concretely, we consider the following gapped Hamiltonian
\be
\label{Mass_H}
\small
H=-\frac{1}{2}\sum_{1\le j\le L} c_j^\dag c_{j+1}+
h c_{L/2}^\dag c_{L/2+1}^\dag +h.c.+m \sum_j (-1)^j c_j^\dag c_j, 
\ee
with periodical boundary conditions. Here $m>0$ controls the correlation length. As $m$ decreases, the correlation length $\xi$ will increase accordingly. In particular, as $m$ approaches to zero, one has $\xi\propto 1/m$. The $U(1)$ symmetry in \eqref{Mass_H} is broken by introducing a defect in the middle of the chain. Then we study how the type-I string order parameter scales with the total length of the system.

\begin{figure}[t]
\centering
\begin{tikzpicture}

    \node[inner sep=0pt] (russell) at (15pt,-85pt)
    {\includegraphics[width=.26\textwidth]{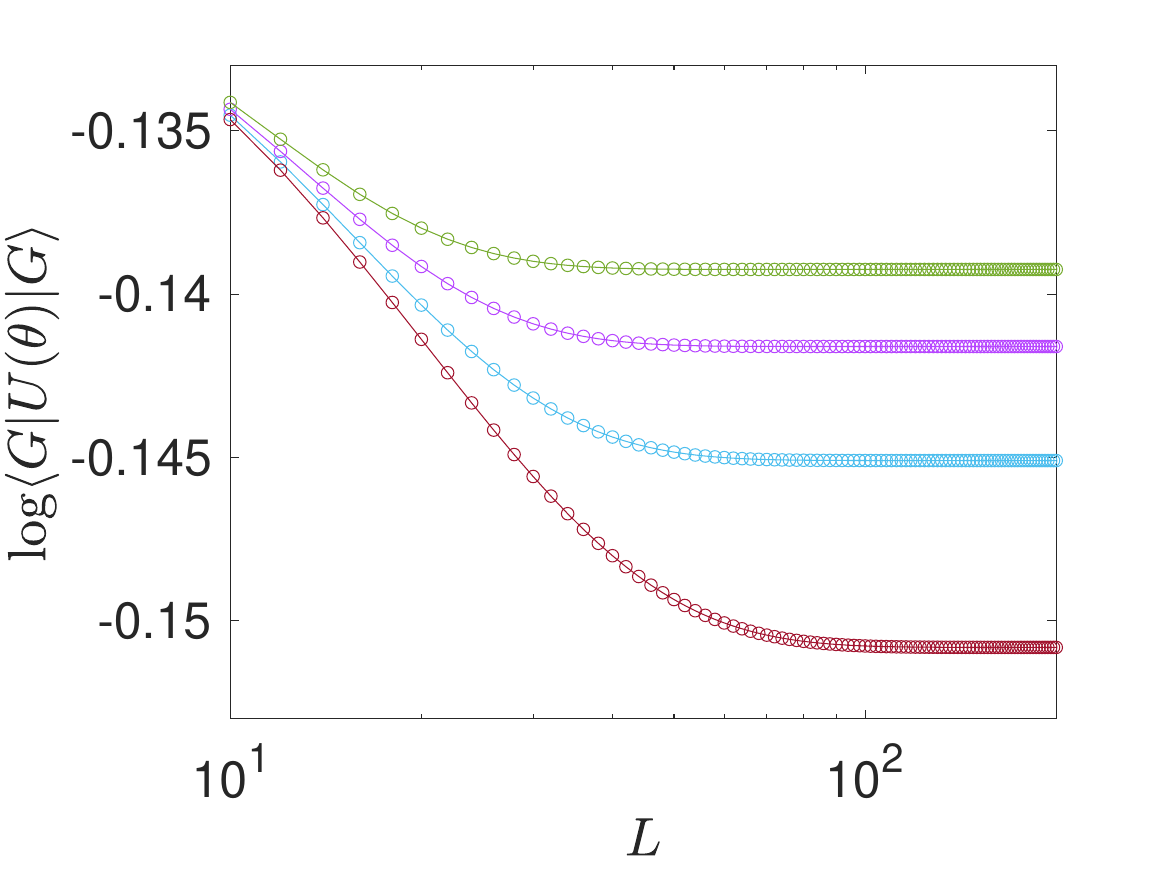}};
       
    \node[inner sep=0pt] (russell) at (140pt,-85pt)
    {\includegraphics[width=.26\textwidth]{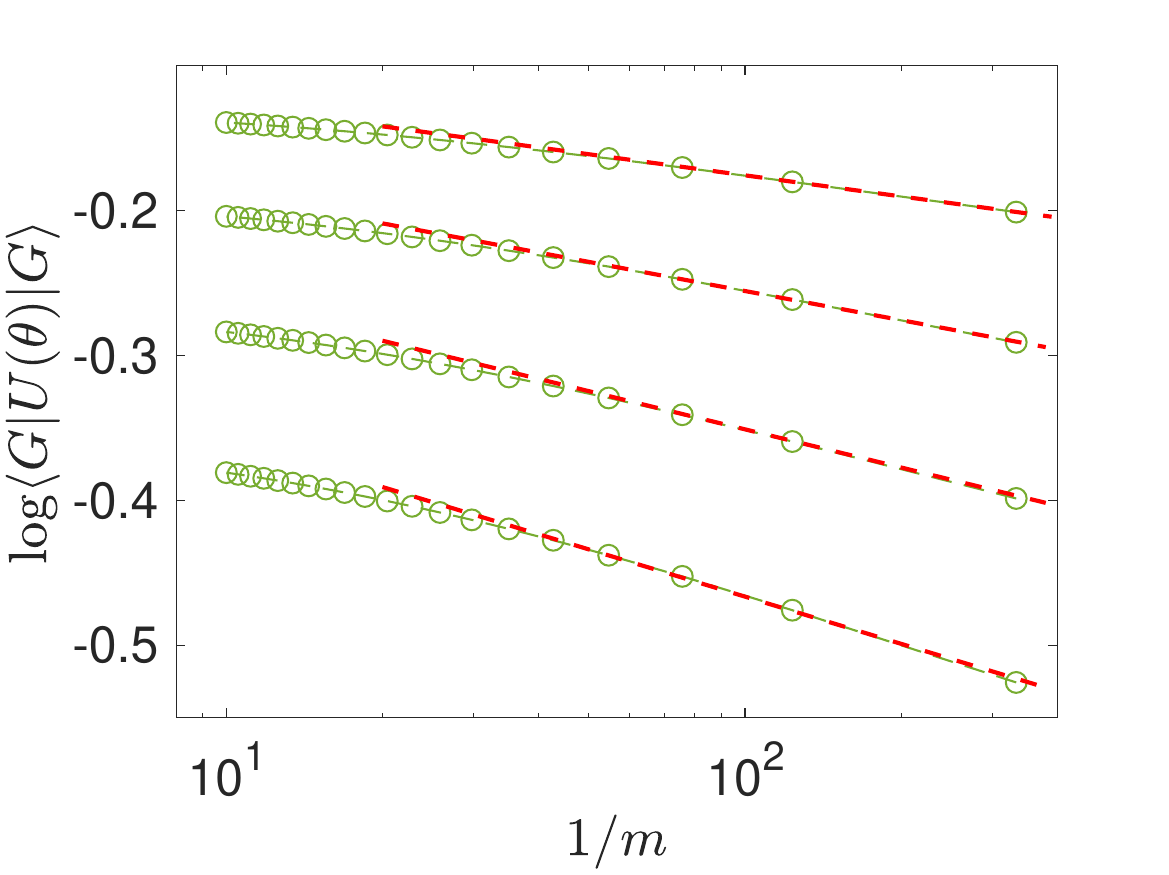}};

             \node at (-15pt, -105pt){(a)};
            \node at (110pt, -105pt){(b)};

    \end{tikzpicture}
    \caption{
    Lattice calculation of type-I string order parameter 
    in the ground state of the gapped Hamiltonian in \eqref{Mass_H} with an interface symmetry-breaking. Here we choose $h=0.6$ in the symmetry breaking term in the middle of the chain.
    (a) From top to bottom, we have $m=0.1$, $0.08$, $0.06$, and $0.04$, while fixing $\theta=0.5$. 
    (b) The type-I string order parameter as a function of $1/m$. Here we fix the total length to be $L=800$.
    From top to bottom, we take $\theta=0.5$, $0.6$, $0.7$, and $0.8$.
    The red dashed lines are fitting lines of the form $y=-a \log (1/m)+b$.}
    \label{Fig:Ground_Off_Critical} 
\end{figure}

The lattice results are shown in Fig.\ref{Fig:Ground_Off_Critical}.
First, by fixing the mass term and increasing the total length $L$, one can find that $\log|\langle G|U(\theta)|G\rangle|$ decays as $-\log L$ first and then saturates. In particular, $\log|\langle G|U(\theta)|G\rangle|$ saturates at a smaller value for a larger correlation length. Second, we study how the string order parameter scales with $1/m$ as $m$ approaches zero. As shown in Fig.\ref{Fig:Ground_Off_Critical} (right), it is found that 
\be
\label{log_m}
\log |\langle G|U(\theta)|G\rangle|\sim -\log \frac{1}{m}
\sim -\log \xi,
\ee
where we take $\xi \ll L$.
This result is consistent with Ref.\cite{2017Ryu}, where it was found that for a gapped system near the critical point that is described by a CFT,  the low-lying entanglement spectrum of the gapped theory is the finite-size spectrum of a boundary conformal field theory defined in a finite interval of length $L=\log(\xi/a)$, where $\xi$ is the correlation length of the gapped theory, and $a$ is the microscopic lattice constant.

\subsection{Type-II string order parameter}
\label{TypeII_ground}

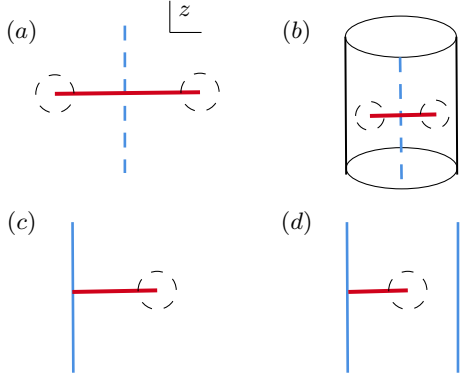
\begin{figure}
\centering
\begin{tikzpicture}[x=0.75pt,y=0.75pt,yscale=-0.8,xscale=0.8]

\draw [color={rgb, 255:red, 74; green, 144; blue, 226 }   ,draw opacity=1 ][line width=1.0]  [dash pattern={on 5.63pt off 4.5pt}]  (145,9) -- (145,102) ;

\draw [color={rgb, 255:red, 208; green, 2; blue, 27 }  ,draw opacity=1 ][line width=1.5]    (100.87,52.05) -- (192.37,51.05) ;
\draw  [dash pattern={on 4.5pt off 4.5pt}] (180.34,51.05) .. controls (180.34,44.41) and (185.73,39.03) .. (192.37,39.03) .. controls (199.01,39.03) and (204.4,44.41) .. (204.4,51.05) .. controls (204.4,57.7) and (199.01,63.08) .. (192.37,63.08) .. controls (185.73,63.08) and (180.34,57.7) .. (180.34,51.05) -- cycle ;
\draw  [dash pattern={on 4.5pt off 4.5pt}] (88.84,52.05) .. controls (88.84,45.41) and (94.23,40.03) .. (100.87,40.03) .. controls (107.51,40.03) and (112.9,45.41) .. (112.9,52.05) .. controls (112.9,58.7) and (107.51,64.08) .. (100.87,64.08) .. controls (94.23,64.08) and (88.84,58.7) .. (88.84,52.05) -- cycle ;

\draw [color={rgb, 255:red, 74; green, 144; blue, 226 } ][line width=1.0]    (112,133) -- (112.5,228) ;

\draw [color={rgb, 255:red, 208; green, 2; blue, 27 }  ,draw opacity=1 ][line width=1.5]    (111.87,177.05) -- (165.37,176.05) ;
\draw  [dash pattern={on 4.5pt off 4.5pt}] (153.34,176.05) .. controls (153.34,169.41) and (158.73,164.03) .. (165.37,164.03) .. controls (172.01,164.03) and (177.4,169.41) .. (177.4,176.05) .. controls (177.4,182.7) and (172.01,188.08) .. (165.37,188.08) .. controls (158.73,188.08) and (153.34,182.7) .. (153.34,176.05) -- cycle ;

\draw [color={rgb, 255:red, 74; green, 144; blue, 226 }][line width=1.0]    (285,133) -- (285.5,228) ;

\draw [color={rgb, 255:red, 208; green, 2; blue, 27 }  ,draw opacity=1 ][line width=1.5]    (285.87,177.05) -- (323.37,176.05) ;
\draw  [dash pattern={on 4.5pt off 4.5pt}] (311.34,176.05) .. controls (311.34,169.41) and (316.73,164.03) .. (323.37,164.03) .. controls (330.01,164.03) and (335.4,169.41) .. (335.4,176.05) .. controls (335.4,182.7) and (330.01,188.08) .. (323.37,188.08) .. controls (316.73,188.08) and (311.34,182.7) .. (311.34,176.05) -- cycle ;


\draw [color={rgb, 255:red, 74; green, 144; blue, 226 }][line width=1.0]    (355,133) -- (355,228) ;

\draw [color={rgb, 255:red, 0; green, 0; blue, 0 }  ,draw opacity=1 ][line width=0.75]    (284,16) -- (284.5,103) ;
\draw [color={rgb, 255:red, 0; green, 0; blue, 0 }  ,draw opacity=1 ][line width=0.75]    (354,16) -- (354.5,103) ;
\draw   (284,16) .. controls (284,8.82) and (299.67,3) .. (319,3) .. controls (338.33,3) and (354,8.82) .. (354,16) .. controls (354,23.18) and (338.33,29) .. (319,29) .. controls (299.67,29) and (284,23.18) .. (284,16) -- cycle ;
\draw   (284.5,99) .. controls (284.5,91.82) and (300.17,86) .. (319.5,86) .. controls (338.83,86) and (354.5,91.82) .. (354.5,99) .. controls (354.5,106.18) and (338.83,112) .. (319.5,112) .. controls (300.17,112) and (284.5,106.18) .. (284.5,99) -- cycle ;

\draw [color={rgb, 255:red, 74; green, 144; blue, 226 }   ,draw opacity=1 ][line width=1.0]  [dash pattern={on 5.63pt off 4.5pt}]  (318.5,29) -- (319.5,112) ;


\draw [color={rgb, 255:red, 208; green, 2; blue, 27 }  ,draw opacity=1 ][line width=1.5]    (299.5,66) -- (341.37,65.05) ;

\draw  [dash pattern={on 4.5pt off 4.5pt}] (331.34,65.04) .. controls (331.34,60.05) and (335.39,56) .. (340.38,56) .. controls (345.38,56) and (349.43,60.05) .. (349.43,65.04) .. controls (349.43,70.03) and (345.38,74.08) .. (340.38,74.08) .. controls (335.39,74.08) and (331.34,70.03) .. (331.34,65.04) -- cycle ;
\draw  [dash pattern={on 4.5pt off 4.5pt}] (290.46,66) .. controls (290.46,61.01) and (294.51,56.96) .. (299.5,56.96) .. controls (304.49,56.96) and (308.54,61.01) .. (308.54,66) .. controls (308.54,70.99) and (304.49,75.04) .. (299.5,75.04) .. controls (294.51,75.04) and (290.46,70.99) .. (290.46,66) -- cycle ;

\node at (60pt,10pt){$(a)$};
\node at (190pt,10pt){$(b)$};
\node at (60pt,100pt){$(c)$};
\node at (190pt,100pt){$(d)$};

\begin{scope}[xshift=-10pt, yshift=-30pt]
\draw (140pt,40pt)--(155pt,40pt);
\draw (140pt,40pt)--(140pt,25pt);
\node at (147pt,30pt){$z$};
\end{scope}


\end{tikzpicture}
\caption{Type-II string order parameter $\langle G|\mathcal L_A|G\rangle$ over a subsystem $A$ in the ground state. (a) Finite interval in an infinite system.  (b) Finite interval in a finite system with periodical boundary conditions.  (c) Finite interval at the end of a semi-infinite system. (d) Finite interval at the end of a finite system with open boundaries with the same boundary condition. There could be a symmetry breaking at the interface (blue dashed line) or at the boundary (blue solid line). The small discs of radius $\epsilon$ are removed to introduce a UV cutoff.
}
\label{Ground_partialString}
\end{figure}

For the type-II string operator, we consider the following  
different setups where the system is always prepared in the ground state, as shown in Fig.\ref{Ground_partialString}:
\begin{enumerate}
\item The system is infinite, with the string operator defined in a finite interval. There could be a symmetry-breaking conformal interface in the middle of the interval.

\item The system is finite with periodical boundary conditions, and the string operator is defined in a finite interval. There could be a symmetry-breaking conformal interface in the middle of the interval.

\item The system is semi-infinite, and the string operator is defined in a finite interval at the end of the system. There could be a symmetry breaking at the physical boundary. 

\item The system is finite with open boundaries, and the boundary conditions at the two ends are same. The string operator is defined in a finite interval at one end of the system. There could be a symmetry breaking at both boundaries.

\end{enumerate}

Next, we study the scaling behaviors of the type-II string order parameters in these cases.

\subsubsection{Finite interval in an infinite system}

As shown in Fig.\ref{Ground_partialString} (a),
we consider an infinite system in the ground state $|G\rangle$, with the string operator defined in a finite interval $[-R,R]$ of length $l=2R$. Our discussion below applies to two cases: (i) The whole system preserves the global symmetry. (ii) There is a symmetry breaking in the middle of the subsystem. 
It is noted that case (i) can be viewed as a specific situation of (ii), when there is no symmetry-breaking term added. In the following, we will discuss the more general case (ii).

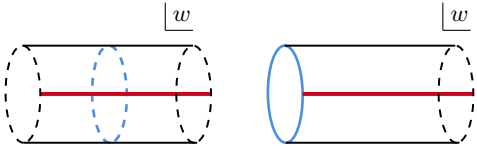
\begin{figure}
\centering
\begin{tikzpicture}[x=0.75pt,y=0.75pt,yscale=-0.7,xscale=0.7]

\begin{scope}[xshift=-95pt]
\draw  [color={rgb, 255:red, 74; green, 144; blue, 226 }  ,draw opacity=1 ][line width=1.0][dashed]  (220.3,147.71) .. controls (213.39,147.81) and (207.56,132.22) .. (207.28,112.89) .. controls (207,93.56) and (212.39,77.81) .. (219.3,77.71) .. controls (226.21,77.61) and (232.04,93.2) .. (232.32,112.53) .. controls (232.6,131.86) and (227.22,147.61) .. (220.3,147.71) -- cycle ;
\end{scope}

\draw [color={rgb, 255:red, 208; green, 2; blue, 27 }  ,draw opacity=1 ][line width=1.5]    (43.32,112.53) -- (166.32,112.53) ;

\draw  [dashed][color={rgb, 255:red, 0; green, 0; blue, 0 }  ,draw opacity=1 ][line width=0.75]  (31.3,147.71) .. controls (24.39,147.81) and (18.56,132.22) .. (18.28,112.89) .. controls (18,93.56) and (23.39,77.81) .. (30.3,77.71) .. controls (37.21,77.61) and (43.04,93.2) .. (43.32,112.53) .. controls (43.6,131.86) and (38.22,147.61) .. (31.3,147.71) -- cycle ;

\draw [color={rgb, 255:red, 0; green, 0; blue, 0 }  ,draw opacity=1 ][line width=0.75]    (30.3,77.71) -- (153.3,77.71) ;

\draw [color={rgb, 255:red, 0; green, 0; blue, 0 }  ,draw opacity=1 ][line width=0.75]    (31.3,147.71) -- (156.3,147.71) ;

\draw  [color={rgb, 255:red, 74; green, 144; blue, 226 }  ,draw opacity=1 ][line width=1.0]  (220.3,147.71) .. controls (213.39,147.81) and (207.56,132.22) .. (207.28,112.89) .. controls (207,93.56) and (212.39,77.81) .. (219.3,77.71) .. controls (226.21,77.61) and (232.04,93.2) .. (232.32,112.53) .. controls (232.6,131.86) and (227.22,147.61) .. (220.3,147.71) -- cycle ;

\draw  [dashed][color={rgb, 255:red, 0; green, 0; blue, 0 }  ,draw opacity=1 ][line width=0.75]  (154.3,147.71) .. controls (147.39,147.81) and (141.56,132.22) .. (141.28,112.89) .. controls (141,93.56) and (146.39,77.81) .. (153.3,77.71) .. controls (160.21,77.61) and (166.04,93.2) .. (166.32,112.53) .. controls (166.6,131.86) and (161.22,147.61) .. (154.3,147.71) -- cycle ;

\draw [color={rgb, 255:red, 208; green, 2; blue, 27 }  ,draw opacity=1 ][line width=1.5]    (232.32,112.53) -- (355.32,112.53) ;

\draw [color={rgb, 255:red, 0; green, 0; blue, 0 }  ,draw opacity=1 ][line width=0.75]    (219.3,77.71) -- (342.3,77.71) ;
\draw [color={rgb, 255:red, 0; green, 0; blue, 0 }  ,draw opacity=1 ][line width=0.75]    (220.3,147.71) -- (345.3,147.71) ;
\draw  [dashed][color={rgb, 255:red, 0; green, 0; blue, 0 }  ,draw opacity=1 ][line width=0.75]  (343.3,147.71) .. controls (336.39,147.81) and (330.56,132.22) .. (330.28,112.89) .. controls (330,93.56) and (335.39,77.81) .. (342.3,77.71) .. controls (349.21,77.61) and (355.04,93.2) .. (355.32,112.53) .. controls (355.6,131.86) and (350.22,147.61) .. (343.3,147.71) -- cycle ;

\begin{scope}[xshift=-40pt,yshift=10pt]
\draw (140pt,40pt)--(155pt,40pt);
\draw (140pt,40pt)--(140pt,25pt);
\node at (149pt,32pt){$w$};
\end{scope}

\begin{scope}[xshift=110pt,yshift=10pt]
\draw (140pt,40pt)--(155pt,40pt);
\draw (140pt,40pt)--(140pt,25pt);
\node at (149pt,32pt){$w$};
\end{scope}

\end{tikzpicture}
\caption{Left: The configurations in Fig.\ref{Ground_partialString} (a) and (b) after the conformal mapping in \eqref{eq:ground_a} and \eqref{eq:ground_b}.
Right: Fig.\ref{Ground_partialString} (c) and (d) after the conformal mapping in \eqref{eq:ground_c} and \eqref{eq:ground_d}. There is a symmetry breaking along the conformal interface (dashed blue line on the left) or along the conformal boundary (solid blue line on the right). For both cylinders, the width in the (vertical) $\Im(w)$ direction is $2\pi$.
}
\label{Fig:TypeII_Ground}
\end{figure}

For the path integral in Fig.\ref{Fig:TypeII_Ground} (a), to introduce a UV cutoff in our later calculation, we remove two small discs of radius $\epsilon$ from the two ends of the string operator, and impose two conformal boundary conditions $|b_1\rangle$ and $|b_2\rangle$ along the boundaries.
Then, by considering the following conformal mapping
\be
\label{eq:ground_a}
w=f(z)=\log\left(\frac{z+R}{R-z}\right),
\ee
the $z$-plane in Fig.\ref{Ground_partialString} (a) is mapped to a $w$-cylinder in Fig.\ref{Fig:TypeII_Ground} (a). The width of this $w$-cylinder in the $\Im(w)$ direction is $2\pi$, and its length in $\Re(w)$ direction is 
\be
\label{Width_ground_1}
W=f(R-\epsilon)-f(-R+\epsilon)=2\log(l/\epsilon)+\mathcal O(\epsilon),
\ee
where $l=2R$.
The conformal interface in Fig.\ref{Ground_partialString} (a) is now mapped to a circle on the $w$-cylinder.

Then the type-II string order parameter can be evaluated as
\be
\label{typeII_ground}
\langle G|\mathcal L_A|G\rangle=\frac{Z_{\text{defect}}}{Z_0}=\frac{\langle b_1|e^{-\frac{W}{2}H_{\text{defect}}}I e^{-\frac{W}{2}H_{\text{defect}}} |b_2\rangle}{ \langle b_1|e^{-\frac{W}{2}H_{0}}I e^{-\frac{W}{2}H_{0} }|b_2\rangle}
\ee
where $I$ denotes the conformal interface operator inserted along the circle in Fig.\ref{Fig:TypeII_Ground} (left), and $H_{\text{defect}}$ ($H_0$) denotes the Hamiltonian which is defined along the $\Im(w)$ direction of the $w$-cylinder with (without) an insertion of the string operator. Pictorially, we have defined:
\be
\label{Ground_Zdefect_plot}
\begin{tikzpicture}[x=0.75pt,y=0.75pt,yscale=-0.7,xscale=0.7]

\draw  [dash pattern={on 4.5pt off 4.5pt}] (100,75.5) .. controls (100,58.1) and (106.16,44) .. (113.75,44) .. controls (121.34,44) and (127.5,58.1) .. (127.5,75.5) .. controls (127.5,92.9) and (121.34,107) .. (113.75,107) .. controls (106.16,107) and (100,92.9) .. (100,75.5) -- cycle ;
\draw [color={rgb, 255:red, 0; green, 0; blue, 0 }  ,draw opacity=1 ][line width=0.75]    (113.75,44) -- (259.75,44) ;
\draw [color={rgb, 255:red, 0; green, 0; blue, 0 }  ,draw opacity=1 ][line width=0.75]    (113.75,107) -- (259.75,107) ;
\draw  [color={rgb, 255:red, 245; green, 166; blue, 35 }  ,draw opacity=1 ] (133,75.5) .. controls (133,58.1) and (139.16,44) .. (146.75,44) .. controls (154.34,44) and (160.5,58.1) .. (160.5,75.5) .. controls (160.5,92.9) and (154.34,107) .. (146.75,107) .. controls (139.16,107) and (133,92.9) .. (133,75.5) -- cycle ;
\draw  [dash pattern={on 4.5pt off 4.5pt}] (246,75.5) .. controls (246,58.1) and (252.16,44) .. (259.75,44) .. controls (267.34,44) and (273.5,58.1) .. (273.5,75.5) .. controls (273.5,92.9) and (267.34,107) .. (259.75,107) .. controls (252.16,107) and (246,92.9) .. (246,75.5) -- cycle ;
\draw [color={rgb, 255:red, 208; green, 2; blue, 27 }  ,draw opacity=1 ][line width=1.5]    (126.75,77) -- (272.75,77) ;

\draw  [color={rgb, 255:red, 74; green, 144; blue, 226 }  ,draw opacity=1 ][dash pattern={on 5.63pt off 4.5pt}][line width=1.0]  (173,75.5) .. controls (173,58.1) and (179.16,44) .. (186.75,44) .. controls (194.34,44) and (200.5,58.1) .. (200.5,75.5) .. controls (200.5,92.9) and (194.34,107) .. (186.75,107) .. controls (179.16,107) and (173,92.9) .. (173,75.5) -- cycle ;
\draw  [dash pattern={on 4.5pt off 4.5pt}] (102,166.5) .. controls (102,149.1) and (108.16,135) .. (115.75,135) .. controls (123.34,135) and (129.5,149.1) .. (129.5,166.5) .. controls (129.5,183.9) and (123.34,198) .. (115.75,198) .. controls (108.16,198) and (102,183.9) .. (102,166.5) -- cycle ;
\draw [color={rgb, 255:red, 0; green, 0; blue, 0 }  ,draw opacity=1 ][line width=0.75]    (115.75,135) -- (261.75,135) ;
\draw [color={rgb, 255:red, 0; green, 0; blue, 0 }  ,draw opacity=1 ][line width=0.75]    (115.75,198) -- (261.75,198) ;
\draw  [color={rgb, 255:red, 245; green, 166; blue, 35 }  ,draw opacity=1 ] (135,166.5) .. controls (135,149.1) and (141.16,135) .. (148.75,135) .. controls (156.34,135) and (162.5,149.1) .. (162.5,166.5) .. controls (162.5,183.9) and (156.34,198) .. (148.75,198) .. controls (141.16,198) and (135,183.9) .. (135,166.5) -- cycle ;
\draw  [dash pattern={on 4.5pt off 4.5pt}] (248,166.5) .. controls (248,149.1) and (254.16,135) .. (261.75,135) .. controls (269.34,135) and (275.5,149.1) .. (275.5,166.5) .. controls (275.5,183.9) and (269.34,198) .. (261.75,198) .. controls (254.16,198) and (248,183.9) .. (248,166.5) -- cycle ;
\draw  [color={rgb, 255:red, 74; green, 144; blue, 226 }  ,draw opacity=1 ][dash pattern={on 5.63pt off 4.5pt}][line width=1.0]  (175,166.5) .. controls (175,149.1) and (181.16,135) .. (188.75,135) .. controls (196.34,135) and (202.5,149.1) .. (202.5,166.5) .. controls (202.5,183.9) and (196.34,198) .. (188.75,198) .. controls (181.16,198) and (175,183.9) .. (175,166.5) -- cycle ;

\begin{scope}[yshift=-55pt]
\node at (10pt,109pt){$Z_{\text{defect}}=$};
\node at (25pt,182pt){$Z_{0}=$};

\small

\node at (58pt,109pt){$\langle b_1|$};
\node at (222pt,109pt){$|b_2\rangle$};

\node at (58pt,180pt){$\langle b_1|$};
\node at (222pt,180pt){$|b_2\rangle$};

\node at (112pt,185pt){\textcolor{orange}{$H_0$}};
\node at (123pt,99pt){\textcolor{orange}{$H_{\text{defect}}$}};
\end{scope}

\end{tikzpicture}
\ee
Note that the length of the above cylinders is $W\simeq 2\log(l/\epsilon)\gg 1$. Therefore, $\langle U(\theta)\rangle$ in \eqref{typeII_ground} is mainly contributed by the ground states $|G_d\rangle$ and $|G_0\rangle$ of $H_{\text{defect}}$
and $H_0$ as follows:
\be
\small
\langle G|\mathcal L_A|G\rangle\simeq
\frac{\langle  b_1|G_d\rangle \langle G_d|I|G_d\rangle \langle G_d|b_2 \rangle}{\langle  b_1|G_0\rangle \langle G_0|I|G_0\rangle \langle G_0|b_2\rangle}
e^{-W(E^0_{\text{defect}}-E_0)}.
\ee
Noting that $W\simeq 2\log(l/\epsilon)$, one can obtain
\be
\label{StringII_ground_b}
\log |\langle G|\mathcal L_A|G\rangle| \simeq -\kappa\log l,
\ee
up to a constant sub-leading term. Similar as before, here the coefficient $\kappa=2(E^0_{\text{defect}}-E_0)$ is not universal and depends on the details of the string operator.

\subsubsection{Finite interval in a finite system}

Now we consider a finite system of length $L$ in the ground state $|G\rangle$ with periodic boundary conditions. The string operator is defined in a finite interval $[-R,\,R]$, and in addition there is a conformal interface that breaks the global symmetry in the middle of this interval, as shown in Fig.\ref{Ground_partialString} (b).

The evaluation of $\langle G|\mathcal L_A|G\rangle$ is similar to the case of a finite interval in an infinite system in Fig.\ref{Ground_partialString} (a). The difference is that now we consider the following conformal mapping 
\be
\label{eq:ground_b}
w=f(z)=\log\left(
\frac{e^{2\pi i z/L}-e^{-2\pi i R /L}}{
e^{2\pi i R/L}-e^{2\pi i z/L}
}
\right),
\ee
which maps the $z$-plane in Fig.\ref{Ground_partialString} (b) to the same $w$-cylinder in Fig.\ref{Fig:TypeII_Ground} (left plot). Now the length of the $w$-cylinder becomes
\be
\label{Width_ground_2}
W=2\log\left[\frac{L}{\pi \epsilon}\sin\left(\frac{\pi l}{L}\right)\right]+\mathcal O(\epsilon),
\ee
where $l=2R$.
Note that in the limit $L\gg l$, $W$ becomes the same as \eqref{Width_ground_2}.
By repeating the procedure in the previous case, one can find 
\be
\label{Ground_typeII_case2}
\log |\langle G|\mathcal L_A|G\rangle| \simeq -\kappa\log \left[\frac{L}{\pi}\sin\left(\frac{\pi l}{L}\right)\right],
\ee
up to a constant sub-leading term. Here the coefficient $\kappa$ is the same as that in \eqref{StringII_ground_b}.

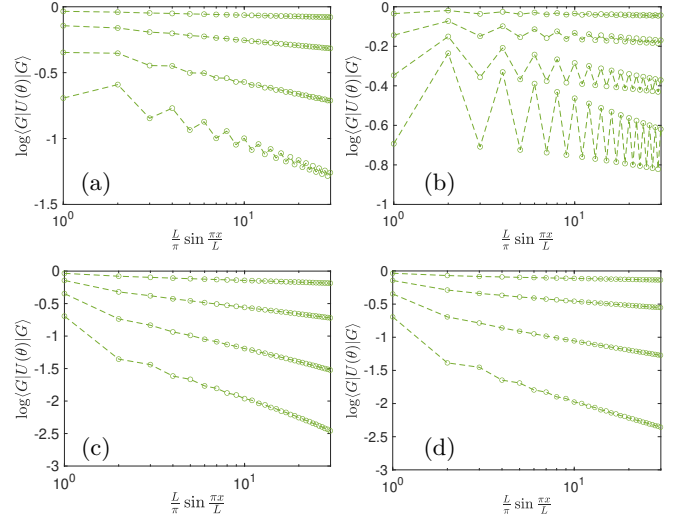
\begin{figure}[t]
\centering
\begin{tikzpicture}

    \node[inner sep=0pt] (russell) at (15pt,-85pt)
    {\includegraphics[width=.26\textwidth]{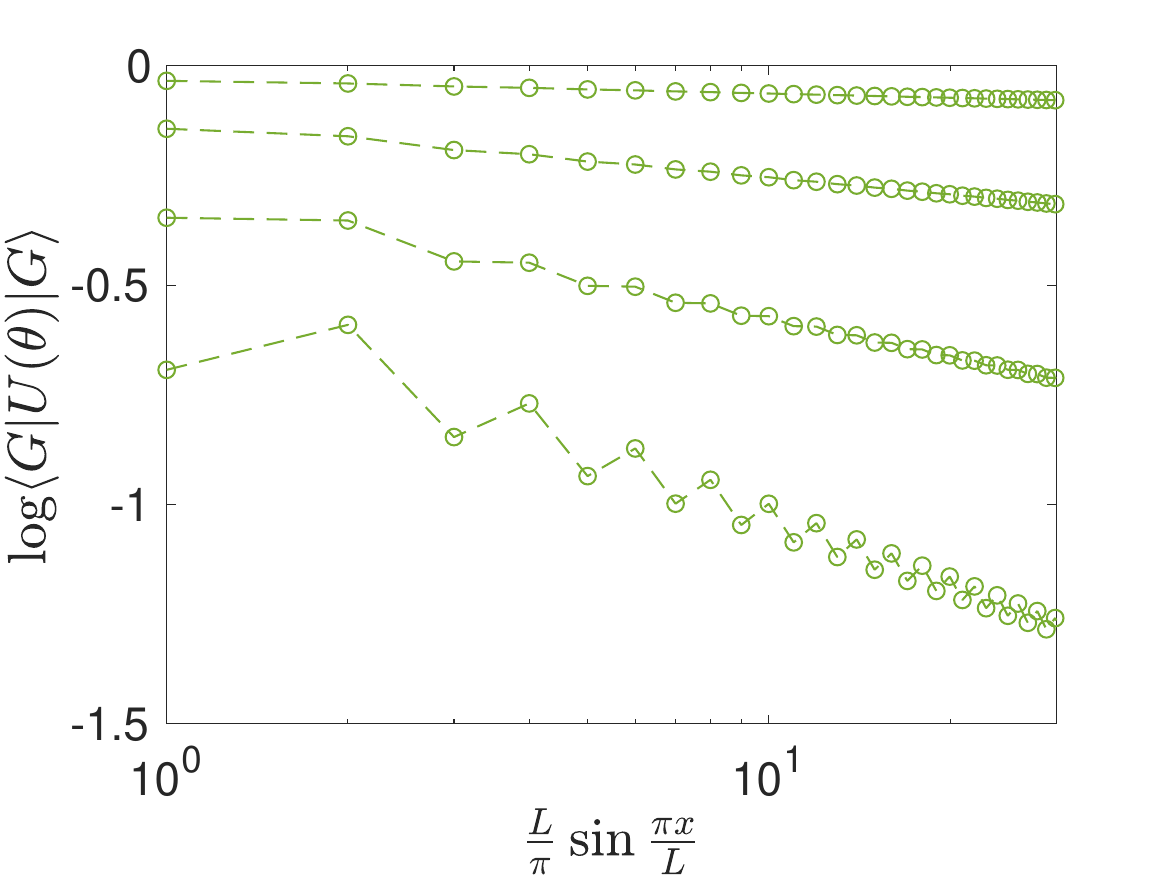}};

    \node[inner sep=0pt] (russell) at (140pt,-85pt)
    {\includegraphics[width=.26\textwidth]{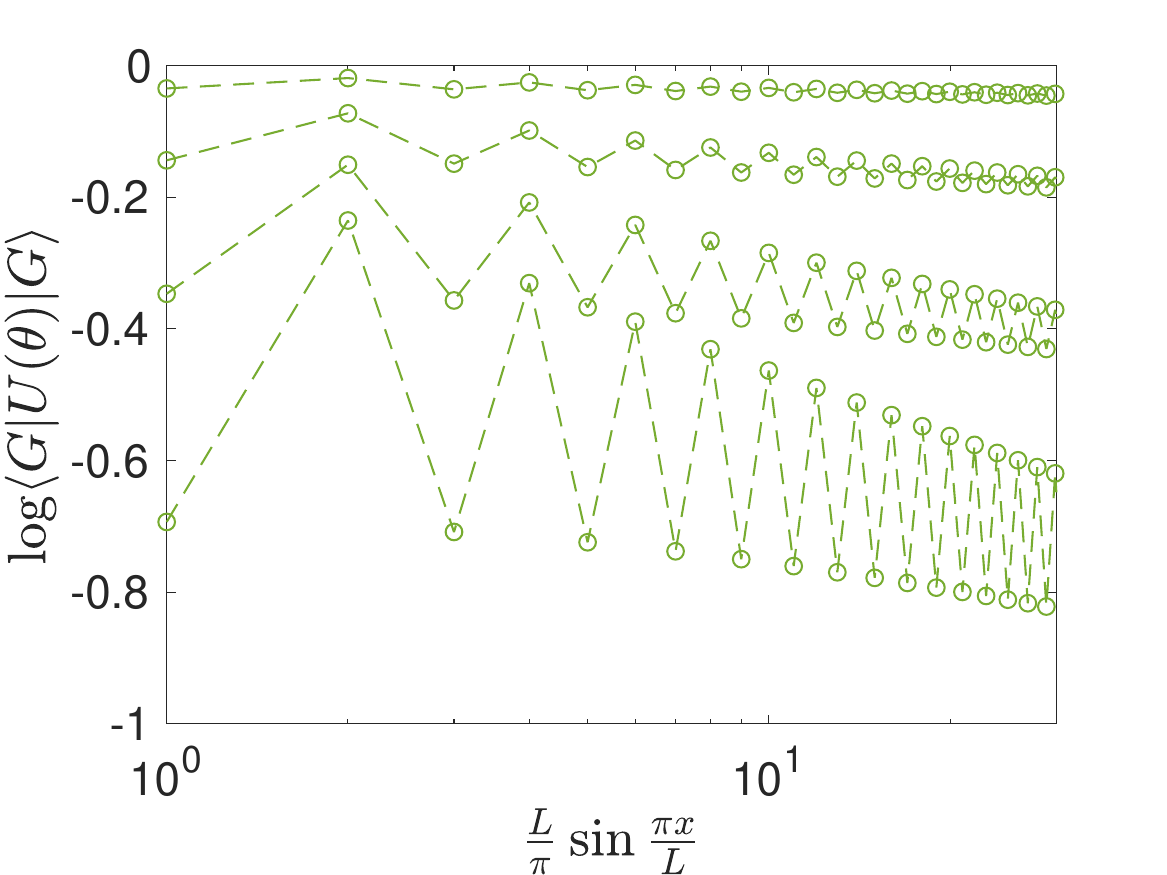}};
 
        \node[inner sep=0pt] (russell) at (15pt,-185pt)
    {\includegraphics[width=.26\textwidth]{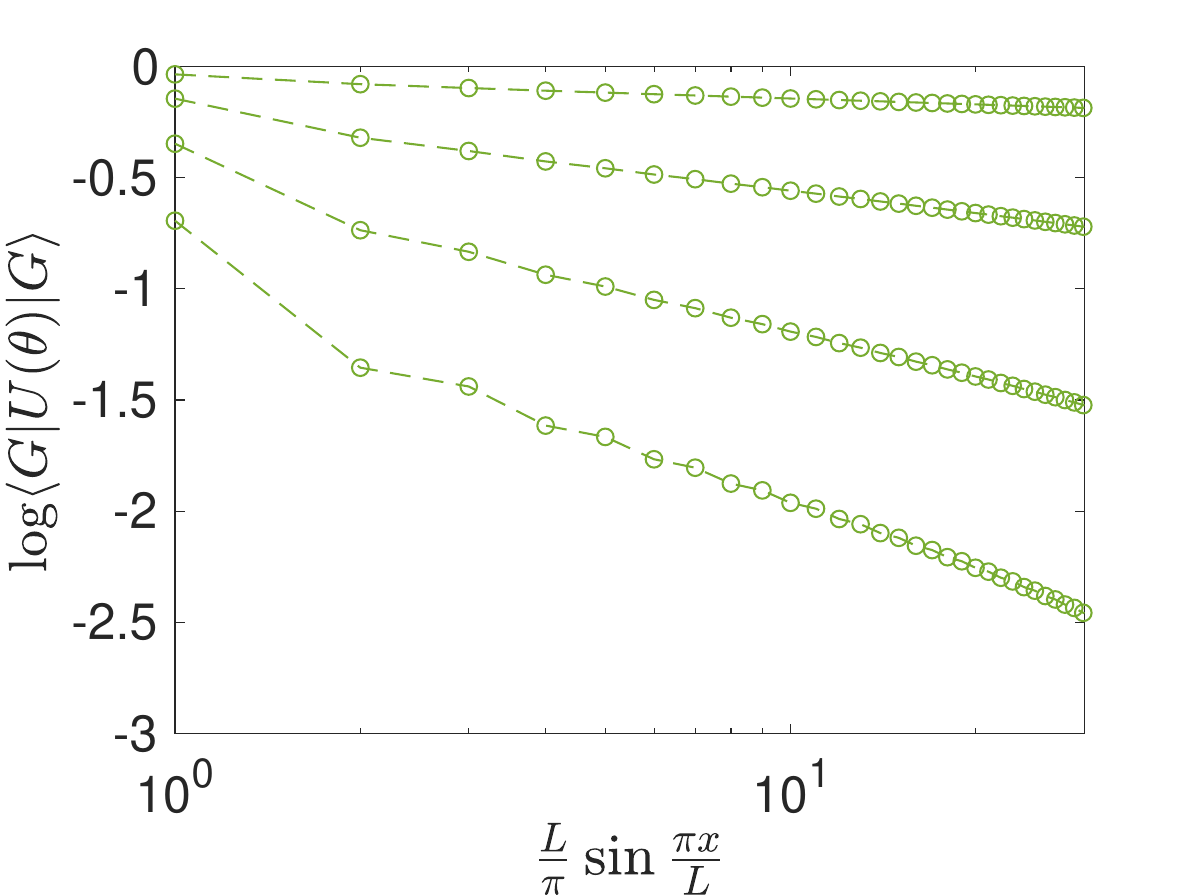}};

         \node[inner sep=0pt] (russell) at (140pt,-185pt)
    {\includegraphics[width=.26\textwidth]{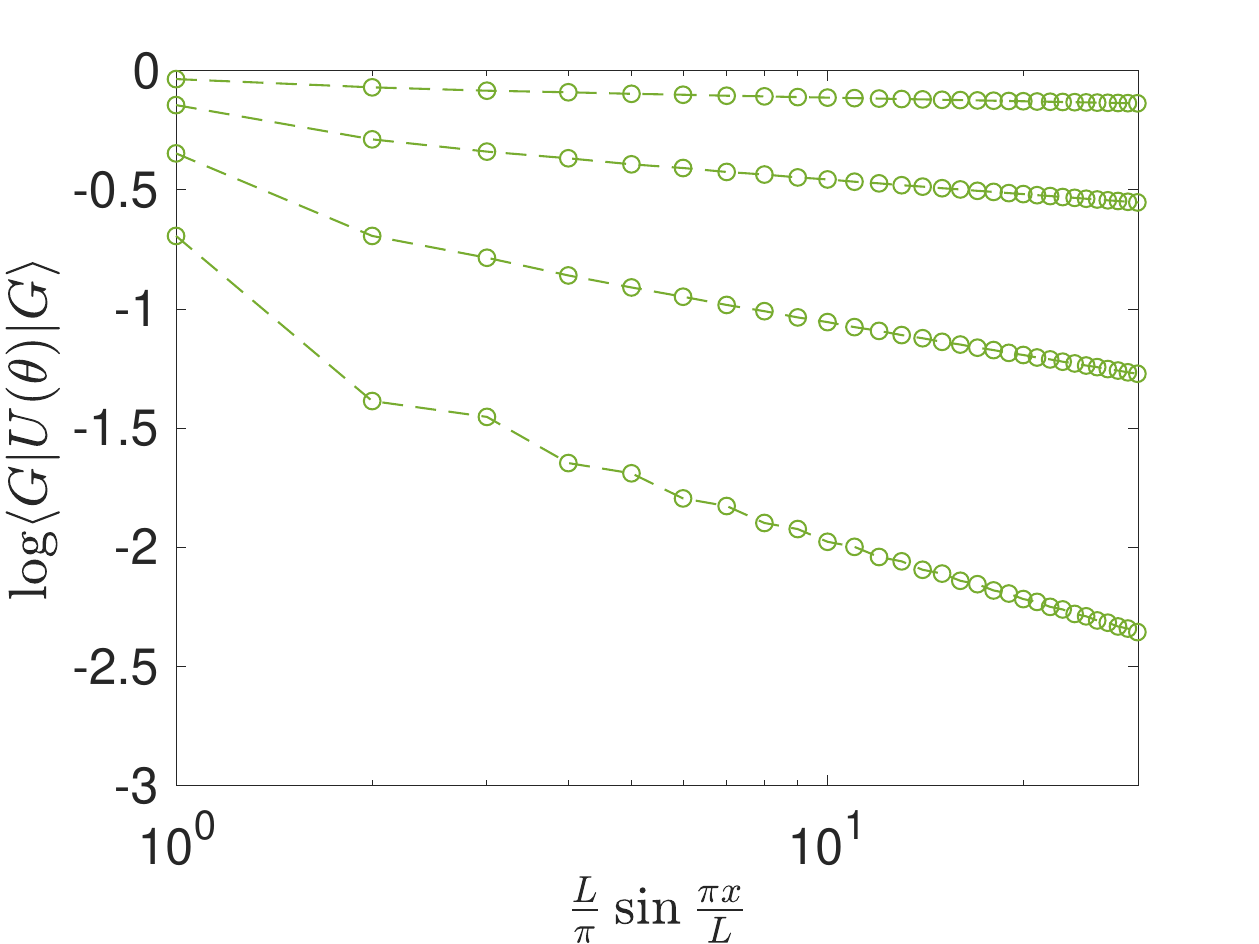}};

             \node at (-20pt, -110pt){(a)};
            \node at (110pt, -110pt){(b)};
            
            \node at (-20pt, -210pt){(c)};  
            \node at (110pt, -210pt){(d)};                       
    \end{tikzpicture}
\caption{Lattice calculation of type-II string order parameter in the interval $[0,\, x]$ for a finite system defined over $[0,\,L]$ in the ground states $|G\rangle$.
We consider (a) periodical boundary conditions and (b) open boundary conditions where the $U(1)$ symmetry is preserved over the whole system. 
In (c) and (d), we introduce the symmetry-breaking interface/boundary. 
(c) Periodic boundary conditions with a symmetry breaking interface located at $x/2$.
(d) Open boundary conditions with a symmetry breaking at the two boundaries. From top to bottom in each plot, we take $\theta=\pi/6$, $2\pi/6$, $3\pi/6$, and $4\pi/6$, respectively, and 
the total length of the lattice is fixed as $L=600$.
}
\label{Fig:TypeII_Ground_lattice}
\end{figure}

The lattice results based on the Hamiltonian in \eqref{Global_defect_middle} are shown in Fig.\ref{Fig:TypeII_Ground_lattice} (a) and (c), where one can see the scaling behavior in \eqref{Ground_typeII_case2}.

\subsubsection{Interval at the end of a semi-infinite line}

Now we consider a semi-infinite system in $[0,+\infty)$, with a symmetry breaking at the boundary $x=0$. The conformal boundary state corresponding to this
symmetry-breaking boundary is denoted as $|B\rangle$.
The total system is in its ground state $|G\rangle$, and the string operator is defined in the interval $[0,l]$, as shown in Fig.\ref{Ground_partialString} (c).

To evaluate $\langle G|\mathcal L_A|G\rangle$ in the path integral, we first remove a small disc of radius $\epsilon$ at $z=l+i0$, and impose a conformal boundary condition $|b\rangle$ along the boundary. Then, by considering the following conformal mapping 
\be
\label{eq:ground_c}
w=f(z)=\log\left(\frac{l+z}{l-z}\right),
\ee
the $z$-plane in Fig.\ref{Ground_partialString} (c) is mapped to 
a $w$-cylinder in Fig.\ref{Fig:TypeII_Ground} (right plot). The length of this $w$-cylinder in $\Re(w)$ direction is $W=\log(2l/\epsilon)+\mathcal O(\epsilon)$.
Then one has
\be
\label{TypeII_ground_C}
\langle G|\mathcal L_A|G\rangle=\frac{Z_{\text{defect}}}{Z_0}
=\frac{\langle B|e^{-H_{\text{defect}}W }|b\rangle}{
\langle B|e^{-H_{0}W }|b\rangle
},
\ee
where the Hamiltonians $H_{\text{defect}}$ and $H_0$ have the same definitions as those in \eqref{Ground_Zdefect_plot}.
Since $W\gg 1$, $\langle G|\mathcal L_A|G\rangle$ is mainly dominated by the ground states of $H_{\text{defect}}$ and $H_0$ as follows:
\be
\langle G|\mathcal L_A|G\rangle\simeq \frac{\langle B|G_d\rangle \langle G_d|b\rangle}{
\langle B|G_0\rangle \langle G_0|b\rangle
}\,
e^{-(E_\text{defect}^0-E^0) W},
\ee
where $|G_d\rangle$ and $|G_0\rangle$ are the ground states 
of $H_{\text{defect}}$ and $H_0$ respectively. Then one can obtain
\be
\label{TypeII_ground_C_scale}
\log |\langle G|\mathcal L_A|G\rangle| \simeq -\kappa \log (2l),
\ee
up to a constant sub-leading term. Here the coefficient $\kappa=E_\text{defect}^0-E^0$ is not universal and it depends on the details of the string operator.

\subsubsection{Interval at the end of a finite line}

Now, let us consider the case in Fig.\ref{Ground_partialString} (d), i.e., a finite system $[0, \,L]$ with open boundary conditions in the ground state. We consider the case that the boundary conditions at the two ends are the same, and more generally there could be a symmetry breaking at the two ends. The string operator is defined over the sub-region $[0,l]$.

In the path integral in Fig.\ref{Ground_partialString} (d), similar as before, we remove a small disc of radius $\epsilon$ to introduce a UV cutoff. Then, by considering the following conformal mapping
\be
\label{eq:ground_d}
w=f(z)=\log\left(
\frac{\sin[\pi(z-l)/2L]}{\cos[\pi(z+l)/2L]}
\right),
\ee
we map the $z$-plane in Fig.\ref{Ground_partialString} (d) to the $w$-cylinder in Fig.\ref{Fig:TypeII_Ground} (right plot). The length of this cylinder in the $\Re(w)$ direction
is
\be
W=\log\left[\frac{2L}{\pi \epsilon}\sin\left(\frac{\pi l}{L}\right)\right]+\mathcal O(\epsilon).
\ee
Note that in the limit $L\gg l$, the length becomes $W=\log(2l/\epsilon)+\mathcal O(\epsilon)$, which becomes the same as that in the case of Fig.\ref{Ground_partialString} (c). By repeating a similar procedure in Eqs.\eqref{TypeII_ground_C}$\sim$\eqref{TypeII_ground_C_scale}, one can find
\be
\label{Ground_typeII_case4}
\log |\langle G|\mathcal L_A|G\rangle| \simeq -\kappa\, \log \left[\frac{2L}{\pi}\sin\left(\frac{\pi l}{L}\right)\right],
\ee
up to a constant sub-leading term. Here the coefficient $\kappa$ is the same as that in \eqref{TypeII_ground_C_scale}.

The lattice results are shown in Fig.\ref{Fig:TypeII_Ground_lattice} (b) and (d), where one can see the scaling behavior as predicted in \eqref{Ground_typeII_case4}.

\section{More on type-I and type-II string order parameters after a global quantum quench}
\label{Appendix:TypeI/II}

In this appendix, we give more details on the type-I and type-II string order parameters after a global quantum quench, in addition to the main results in Sec.\ref{Sec:BSB} and Sec.\ref{Sec:Partial_String}.

\subsection{Initial states that break the global symmetry}

In our CFT study of the global quench in the main text, the initial state is prepared as
\be
\label{SymBreak_initial}
|\psi_0\rangle=e^{-\frac{\beta}{4}H_{\text{CFT}}}|B\rangle,
\ee
where $|B\rangle$ is a conformal boundary state.
For our configurations of path integrals in Fig.\ref{Fig:Global1}, Fig.\ref{Fig:GlobalQuench2}, and
Fig.\ref{Fig:PartialString_Global}, it is noted that we introduce the boundary symmetry breaking along the whole line along $C=\{\Re(z)=0,\,\Im(z)\in[-\beta/4,\beta/4]\}$.
Rigorously speaking, this means we should also introduce the boundary symmetry breaking in $H_{\text{CFT}}$ in the regulator in \eqref{SymBreak_initial}.

For the lattice calculation in the main text, we consider the initial state as the ground state of a gapped Hamiltonian where the symmetry is preserved over the whole system. Here we consider the initial state as the ground state of a gapped Hamiltonian with a symmetry-breaking boundary. It is found that our conclusion in the main text does not change.

As an illustration, let us consider the case of global quench in the setup of Sec.\ref{Sec:Global1}.
We consider the initial state as the ground state of the following gapped Hamiltonian:
\be
\label{H0_initial}
H_0=-\frac{1}{2}\sum_{1\le j\le L-1} c_j^\dag c_{j+1}+h\,c_1^\dag c_2^\dag+h.c.
+m \sum_j (-1)^j c_j^\dag c_j.
\ee
Comparing to \eqref{H0_globalQuench}, here we have added
the second term which breaks the $U(1)$ symmetry at the boundary.

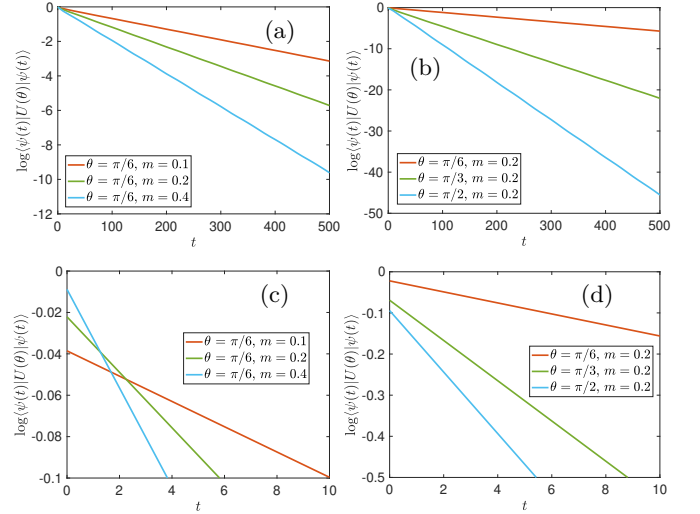
\begin{figure}[t]
\centering
\begin{tikzpicture}

    \node[inner sep=0pt] (russell) at (15pt,-85pt)
    {\includegraphics[width=.26\textwidth]{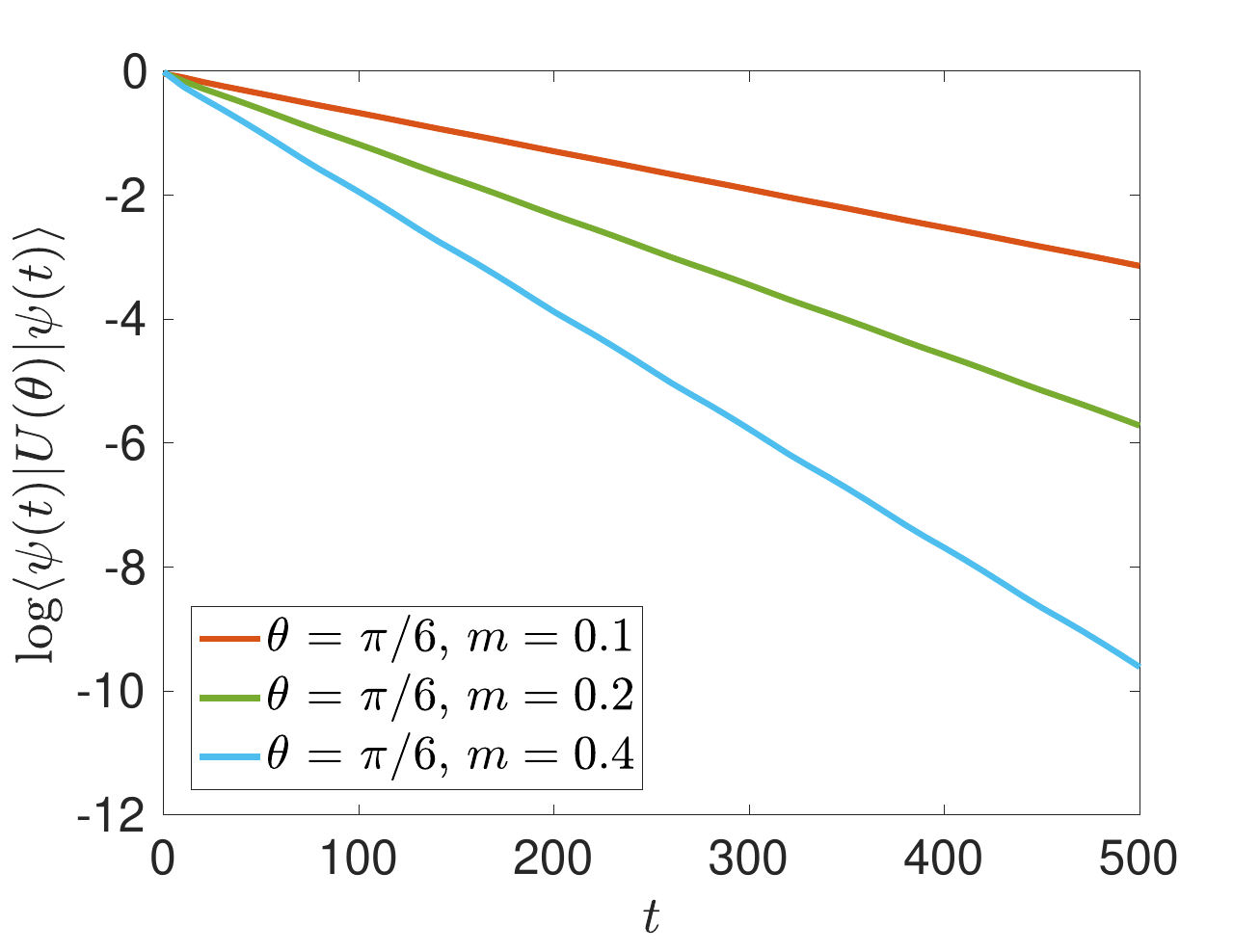}};
        \node[inner sep=0pt] (russell) at (140pt,-85pt)
    {\includegraphics[width=.26\textwidth]{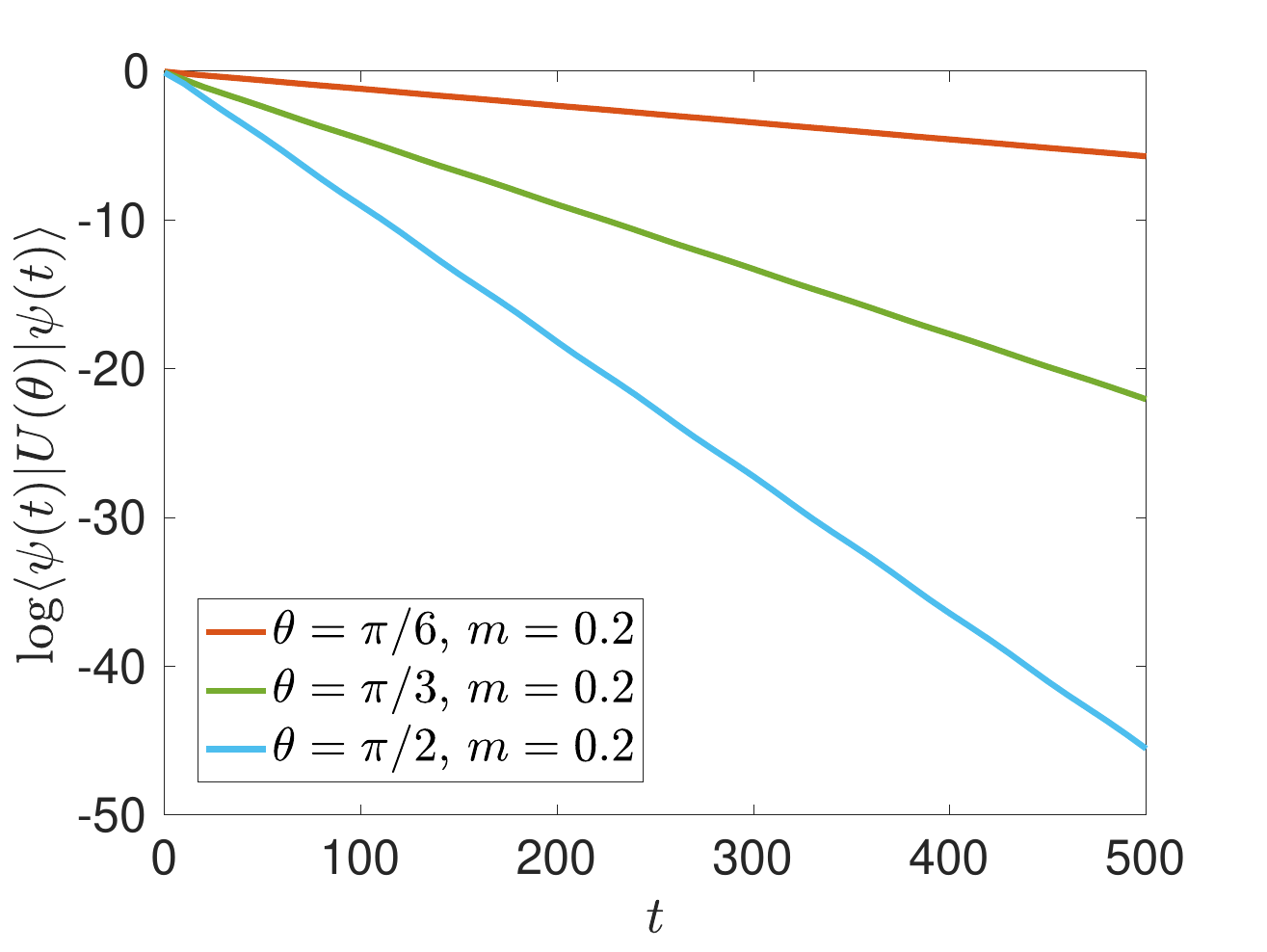}};

    \node[inner sep=0pt] (russell) at (15pt,-185pt)
    {\includegraphics[width=.26\textwidth]{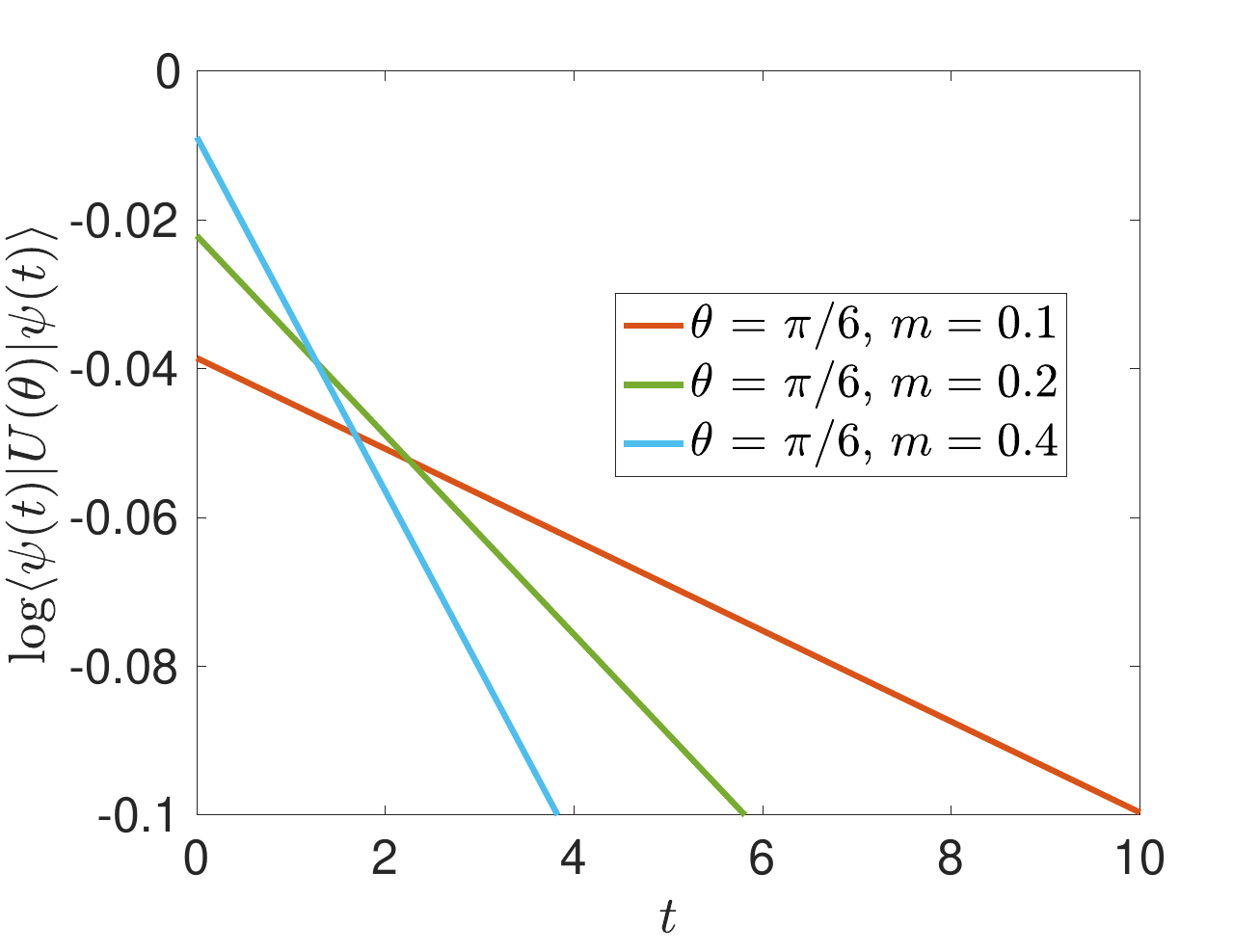}};
    
        \node[inner sep=0pt] (russell) at (140pt,-185pt)
    {\includegraphics[width=.26\textwidth]{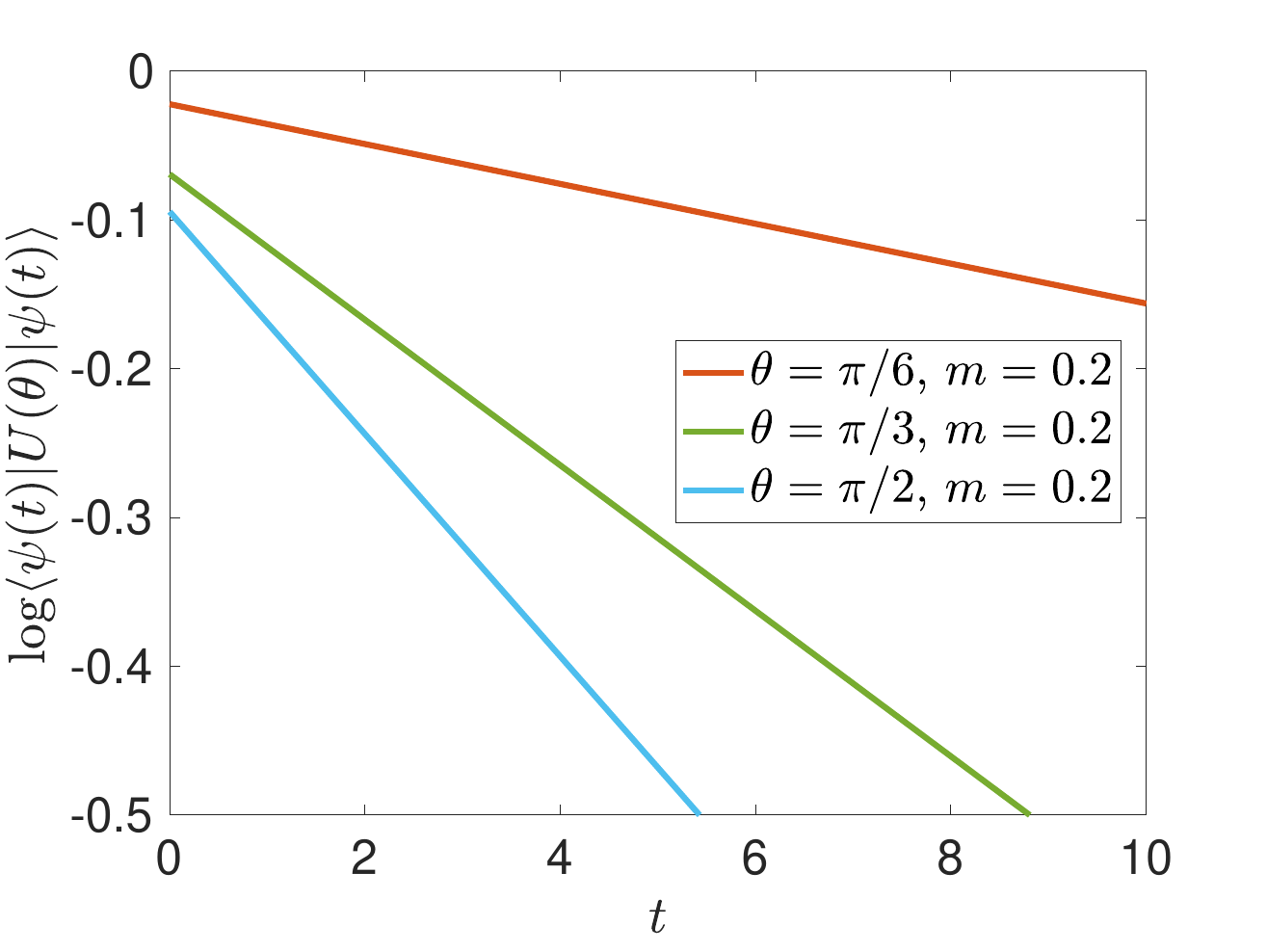}};
  
             \node at (50pt, -52pt){(a)};
            \node at (105pt, -66pt){(b)};
             \node at (48pt, -152pt){(c)};          
             \node at (170pt, -152pt){(d)};                
                          
    \end{tikzpicture}
\caption{
Lattice calculation of type-I string order parameter evolution after a global quantum quench.
The parameters are the same as those in Fig.\ref{Fig:SymBreak_Global_finite}, except that 
now the initial state is chosen as the ground state of \eqref{H0_initial}, with $h=0.5$. Here (c) and (d) are zoom-in plot of (a) and (b) near $t=0$, respectively.
}
\label{Fig:SymBreakInitial_Global}
\end{figure}

The lattice results after a global quench are shown in Fig.\ref{Fig:SymBreakInitial_Global}. All the parameters are the same as those in Fig.\ref{Fig:SymBreak_Global_finite}, except that now we have a symmetry-breaking initial state. One can find the general features in Fig.\ref{Fig:SymBreakInitial_Global} and 
Fig.\ref{Fig:SymBreak_Global_finite} are the same, i.e., the type-I string order parameters decay exponentially in time.
The subtle difference can be found near $t=0$, as follows:

(i) At $t=0$, one can find $\langle \psi(t)|U(\theta)|\psi(t)\rangle$ or equivalently $\langle G|U(\theta)|G\rangle$ in Fig.\ref{Fig:SymBreak_Global_finite} is exactly zero, because the initial state, which is the ground state of \eqref{H0_globalQuench} , preserves the $U(1)$ symmetry.
In Fig.\ref{Fig:SymBreakInitial_Global} (c) and (d), one can find that $\langle \psi(t)|U(\theta)|\psi(t)\rangle$ at $t=0$ is not zero, because the initial state breaks the $U(1)$ symmetry.

(ii) For different mass terms $m$ and $\theta$ in Fig.\ref{Fig:SymBreakInitial_Global} (c) and (d), $\langle \psi(t)|U(\theta)|\psi(t)\rangle$ at $t=0$ have different values. This can be understood based on our discussion in Appendix \ref{Appendix:off-critical}, where we analyze the features of 
type-I string order parameters in the ground state of a gapped Hamiltonian with a symmetry-breaking boundary. See   Fig.\ref{Fig:Ground_Off_Critical} for details.

\subsection{Dependence of decaying rates on the mass term after a global quench }
\label{Appendix:Mass_kappa}

In this appendix, we study in detail how the decaying rates $\kappa$ 
of type-I string order parameters after a global quench
depend on the mass term in the initial state. More concretely, we consider the setup in Sec.\ref{Sec:Global1}, and study how the decaying rates in Fig.\ref{Fig:SymBreak_Global_finite} depend on the mass term in \eqref{H0_globalQuench}.

First, from the CFT result in \eqref{kappa_global1}, we know that the decaying rates $\kappa\propto 1/\beta$, where $\beta$ is used to define the initial state $|\psi_0\rangle=e^{-\frac{\beta}{4}H_{\text{CFT}}}|B\rangle$. Physically, $\beta$ characterizes the correlation length of the state as $\beta\propto \xi$. On the other hand, we know that for a small mass term in \eqref{H0_globalQuench},
\footnote{Here when we talk about a small mass $m$, it is compared to the bandwidth of the system -- it is determined by the hopping strength which is $1/2$ in \eqref{H0_globalQuench}.
For a small mass, the low energy physics can be viewed as a Dirac fermion CFT perturbed by a relevant mass term. 
} the correlation length of the ground state is proportional to $1/m$. 
Therefore, we would expect $\kappa \propto m$.
In fact, as discussed in Appendix C of \cite{2018_Wen_Wang_Ryu}, when $m$ is small in the lattice model, one can show that 
\be
\beta=1/m.
\ee
This indeed indicates that one would have $\kappa\propto m$ for a small $m$.

\begin{figure}[h]
\centering
{\includegraphics[width=.32\textwidth]{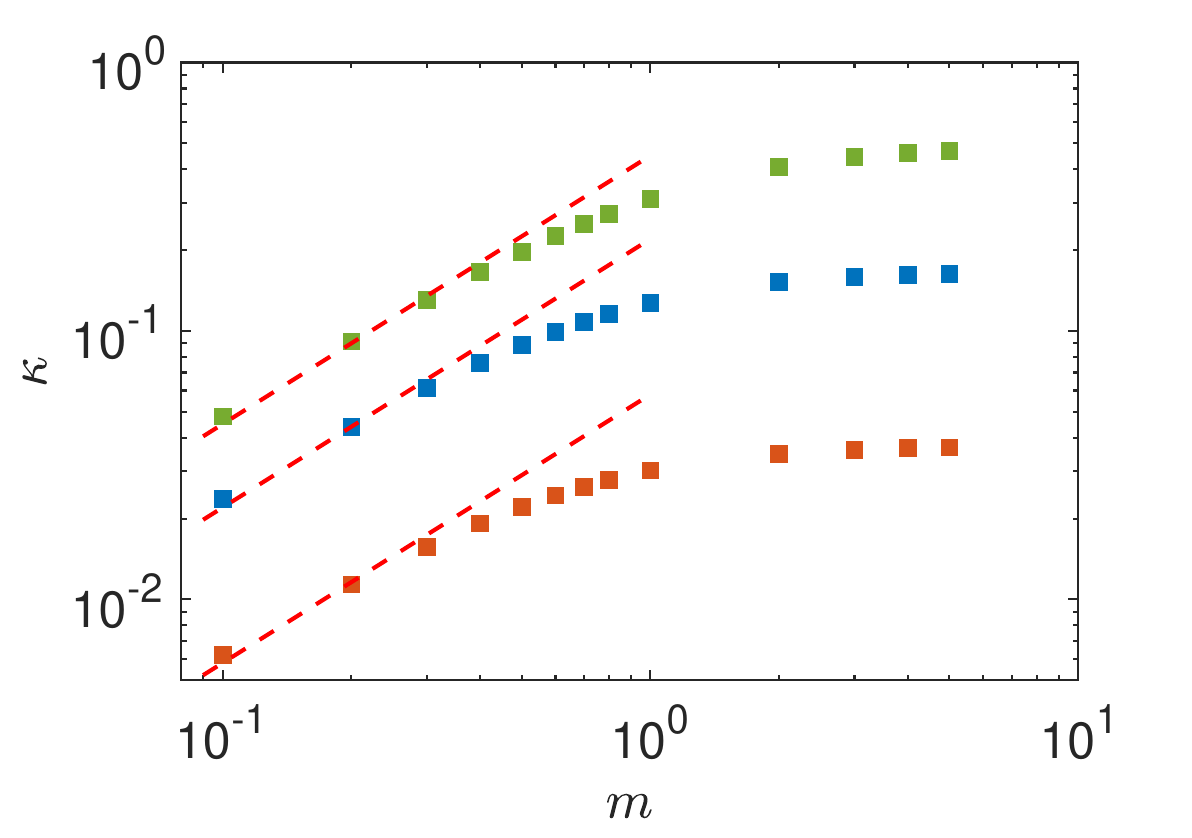}}
    \caption{
The dependence of decaying rates $\kappa$ in Fig.\ref{Fig:SymBreak_Global_finite} as a function of the mass term $m$ in the initial state.
From top to bottom, we take $\theta=\pi/2$, $\pi/3$, and $\pi/6$, respectively.
The red dashed lines are guiding lines with $y\propto x$.
}
    \label{Fig:mass_rate} 
\end{figure}

As shown in Fig.\ref{Fig:mass_rate}, we study the decaying rates $\kappa$ in Fig.\ref{Fig:SymBreak_Global_finite} as a function of $m$.
For $m<1/2$, where $1/2$ is the strength of the hopping term in the lattice, one can find that $\kappa\propto m$, as expected. 
For $m>1/2$, the mass term is comparable to the energy-bandwidth of the system, then one can find that $\kappa$ saturates as $m$ increases. Physically, for a very large $m$, the initial state is approximated by a product state that is independent of the concrete value of $m$.

\subsection{Type-II string order parameter after a global quench}

For the setup considered in Sec.\ref{Sec:TypeII_GlobalQuench},
where we study a finite interval at the end of a semi-infinite system after a global quench,
there is an interesting feature in the time evolution of the type-II string order parameter, as shown in Fig.\ref{Fig:NoSymBreak_Global}. That is, when the boundary is symmetry-breaking, one can observe a dip near $t=l/2$. Here we give more details on this feature.

\begin{figure}[h]
\centering
\begin{tikzpicture}

    \node[inner sep=0pt] (russell) at (15pt,-85pt)
    {\includegraphics[width=.26\textwidth]{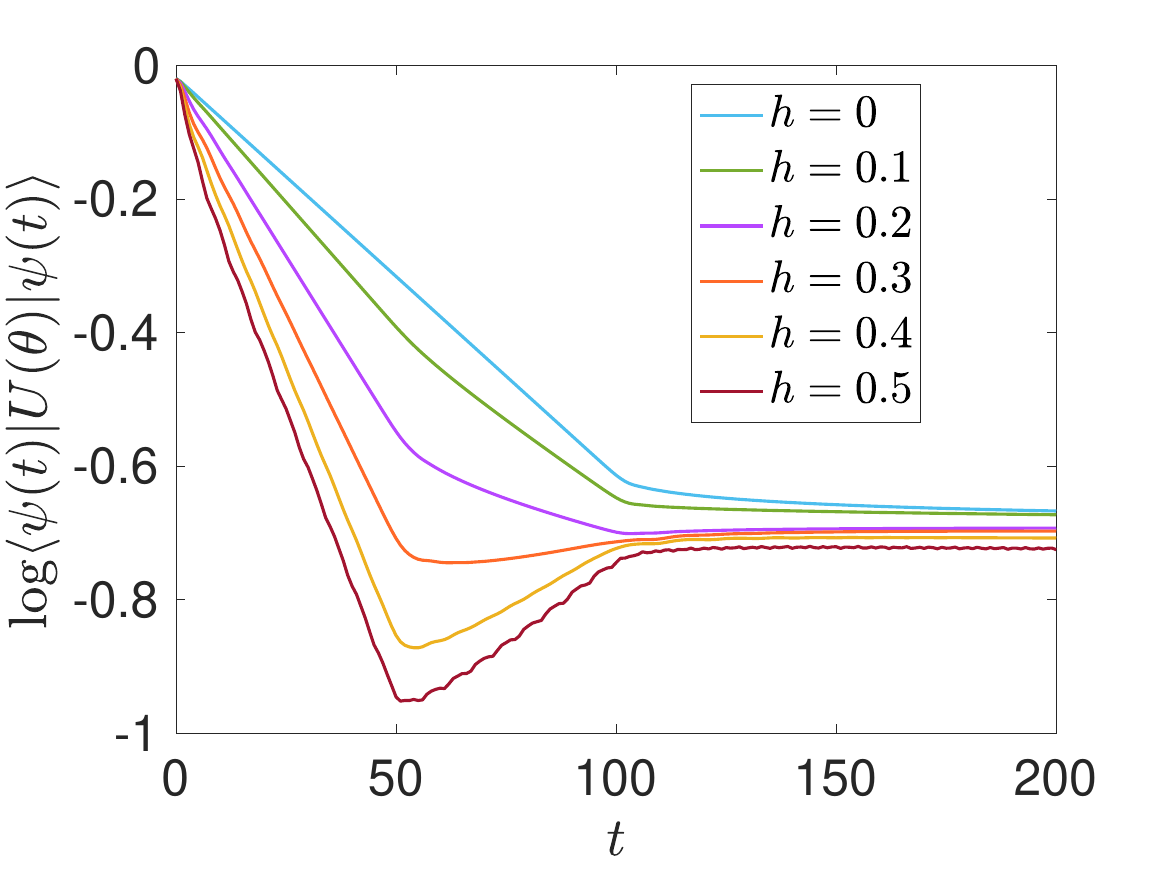}};
       
    \node[inner sep=0pt] (russell) at (140pt,-85pt)
    {\includegraphics[width=.26\textwidth]{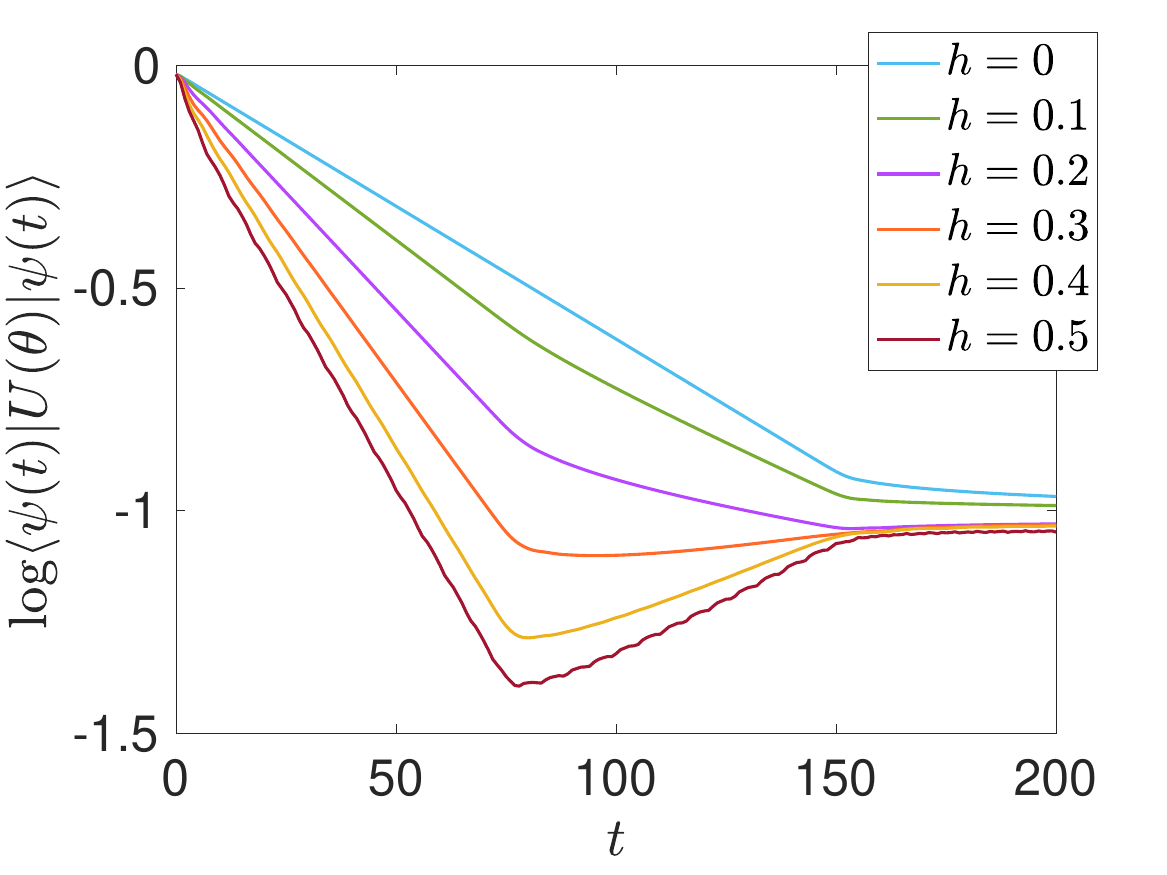}};

             \node at (-20pt, -110pt){(a)};
            \node at (110pt, -110pt){(b)};

    \end{tikzpicture}
\caption{Lattice calculation of type-II string order parameters in the interval $[0,\, l]$ at the end of a finite system defined over $[0,\,L]$ after a global quench. Here we take $L=400$, and $m=0.2$ in the initial state in \eqref{H0_globalQuench}.
We consider different strengths $h$ in the boundary symmetry breaking term in \eqref{H1_Global_semi}.
The subsystem size is chosen as (a) $l=100$, and (b) $l=150$.
We fix $\theta=\pi/6$ in the operator $U(\theta)$.
}
\label{Fig:TypeII_changeH}
\end{figure}

The lattice results are shown in Fig.\ref{Fig:TypeII_changeH}, where we consider two different subsystem sizes $l$.
The first feature one can observe is that the ``dip'' feature near $t=l/2$ is not universal. As one changes the magnitude of the boundary symmetry-breaking term, the dip feature could vanish. In particular, for $h=0$, which corresponds to the symmetry preserving boundary, the lattice result agrees with the CFT results in \eqref{Decay_typeII_Global2}. For nonzero $h$, which correspond to a symmetry-breaking boundary, the universal feature one can observe is that the type-II string order parameters decay exponentially for $t\in[0,\,l/2]$ and saturate for $t>l$. For the feature in $t\in[l/2,l]$, as we remarked in the main text, currently we do not have a good understanding from the CFT point of view. We hope to come back to this feature in the future.

\section{Details on the lattice calculation}
\label{Appendix:Lattice}

In this appendix, we give details on the numerical calculations of string order parameters in the free fermion lattice models.

Let us consider the general case of quantum quenches. 
That is, we prepare the initial state as the ground state of $H_0$, and evolve it with the quenched Hamiltonian $H_1$, where $[H_0,\,H_1]\neq 0$.
For example, $H_0$ corresponds to \eqref{H0_globalQuench} in the global quench and \eqref{H0_local_quench} in the local quench.
$H_1$ correspond to \eqref{H1_Global_semi} or \eqref{Global_defect_middle}, depending on where we define the symmetry-breaking term. 
In the following, we consider two slightly different approaches to evaluate the string order parameters, which give the same result.

\subsection{The string order parameter as an exponent of fermion bilinears}

To compute the time evolution with boundary symmetry-breaking terms, we use the Bogoliubov transformation 
\be
\label{c_f}
c_{j} =  \sum_{m}U_{jm}f_{m} + V_{jm}^{\ast}f_{m}^{\dagger}, \,\, f_{m} =  \sum_{i}(U^{\dagger})_{mi}c_{i} + (V^{\dagger})_{mi}c_{i}^{\dagger} .
\ee
to diagonalize the driven Hamiltonian $H_1$. 
Based on this we can rewrite $H_{1}$ as 
\be 
H_{1} =  \sum_{m=1}^{L}\varepsilon^{1}_{m}(f^{\dagger}_{m}f_{m} - f_{m}f^{\dagger}_{m})  = \sum_{m=1}^{L}2\varepsilon^{1}_{m}f^{\dagger}_{m}f_{m} - E_{1},
\ee 
where $E_{1} = \sum\limits_{m=1}^{L}\varepsilon^{1}_{m}$ and $\varepsilon^{1}_{m} > 0$, $\forall m$ .

The string operator we consider is of the form
\be
U(\theta) = e^{i\theta\sum_{j\in A}c^{\dagger}_{j}c_{j}}, \quad \theta\in[0,2\pi),
\ee
which corresponds to the type-I string operator if $A$ is chosen as the whole system and the type-II string operator if $A$ is a subsystem.

After the quantum quench, the string order parameter evovles in time as
\be
\label{SOP_appendix_1}
\langle\psi(t)|U(\theta)|\psi(t)\rangle = \langle\psi_{0}|e^{iH_{1}t}e^{i\theta\sum_{j\in A}c^{\dagger}_{j}c_{j}}e^{-iH_{1}t}|\psi_{0}\rangle,
\ee
where $|\psi_0\rangle$ is the ground state of $H_0$.
To evaluate \eqref{SOP_appendix_1}, let us first consider the time evolution of $c_j$ ($c_j^\dag$). By considering 
\be 
\begin{split}
 c_{j}(t) & = e^{iH_{1}t}c_{j}e^{-iH_{1}t} \\
        & = e^{iH_{1}t}\left(\sum_{m}U_{jm}f_{m} + V^{\ast}_{jm}f_{m}^{\dagger}\right)e^{-iH_{1}t}
\end{split}
\ee
and 
\be
e^{iH_{1}t}f_{m}e^{-iH_{1}t} = e^{{-i2\varepsilon_{m}t}}f_{m}, \quad e^{iH_{1}t}f_{m}^{\dagger}e^{-iH_{1}t} = e^{{i2\varepsilon_{m}t}}f_{m}^{\dagger}
\ee
we can obtain
\be
\label{c_time}
 c_{j}(t) = \sum_{m}U_{jm}e^{{-i2\varepsilon^{1}_{m}t}}f_{m} + V_{jm}^{\ast}e^{{i2\varepsilon^{1}_{m}t}}f_{m}^{\dagger}.
\ee
Based on \eqref{c_f} and \eqref{c_time}, one can further write $c_j(t)$ as
\be
c_{j}(t) = \sum\limits_{i}A_{ji}(t)c_{i} + B_{ji}(t)c_{i}^{\dagger},
\ee
where we have defined
\be
\begin{split}
A_{ji}(t) =& \sum\limits_{m}U_{jm}e^{{-i2\varepsilon^{1}_{m}t}}(U^{\dagger})_{mi} + V_{jm}^{\ast}e^{{i2\varepsilon^{1}_{m}t}}V_{mi}^{T},
\\
B_{ji}(t) =& \sum\limits_{m}U_{jm}e^{{-i2\varepsilon^{1}_{m}t}}(V^{\dagger})_{mi} + V_{jm}^{\ast}e^{{i2\varepsilon^{1} _{m}t}}U_{mi}^{T}.
\end{split}
\ee

Thus, we can rewrite string order parameter evolution in \eqref{SOP_appendix_1} as
\be
\label{SOP_appendix_2}
\langle\psi(t)|U(\theta)|\psi(t)\rangle = \langle\psi_{0}|e^{i\theta\sum_{j\in A}c_{j}^{\dagger}(t)c_{j}(t)}|\psi_{0}\rangle.
\ee
It is convenient to switch to the Majorana fermion basis by defining
\be
c_{k} = \frac{(a_{2k-1}-ia_{2k})}{2}, \quad c^{\dagger}_{k} = \frac{(a_{2k-1}+ia_{2k})}{2},
\ee
where $\{a_i, \,a_j\}=2\delta_{ij}$,
and transform $c_{j}^{\dagger}(t)$ and $c_{j}(t)$ to the Majorana fermion basis 
\be
c_{j}(t) = \sum_{m}M_{jm}(t)\, a_{2m-1} + N_{jm}(t)\, a_{2m},
\ee
where
$M_{jm} = \frac{A_{jm}(t)+B_{jm}(t)}{2}$ and
$N_{jm} = \frac{i(B_{jm}(t)-A_{jm}(t))}{2}$.
Furthermore, let us write $c_j(t)$ as:
\be
c_{j}(t) = \sum_{m}Q_{jm}(t)\,a_{m},
\ee
where 
\be
\label{Q_matrix_appendix}
\begin{split}
   &Q_{jm} = M_{j,\frac{m+1}{2}}, \quad m \in \text{odd},\\
   &Q_{jm} = N_{j,\frac{m}{2}}, \quad  m \in \text{even}.
\end{split}
\ee
Then the string order parameter in \eqref{SOP_appendix_2} can be written as
\be
\label{String_Majorana}
\begin{split}
\left\langle e^{i\theta\sum_{j\in A} c_{j}^{\dagger}(t)c_{j}(t)} \right \rangle
=&
e^{\frac{i\theta L_{A}}{2}}
\Big\langle
e^{i\frac{\theta}{2}\sum_{j\in A} c_{j}^{\dagger}(t)c_{j}(t)
-c_{j}(t) c_{j}^{\dagger}(t)
} 
\Big\rangle\\
=&
e^{\frac{i\theta L_{A}}{2}}  \Big \langle e^{ \sum_{i,k} D_{ik}a_{i}a_{k}} \Big \rangle.
\end{split}
\ee
The expectation value in \eqref{String_Majorana} in a thermal ensemble at finite temperature $1/\beta$ has been studied in 
\cite{2014_fermion_Klich}:
\be 
\label{exp_D}
\Big \langle e^{\sum_{i,k}D_{ik}a_{i}a_{k} } \Big \rangle = {\sqrt{\textnormal{det}[n_{\beta}+(1-n_{\beta})e^{4D}]}}
\ee
where 
\be 
D = \frac{i\theta}{2} \begin{bmatrix}
    Q^{\dagger} & Q^{T}
\end{bmatrix} \begin{bmatrix}
\mathbb{I}_{A} & O \\
O & -\mathbb{I}_{A} \\
\end{bmatrix}
\begin{bmatrix}
    Q\\
    Q^{\ast}
\end{bmatrix},
\ee 
with $\mathbb I_A$ being the $L$ by $L$ identity matrix with first $L_{A}$ entries as 1 and rest are 0 and $Q$ defined in \eqref{Q_matrix_appendix}, and
\be
n_{\beta} = \frac{1}{1+e^{4\beta \tilde{H}_{0}}}.
\ee
In the Majorana fermion basis, $H_{0} = \sum_{i,j}  (\tilde{H}_{0})_{ij} a_{i}a_{j}$. $\tilde{H}_{0}$ is a purely imaginary anti-symmetric matrix, $\tilde{H}_{0} =i \, O \tilde{E}_{0} O^{T}$ by orthogonal diagonalization, with
\be
\tilde{E}_{0} = \left(\begin{matrix}
0 & \epsilon_{01} \\ -\epsilon_{01} & 0 \\ 
 &  & 0 & \epsilon_{02} \\  &  & -\epsilon_{02} & 0 \\ 
& & & &   \ddots
\end{matrix}\right)  
\ee
$\epsilon_{0i}$ is an eigenvalue of $H_{0}$. In the ground state of $H_0$ ($\beta \to \infty$), one has
\be 
\lim_{\beta\to\infty}n_{\beta} = O \frac{\mathbb{I}_L \otimes (\mathbb{I}_{2} + \sigma_{y})}{2} O^{T}.
\ee
Then, one can obtain the expression in \eqref{exp_D} and therefore the time evolution of string order parameters.

\subsection{The string order parameter as a product of two Gaussian density matrices}

Now we consider a slightly different approach to calculate the string order parameters by using considering the problem of computing string order parameter is equivalent to compute the trace of the product of two (normalized) Gaussian density matrices by considering~\cite{Groha_2018}
\be
\begin{aligned}
\langle e^{i \theta \sum_{i = 1}^l c_i^\dagger c_i} \rangle
= \Tr \left( \rho_A e^{i \theta \sum_{i = 1}^l c_i^\dagger c_i} \right)  
\coloneq \widetilde{Z} \Tr \left( \rho_A \widetilde{\rho} \right) ,
\end{aligned}
\ee
where $l$ denotes the length of the subsystem (or the total system), and
\be
\widetilde{\rho} \coloneq \frac{1}{\widetilde{Z}} e^{i \theta \sum_{i = 1}^l c_i^\dagger c_i} , \quad 
\widetilde{Z} = \Tr e^{i \theta \sum_{i = 1}^l c_i^\dagger c_i} 
= \left( 1 + e^{i \theta} \right)^l . 
\ee
For a generic free fermionic theory, its (reduced) density matrix $\rho = \frac{1}{Z} e^{\frac{1}{4} \sum_{i,j} W_{ij} a_i a_j}$ can be fully addressed to two-point correlation function via Wick's theorem, as
\be
\tanh \frac{W}{2} = \Gamma, \quad 
\Gamma_{ij} = \langle a_i a_j \rangle - \delta_{i,j} . 
\ee

The product of several normalized Gaussian density matrix (including the trace) was studied in Ref.~\cite{Fagotti_2010} for solving entanglement entropy of disjoint regions, which gives
\be
\Tr ( \rho_1 \rho_2 ) = \sqrt{ \det \frac{1 + \Gamma_1 \Gamma_2}{2} } . 
\ee

Applying this to the string order parameter, we have
\be
\label{String_Correlation}
\begin{aligned}
\langle e^{i \theta \sum_{i = 1}^l c_i^\dagger c_i} \rangle 
& = (1 + e^{i \theta})^l \Tr ( \rho_A \widetilde{\rho} ) \\
& = \sqrt{ \det \left[ \frac{1 + e^{i \theta}}{2} (1 + \Gamma_\rho \widetilde{\Gamma}) \right] } . 
\end{aligned}
\ee
Here the matrix $\Gamma_\rho$ can be easily obtained from the correlation matrices $C$ and $F$ of the complex fermions by a basis transformation
\be
a_{2 i} = c_i^\dagger + c_i, \quad 
a_{2i - 1} = i (c_i - c_i^\dagger) . 
\ee
For $\widetilde{\Gamma}$, we have
\be
\begin{aligned}
M_{ij} & = \Tr (\widetilde{\rho} a_i a_j) 
= \frac{1}{\widetilde{Z}} \Tr \left( e^{i \theta \sum_{i=1}^l c_i^\dagger c_i} a_i a_j\right) \\ 
& = \frac{1}{\widetilde{Z}} \Tr 
\left[ \prod_{k=1}^{l} \left( 1 + (e^{i\theta} - 1) c_k^\dagger c_k \right) a_i a_j \right] \\ 
& = \frac{1}{\widetilde{Z}} \Tr \left[ \prod_{k=1}^{l} \left( 1 + (e^{i\theta} - 1) \frac{1 - i a_{2k-1} a_{2k}}{2} \right) a_i a_j \right] \\ 
& = \Tr \left[ \prod_{k=1}^{l} \left( \frac{1}{2} - \frac{i}{2} \tan \frac{\theta}{2} (i a_{2k-1} a_{2k}) \right) a_i a_j \right] .
\end{aligned}
\ee
Note that $a_i a_j$ is traceless when $i \neq j$, therefore the only non-vanishing terms of $M$ are
\be
M_{i,i} = 1, \quad 
M_{2i, 2i-1} = \tan \frac{\theta}{2} . 
\ee
Then we have 
\be
\widetilde{\Gamma} = M - \text{Id} 
= \tan \frac{\theta}{2} 
\left(\begin{matrix}
0 & -1 \\ 1 & 0 \\ 
 &  & 0 & -1 \\  &  & 1 & 0 \\ 
& & & &   \ddots
\end{matrix}\right) . 
\ee
Once we know $\Gamma_\rho$ and $\tilde \Gamma$, we can obtain the string order parameter evolution according to \eqref{String_Correlation}. 

In addition to the above discussed two approaches, we introduce a formula that is useful for evaluating the string order parameter when $U(1)$ symmetry is preserved, e,g, the type-II string order parameters. 
In this case, it is straightforward to find that\cite{2020_free_fermion}
\be
\langle \psi|e^{\sum_{i,j} P_{i,j}c_i^\dag c_j}
|\psi\rangle
=\det\left(\mathbb I+(e^P-\mathbb I) C\right),
\ee
where $C$ is the correlation matrix with elements $C_{ij}=\langle \psi|c_i^\dag c_j |\psi\rangle$.
This formula works for both equilibrium and non-equilibrium states $|\psi\rangle$ that preserves the $U(1)$ symmetry.
%

\bibliography{BoundarySymmetryBreak.bib}

\end{document}